\documentclass[twocolumn]{aastex701}

\usepackage{amsmath}
\usepackage[utf8]{inputenc}
\usepackage{xcolor}
\usepackage{cprotect}
\newcommand*{\kms}{\,km\,s$^{-1}$}

\newcommand{\nmask}{411~}   % NUMBER OF UNIQUE DEIMOS MASKS
\newcommand{\nexp}{1426~}   % NUMBER OF SCIENCE EXPOSURES

\newcommand{\nobj}{78~}   % NUMBER OF UNIQUE OBJECTS
\newcommand{\ngal}{35~}   % NUMBER OF UNIQUE gal or ufc
\newcommand{\ngc}{43~}   % NUMBER OF UNIQUE  GC
\newcommand{\nPI}{40~} % NUMBER OF UNIQUE PIs

\newcommand{\nstar}{22,339~}   % NUMBER OF Unique STARS REPORTED
\newcommand{\nexgal}{2097~}    % NUMBER OF UNIQUE exal galaxies
\newcommand{\nall}{24,436~}    % stars + galaxies

\newcommand{\nmembers}{12,404~} % Number of member stars
\newcommand{\nonmembers}{9,935~} % Number of non-member stars

\newcommand{\nrepeat}{3421~}   % NUMBER OF Unique stars with repeats
\newcommand{\npair}{1209~}   % NUMBER OF Velocity pairs used to determine k_v, floor

\shorttitle{The Keck/DEIMOS Stellar Archive: I.}
\shortauthors{Geha et al.}

\begin{document}

\title{The Keck/DEIMOS Stellar Archive: I.~Uniform Velocities and
Metallicities for \nobj Milky Way Dwarf Galaxies and Globular Clusters}

\author[0000-0002-7007-9725]{Marla~Geha}
\affiliation{Department of Astronomy, Yale University, New Haven, CT 06520, USA}
\email[show]{marla.geha@yale.edu}

\author[0000-0002-3007-0013]{Debora~Pelliccia}
\affiliation{University of California Observatories, University of California, Santa Cruz, 1156 High St., Santa Cruz, CA 95064, USA}
\email{dpelliccia@ucolick.org}

\author[0000-0002-7738-6875]{J.~Xavier~Prochaska}
\affiliation{University of California Observatories, University of California, Santa Cruz, 1156 High St., Santa Cruz, CA 95064, USA}
\affiliation{Kavli Institute for the Physics and Mathematics of the Universe, 5-1-5 Kashiwanoha Kashiwa 277-8583 Japan}
\affiliation{Division of Science, National Astronomical Observatory of Japan, 2-21-1 Osawa, Mitaka Tokyo 181-8588 Japan} 
\email{xavier@ucolick.org}

\author[0000-0003-1697-7062]{William~Cerny}
\affiliation{Department of Astronomy, Yale University, New Haven, CT 06520, USA}
\email{william.cerny@yale.edu}

\author[0000-0003-0821-3644]{Frederick B.~Davies}
\affiliation{Max-Planck-Institut f\"{u}r Astronomie, K\"{o}nigstuhl 17, D-69117 Heidelberg, Germany}
\email{davies@mpia.de}

\author[0000-0002-7054-4332]{Joseph~Hennawi}
\affiliation{University of California, Santa Barbara, CA 95064, USA}
\email{joe@physics.ucsb.edu}

\author[0000-0002-6153-3076]{Brad~Holden}
\affiliation{University of California Observatories, University of California, Santa Cruz, 1156 High St., Santa Cruz, CA 95064, USA}
\email{holden@ucolick.org }

\author{Dusty Reichwein}
\affiliation{University of California Observatories, University of California, Santa Cruz, 1156 High St., Santa Cruz, CA 95064, USA}
\email{dusty@ucolick.org}

\author[0000-0003-1809-6920]{Kyle~B.~Westfall}
\affiliation{University of California Observatories, University of California, Santa Cruz, 1156 High St., Santa Cruz, CA 95064, USA}
\email{westfall@ucolick.org}

\begin{abstract}

We present a homogeneous spectroscopic dataset of \nstar  stars in \nobj Milky Way dwarf galaxy satellites and globular clusters.  All data were taken with the Keck II telescope and Deep Extragalactic Imaging Multiobject
Spectrograph (DEIMOS) spectrograph using the 1200G grating  (spectral resolution R$\simeq$6000).  Based on a uniform data reduction of \nmask DEIMOS masks, we present a catalog of individual stellar radial velocities, equivalent width-based [Fe/H] metallicities, and membership estimates.  The Milky Way satellites range from $M_V = 2$ to $-14$ ($M_{\star} = 10^{1.5}$ to $10^{7.5}\,M_{\odot}$);  the majority of individual stars presented in these systems have magnitudes between $17 > r > 22$.   We additionally present redshifts for \nexgal compact background galaxies and QSOs in the same magnitude range.   The data were reduced to 1D spectra using {\tt PypeIt}, which provides near Poisson statistics-level sky subtraction.  Radial velocities were determined via {\tt dmost}, a forward modeling method first presented here, which combines both synthetic telluric and stellar templates to determine stellar radial velocities.   We assess the accuracy and precision our method via comparison to thousands of repeat measurements and literature values.  We determine a velocity error floor of 1.1\kms\ and a Ca\,II triplet-based metallicity error floor of 0.1\,dex.   We calculate internal velocity dispersions and compare to literature values, demonstrating 20-50\% improved precision over the literature in most cases.  In a companion paper, we use our homogeneous catalogs to explore properties of these Milky Way satellites, including previously unpublished measurements in several systems including Bo\"otes\,II and Draco\,II.  We provide full access to the data catalogs to enable further studies.
\end{abstract}

\keywords{Dwarf galaxies --- Globular Clusters --- Stellar kinematics}

\section{Introduction}\label{sec_intro}
 The Milky Way's dwarf galaxy and globular cluster satellites include, by virtue of proximity, the faintest and lowest mass stellar systems in the known Universe.   These satellites are important objects to test models of galaxy formation and cosmology \citep[for reviews see][]{bullock2017,simon2019, review2022}.  They are compelling targets for understanding the nature of dark matter \citep[e.g.,][]{nadler2021,nadler2024A,Acharyya2024,Ando2025,may2025} and proposed modifications to gravity \citep[e.g.,][]{Haghi2016}.    Orbital histories are crucial for discerning between internal and external processes which have shaped the present satellites.   Proper motion measurements from the Gaia Data Release 3 \citep[DR3; ][]{GaiaDR3}, combined with radial velocity data, allow full orbital solutions for a large fraction of Milky Way satellites \citep[e.g.,][]{Fritz2018,simon2018a,pace2022}. 

The tests above depend heavily on accurate kinematic measurements of individual stars.    While proper motion-only kinematic data have been used to determine dynamical masses of the nearest Milky Way globular clusters \citep[e.g.,][]{Watkins2015,sollima2019}, observational errors typically limit proper motion-based internal kinematics to distances less than $\sim 10$\,kpc (although see \citealt{vitral2024,vitral2025}).   Particularly for the lowest luminosity stellar systems in the Milky Way, radial velocities are currently the only way to obtain a sufficiently large number of stars to determine dynamical masses.  The internal dispersion of the lowest luminosity satellites are comparable to typical velocity errors for individual stars in the current literature ($\sim1-2$\kms).  Thus, small improvements to individual stellar velocity measurements can result in substantial improvements on the above constraints.  Of particular concern is the effect of unresolved binary stars systems, inflating the dynamical masses of Milky Way satellites based on single-epoch radial velocity data \citep{McConnachie2010, Martinez2011, pianta2022,graton2025}. 

%%%%%%%%%%%%%%%%%%%%%%%%%%%%%%%%%%
% Figure:  ALL SKY
\begin{figure*}[htb!]
\centering
 \includegraphics[width=1.0\textwidth]{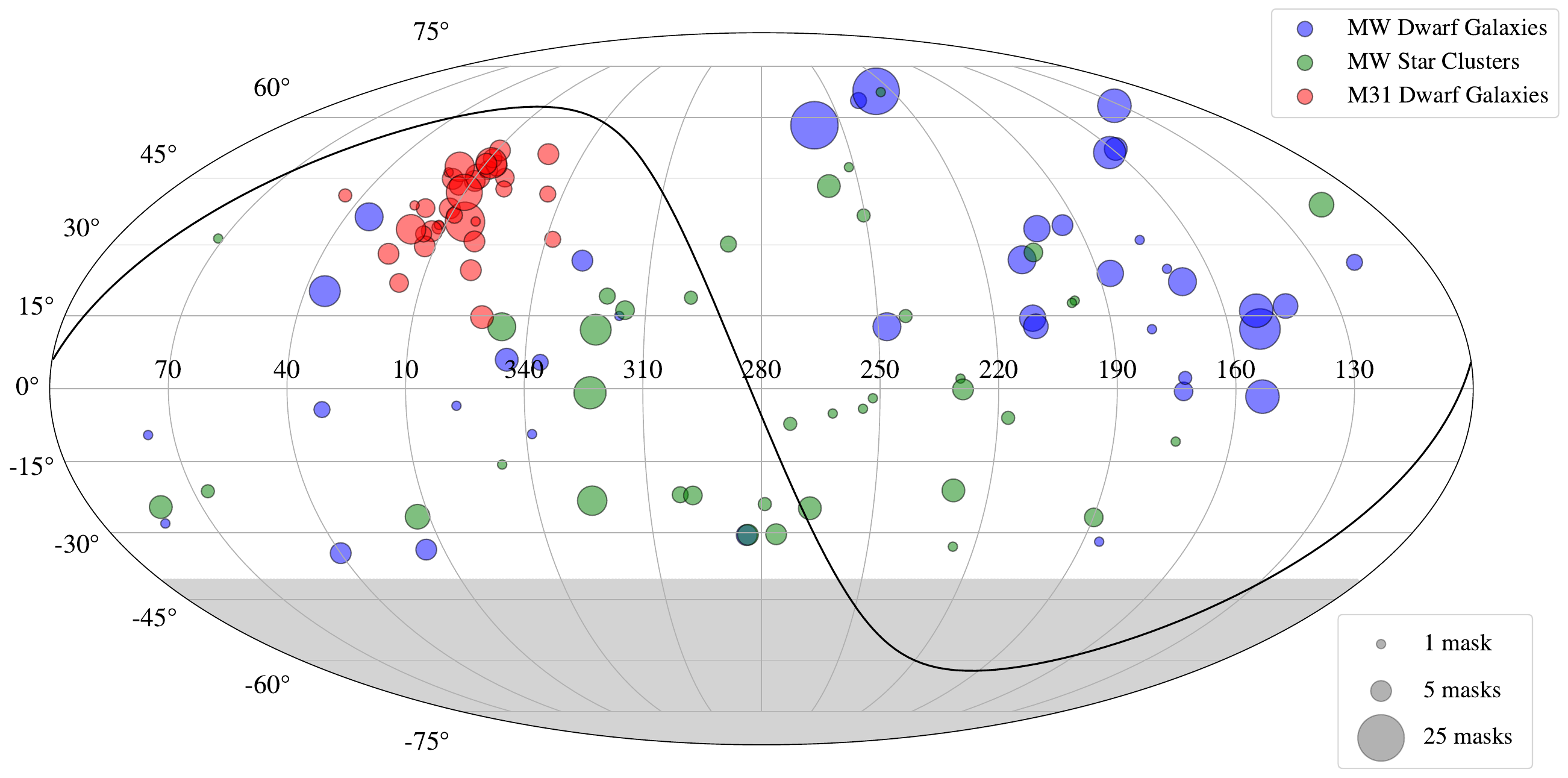}
\caption{Spatial distribution (RA/Dec) of Milky Way (MW) dwarf satellite galaxies (blue), globular and faint star clusters (green), and M31 dwarf galaxy satellites (red) observed with the Keck/DEIMOS instrument and 1200G grating.  The circle size represents the number of 1200G Keck/DEIMOS archival pointings (unique multiobject masks) for each object, ranging between 1 and 26.   The median number of DEIMOS masks per object is three.  The black solid line is the Galactic plane; the gray shaded area is the region of sky not visible from the Keck telescope latitude.   Spectra extracted from all MW systems are presented in this paper. M\,31 will be presented in a future contribution. \label{fig_intro}}
\end{figure*}
%%%%%%%%%%%%%%%%%%%%%%%%%%%%%%%%%%

Many teams have contributed to the measurements of radial velocity data in Milky Way satellites.   Starting with a sample of just three stars in the Draco dwarf spheroidal (dSph) galaxy \citep{Aaronson1983}, early work in the Milky Way's dwarf spheroidal galaxies used single-slit spectroscopy to build up samples of a few tens of stars \citep[for a review see][]{Mateo1998}.   Larger samples of stars were possible in nearby globular clusters \citep{Gunn1979, Armandroff1988}, however, crowding issues limit the number of individual stellar radial velocities particularly toward cluster centers.   Development of multislit and multifiber spectrographs significantly increased the number of radial velocities available in the Milky Way satellites \citep[e.g.,][]{Tolstoy2004,Helmi2006,simon07a,Baumgardt2018, li2019, Cooper2023,walker2023}.  At the same time, the number of Milky Way satellites has been increasing.  The Sloan Digital Sky Survey (SDSS) doubled the number of known Milky Way dwarf galaxies \citep[beginning with][]{willman2005}, and deeper all-sky imaging surveys such as the Dark Energy Survey (DES), DELVE, and UNIONS have continued this trend \citep[e.g.,][]{Bechtol2015,koposov2015,cerny2023,Tan2025,smith2025}, which will accelerate with anticipated discoveries from the Euclid and Roman Space Telescopes, and the Rubin Observatory \citep{Tsiane2025}.   However, to confirm the nature of these systems requires deep spectroscopy of many tens to hundreds of stars. This is particularly true of the faintest systems ($M_V >  -3 $), termed the \textquotedblleft ultra-faint compact systems", where photometric properties appear to overlap with faint star clusters.
 
The Deep Extragalactic Imaging Multi-Object Spectrograph \citep[DEIMOS; ][]{faber03a} on the Keck-II 10-meter telescope provided some of the first evidence that the Milky Way's ultra-faint dwarf galaxies were indeed dark matter-dominated galaxies \citep{simon07a, Martin2007}, and has continued to provide kinematic measurements for newly discovered Local Group satellites \citep[e.g.,][]{smith2024,cerny2024} and improved measurements for known systems \citep[e.g.,][]{pace2020,buttry2022,ou2024}.  The Keck/DEIMOS spectrograph is one of a handful of instruments able to measure resolved star kinematic data in Milky Way satellites.  However, these DEIMOS data have been taken by multiple groups and reduced with different methods \citep[e.g.,][]{simon07a,Martin2007, geha2009, kirby10a,willman2011,kim2016, collins2017,Longeard2020, 2023A&A...672A.131A,Kvasova2024, Tan2025}.    The Keck Observatory Archive (KOA) has opened access to over 20 years of archival DEIMOS data including hundreds of pointings toward Milky Way dwarf galaxies and globular clusters.   Homogeneously reducing the full Keck/DEIMOS archive, including accurate and uniform radial velocities, metallicities, and membership probabilities, will optimize these faint systems as probes of both dark matter and the extreme low-mass end of galaxy formation.

In this paper, we present a catalog of \nstar stars over \nobj Milky Way  satellite systems as observed in \nmask unique Keck/DEIMOS slitmasks.  The data were all taken with the Keck/DEIMOS spectrograph and 1200G grating.   In \S\,\ref{sec_data} we describe the archival dataset and 2D data reduction using {\tt PypeIt}.  In \S\,\ref{sec_1D}  we present {\tt dmost}, a forward modeling method to determine line-of-sight velocities for slit spectrographs, and use this method in \S\,\ref{sec:rvs} to compute radial velocities.  In \S\,\ref{sec:errors_combine}, we take advantage of this large dataset to evaluate the accuracy of our velocity errors, critical to inferring physical properties of Milky Way satellites. In \S\,\ref{sec:validation} we validate our velocity measurements against larger spectroscopic surveys and evaluate velocity variability using multi-epoch observations where possible.   In \S\,\ref{sec: ew}, we determine equivalent width (EW)-based metallicities, determining accurate errors and validating against literature values.   In \S\,\ref{sec_membership} we evaluate membership to a given Milky Way satellite.    Future work will include similar analysis for M\,31 satellite galaxies.  In \S\,\ref{sec_catalogs} we describe the catalog quantities provided as tables in this work, and demonstrate improved internal velocity dispersion precision in MW dwarf satellites.   In a companion paper, we use this database to determine and explore dynamical masses, mean metallicities and other physical properties of the Milky Way's satellite system \citep[Paper II;][]{geha_paper2}.

\section{Data and 2D Data Reduction}\label{sec_data}

Our primary goal is to provide homogeneous velocity measurements of individual
stars in all of the Milky Way satellites observed with the Keck~II
DEIMOS spectrograph.  The DEIMOS spectrograph is well matched to the properties of
Milky Way satellites.   Its highest resolution (1200\,l/mm) red-optimized
grating (1200G) provides velocity errors on a single measurement of
 1-2\,\kms, similar to the expected internal velocity
dispersions of faint Milky Way satellites.   In \S\,\ref{ssec:koa}, we outline the dataset analyzed in this work.  \S\,\ref{ssec:deimos_setup} describes the DEIMOS spectroscopic setup, and \S\,\ref{ssec:pypeit} outlines the 2D data reduction steps.

%%%%%%%%%%%%%%%%%%%%%%%%%%%%%%%%%%
% Figure:  sky lines
\begin{figure*}[hbt!]
\centering
 \includegraphics[width=1.0\textwidth]{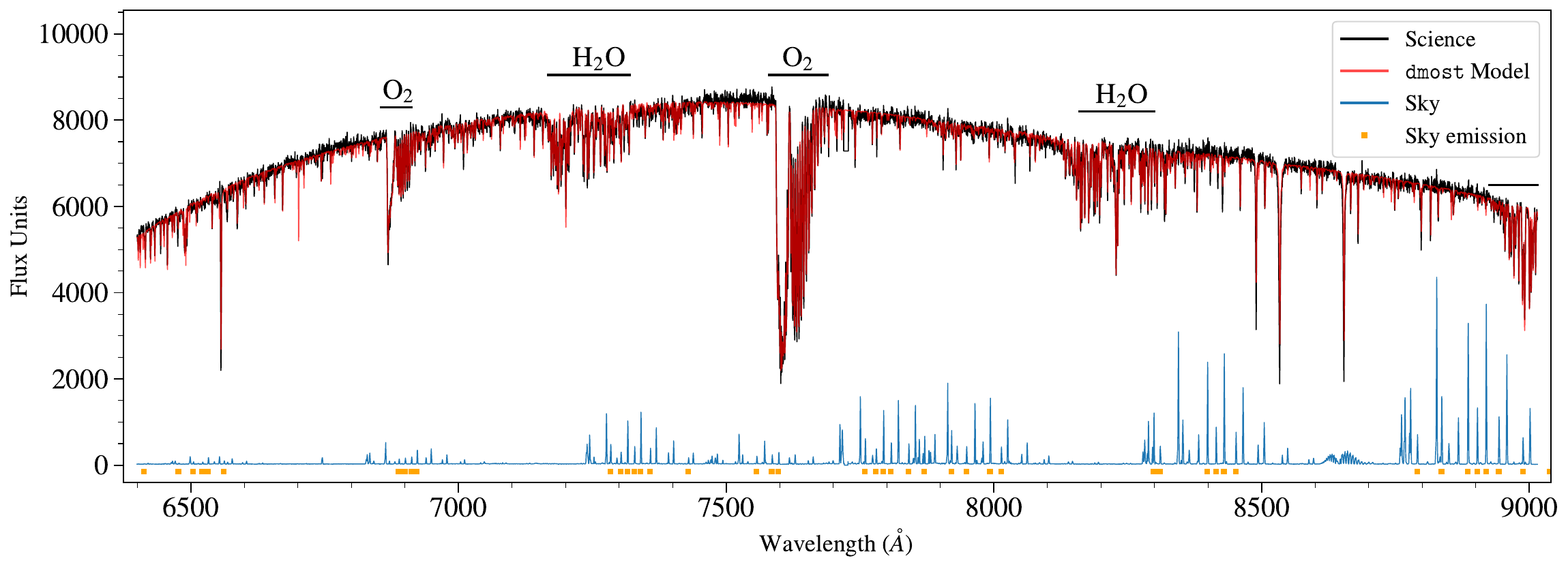}
\caption{An example DEIMOS 1200G science spectrum (black) for a high S/N star in the Draco dSph galaxy ($r=16.5$, $t_{\rm exp} = 1200$\,s, S/N$\sim$80).  We plot the associated sky emission spectrum (blue), scaled down for plotting purposes.  Orange squares indicate relatively isolated sky emission lines used in the flexure fitting (\S\,\ref{ssec:flexure}).   The wavelength region used to fit telluric absorption quantities and the primary telluric absorption species are indicated above each spectral region.   The red line is the best-fitting {\tt dmost} forward model spectrum created by combining a synthetic stellar and synthetic telluric absorption spectrum.   \label{fig:spectrum}}
\end{figure*}
%%%%%%%%%%%%%%%%%%%%%%%%%%%%%%%%%%

\subsection{The Keck Observatory Archive (KOA)}\label{ssec:koa}

The Keck Observatory Archive\footnote{https://koa.ipac.caltech.edu}\,(KOA) provides access to archival data from all Keck instrumentation via an online portal.  Data are publicly available 18 months after observations.  With the goal of improved stellar velocity measurements, we focus only on the DEIMOS 1200G grating, which provides the highest spectral resolution and represents roughly one-third of all data taken with DEIMOS.   We search the KOA archives for DEIMOS observations taken with the 1200G grating in multiobject mode with an exposure time greater than 60\,seconds.   We analyze all Milky Way systems visible at Mauna Kea (${\rm Dec} > -40^{\circ}$; Figure~\ref{fig_intro}) including data taken within three effective radii of an object's center for Milky Way dwarf galaxies and faint star clusters from the Local Volume Database \citep{pace2024}, and globular clusters from \citet{Baumgardt2019}.    We find \nmask unique DEIMOS masks covering \nobj objects (\ngal galaxies and \ngc globular clusters or unclassified systems).  Each  mask is associated with a unique Milky Way system, with the exception of five masks which include both NGC\,6715 (M54) and the Sagittarius (Sgr) dSph (see \S\,\ref{ssec:matching}).   The data were taken over two decades between 2003 and 2023 by \nPI unique Principal Investigators (PIs).  Raw data files, both calibration and science frames, were downloaded directly from the KOA.  We list the Milky Way systems with DEIMOS data reduced in this work in Table~\ref{table_objects} and details for each mask in Table~\ref{table_obs}.

Data presented in this paper were taken for a variety of science purposes.   The targeting prioritizations and photometry used to target stars vary between datasets.   Generally, spectroscopic targeting was done via isochrone fitting, prioritizing stars near a single isochrone chosen to match the expected or known properties of a given system \citep[for example, see \S\,2.1 of ][]{simon07a}.   With rare exception, the stellar samples are not complete at any radius or magnitude.  We provide measurements for all targeted objects, including nonmembers and extragalactic sources (\S\,\ref{sec_membership}).

\subsection{DEIMOS Spectroscopic Setup}\label{ssec:deimos_setup}

The Keck/DEIMOS spectrograph was commissioned in 2002 \citep{faber03a}
and continues to be a primary facility instrument at the Keck Observatory.  DEIMOS's multiobject mode allows for nearly 200 objects to be observed simultaneously over a $4' \times 16'$ field of view.   To maximize homogeneity, we analyze only data taken with the DEIMOS 1200G grating.  The 1200G grating provides a spectral resolving power of R $\sim 6000$ at $8500\mbox{\AA}$, assuming a $0.7''$ slitwidth.  The spectral dispersion of this setup is roughly $0.33\,\mbox{\AA}$ per pixel with a resulting spectral resolution of $\sim 1.37\,\mbox{\AA}$ (FWHM).  The central wavelength, and thus wavelength coverage, varies between datasets, but includes $6600-8800\,\mbox{\AA}$ (Figure~\ref{fig:spectrum}). The spatial scale is $0.12''$~per pixel.  The majority of the data use $0.7''$ slitwidths, although a few masks were observed with slitwidths ranging from $0.75''$ to $1.0''$.   The majority of data were taken with the order-blocking OG550 filter with a cutoff wavelength blueward of $5500\,\mbox{\AA}$, while a small number of masks used a filter with a slightly bluer cutoff wavelength (GG455 or GG495).  

We consider only DEIMOS multislit mask data for reduction (i.e., no long-slit data).  We require individual science exposure times of at least 60\,seconds and slit lengths longer than $4''$.   The first criterion ensures sufficiently bright sky emission lines required for calibration (\S\,\ref{ssec:flexure}); the second criterion allows for accurate local sky subtraction within each slit.   Although calibration frames are rarely missing, a given mask is considered for reduction only if all associated calibration data are available:  three or more quartz lamp exposures for flat fielding and at least one arc lamp (Ne, Ar, Kr, and Xe) for wavelength calibration.  Calibrations are usually taken in the afternoon and are separated in time from the science exposures by several hours.  We do not include bias or dark frames, instead using the overscan region to subtract the bias.   Information for each DEIMOS mask is provided in Table~\ref{table_obs}.

\subsection{2D Data Reduction with {\tt PypeIt}}\label{ssec:pypeit}

Data were reduced to 1D wavelength-calibrated spectra using the open-source Python-based data reduction code {\tt PypeIt} \citep{pypeit2020}.   {\tt PypeIt} is a generalized spectral reduction code for both echelle and lower resolution spectroscopy which operates on a growing number of instruments.  {\tt PypeIt} performs flat fielding, wavelength calibration, 2D sky subtraction, object detection and extraction.    In this work, we use {\tt PypeIt} v.1.10.   

{\tt PypeIt} reduces the eight individual DEIMOS detectors as four mosaic images, where each red-blue pair of detectors is reduced together.   The blue detector image is interpolated into the pixel coordinates of the red detector, including shifts in the spatial and spectral directions, using a fifth-order affine transformation.   This interpolation slightly increases the size of cosmic rays and other detector artifacts in the blue detector resulting in larger masked areas as compared to similar features on the red detector.  This is compensated by improvement in the wavelength solution gained by simultaneously solving across both detectors. 

%%%%%%%%%%%%%%%%%%%%%%%%%%%%%%%%%%
% Figure:  sky lines
\begin{figure*}[ht!]
\centering
 \includegraphics[width=1.0\textwidth]{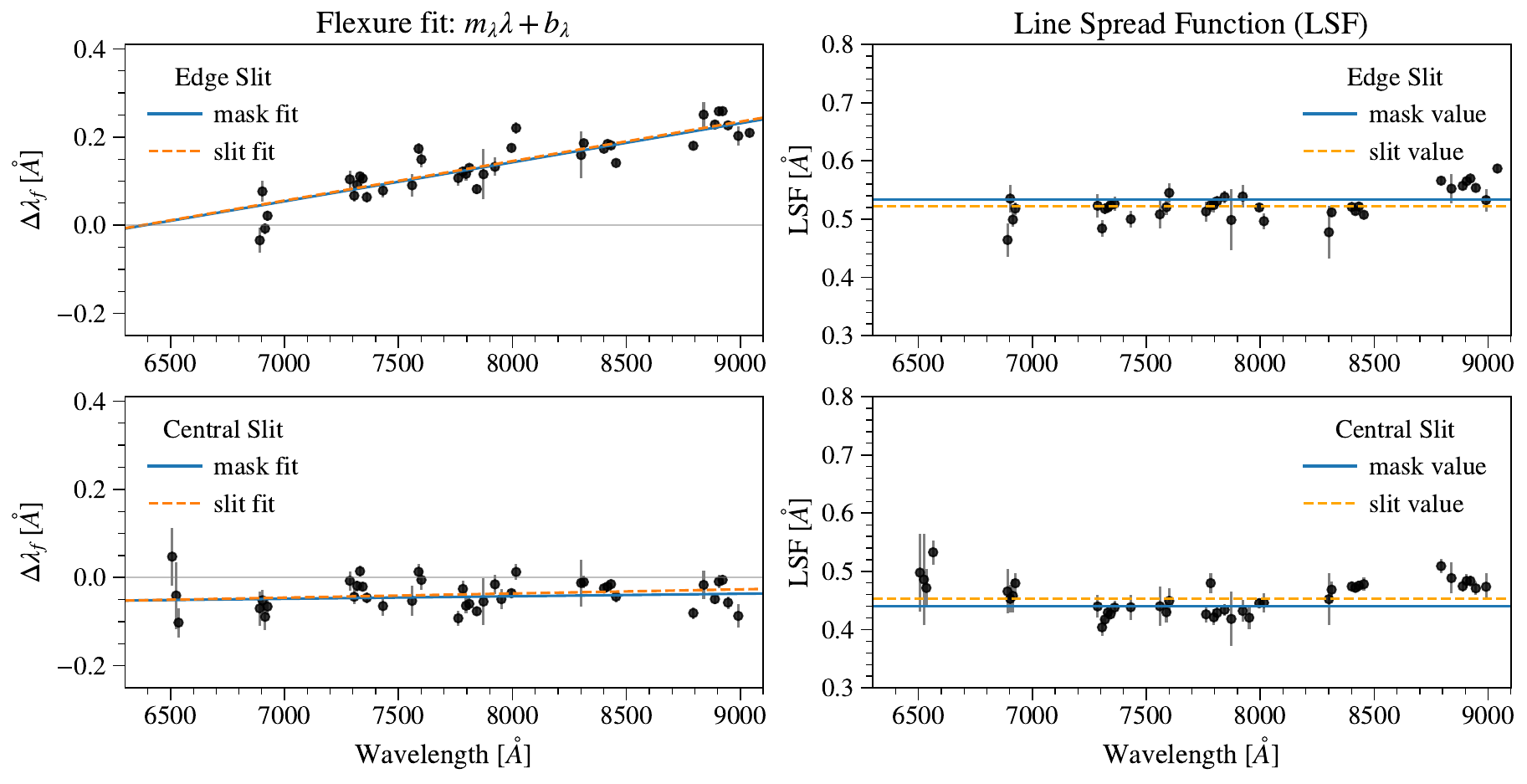}
\vskip -0.2cm
\caption{To correct for wavelength flexure between the afternoon arc calibrations and evening science exposures, we fit Gaussian line profiles to a set of isolated sky emission lines across all slits. {\it Left:\/} The difference between the predicted and observe sky emission line center ($\Delta \lambda_f$) versus wavelength before flexure correction for a two slits on the edge ({\it top}) and center ({\it bottom}) of the mask.   Non-zero wavelength offsets of skylines in the science exposure are corrected during the flexure procedure. {\it Right:\/} Gaussian line width as a function of wavelength.  Although the LSF depends slightly on wavelength, we opt to use the average value as the LSF for each slit.\label{fig:skylines}}
\end{figure*}
%%%%%%%%%%%%%%%%%%%%%%%%%%%%%%%%%%

Wavelength calibration is a critical step in determining accurate stellar velocities.   We use the {\tt PypeIt} \textquotedblleft full template" wavelength calibration algorithm.   To obtain a rough estimate of the wavelength range, {\tt PypeIt} first performs a cross correlation between the observed arc spectrum and an archived calibrated arc spectrum created for the 1200G  grating.  Depending on the wavelength coverage, typically 40-50 arc lines are identified and a sixth-order Legendre  polynomial is fit to obtain the final wavelength solution for each slit.  Across the masks presented in this paper, the median RMS for the wavelength solution is 0.06\,pixels (0.02\mbox{\AA}).  At the central wavelength of 7700\AA, this RMS corresponds to a velocity error of 0.75\kms.  The RMS in the wavelength solution is the primary contribution to the velocity error floor determined in \S\,\ref{sec:errors_combine}.   We remove a handful of slits with failed wavelength solutions, defined as RMS wavelength residuals greater than 0.4\,pixels.  We apply an additional wavelength solution correction to account for flexure between the afternoon arc frame and evening science exposure as described in \S\,\ref{ssec:flexure}.

For each slit, {\tt PypeIt} iterates sky subtraction and object detection twice.   A preliminary sky subtraction is first performed using all pixels in a slit to determine and subtract the sky spectrum.  This sky-subtracted spectrum is collapsed along the wavelength axis and a preliminary object detection is run with a collapsed S/N threshold of 10.  The object FWHM sets the masking width around each object.  A second sky subtraction is performed with object pixels masked and object detection is rerun on the second sky-subtracted image.   Objects are traced and extracted using both boxcar and optimal Horne algorithms;  we use the optimal extracted spectra throughout this paper.   The final sky subtraction is performed after the second object detection is performed.

{\tt PypeIt} uses first principle noise modeling to properly account for sources of uncertainties \citep{pypeit2020}.   To confirm that {\tt PypeIt} achieves Poisson-level statistics in the sky-subtracted 2D science spectra, we examine the distribution of sky-subtracted nonobject pixels divided by the associated pixel errors. If the noise is random and properly predicted, the observed distributions should be Gaussian with an RMS sigma of unity.   For {\tt PypeIt}, we estimate this value separately for each individual slit.  For all slits across all masks, we find the noise is Gaussian distributed with a mean RMS sigma of 0.94.  This is slightly lower than unity and is due to interpolated noise values in the blue detector, which are slightly underestimated.   For comparison, this value in the DEEP2 {\tt spec2d} DEIMOS pipeline was 0.86 as reported by \citet[][see their \S\,10.2]{newman2013a}. 

Spectra extracted from the DEIMOS science frames include both program targets and  serendipitous detections.   To assign sky coordinates and slitmask design identifications, {\tt PypeIt} first matches slits positions to the input DEIMOS design information.  Objects are assigned program target information if the measured distance of the detected object from the left edge of the slit is within $1''$ (5.5 pixels) of the predicted position.   If the measured distance is not within tolerance, the detected object is considered serendipitous and its sky coordinates are computed as offsets from slit center.

As an additional step, we group spectra across exposures and create two different versions of the coadded 1D spectrum.  We use one version to remove extragalactic objects prior to running the stellar velocity pipeline (\S\,\ref{ssec:marz}).  A second version is corrected for flexure in the individual exposures and used to determine stellar templates (\S\,\ref{ssec:chi2}).  In both cases,  we coadd our 1D spectra using {\tt PypeIt}'s {\tt pypeit\_collate\_1d} which groups the 1D spectra by object using its coordinates, and coadds all matching spectra by performing a weighted average on the optimally extracted individual 1D spectra.  However when possible, line-of-sight velocities are determined from the single, uncoadded exposures.

When reducing data for this paper, we turn off {\tt PypeIt}'s default heliocentric and flexure corrections.  We correct our velocity values to the heliocentric frame after full forward modeling.  For flexure, {\tt PypeIt} v.1.10 determines a single value wavelength shift for each slit.  Based on fits to sky emission lines, we find that a linear function with wavelength is a better flexure correction for DEIMOS (Figure~\ref{fig:skylines}).  We thus opt to do our own flexure fits (\S\,\ref{ssec:flexure}).  Finally, {\tt PypeIt} determines wavelength solutions for vacuum, which we transform into air wavelengths throughout the forward modeling pipeline.

%%%%%%%%%%%%%%%%%%%%%%%%%%%%%%%%%%
% Figure:  FLEXURE Correction
\begin{figure*}[ht!]
%\centering
 \includegraphics[width=1.03\textwidth]{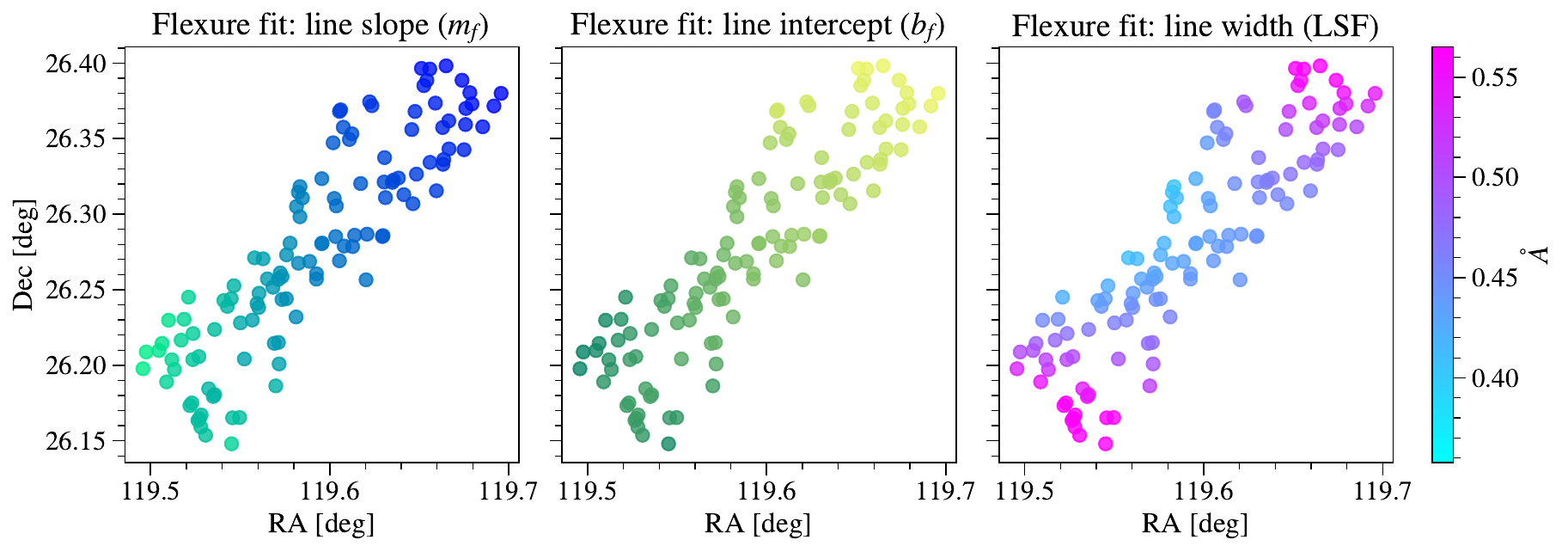}
%\vskip -0.1cm
\caption{The flexure fit values for all slits in a single DEIMOS exposure based on sky emission lines.  We plot the slope ($m_f$, {\it left}), intercept ($b_f$, {\it middle}) and average line width (LSF, {\it right}) as a function of mask position.   The pattern in the left two panels is due to flexure between the afternoon wavelength calibration and science exposure, while the rightmost panel is due to variation in image quality across the spectrograph.   \label{fig:flexure}}
\end{figure*}
%%%%%%%%%%%%%%%%%%%%%%%%%%%%%%%%%%

\section{1D Data Reduction}\label{sec_1D}

The {\tt PypeIt} pipeline reduction produces optimally extracted 1D spectra  for each slit in each science exposure with associated native wavelength solution and variance per pixel.  Before measuring line-of-sight velocities, we first determine a flexure correction (\S\,\ref{ssec:flexure}) and LSF (\S\,\ref{ssec:LSF}) for each slit, using information from the full mask.  We also identify nonstellar sources  and measure extragalactic redshifts (\S\,\ref{ssec:marz}), removing these sources from the stellar velocity pipeline. Stellar radial velocities are computed in \S\,\ref{sec:rvs}.

\subsection{Wavelength Flexure Correction}\label{ssec:flexure}

Although the DEIMOS Flexure Compensation System (FCS) maintains a relatively stable wavelength calibration throughout the night \citep{faber03a}, there are small flexure shifts below the accuracy of the FCS pixel scale ($0.2\mbox{\AA}$), which can be corrected using the sky emission lines in the science exposure.   To account for small shifts in the wavelength solution between afternoon arc calibration frames and nighttime science exposures, we determine a flexure correction to the wavelength solution for each science spectrum.   

We fit a Gaussian line profile to a set of isolated sky emission lines.  Shown as orange squares in Figure~\ref{fig:spectrum}, we chose sky lines whose peaks are separated by at least 5 times the Gaussian sigma line width from another strong sky line, but relax this criterion to 3-sigma in regions where there are fewer lines to maintain a somewhat constant distribution of sky lines with wavelength.    In the left panels of Figure~\ref{fig:skylines}, we plot the difference between the Gaussian line center and the expect sky emission line center, $\Delta \lambda_f$, as a function of wavelength for two example slits. This difference is a smooth function of wavelength  and we fit a linear function:  $\Delta \lambda_f = m_f \lambda + b_f$.   Although the slope ($m_f$) and offset ($b_f$) of this flexure fit changes significantly both across and between masks, we find that a linear fit is adequate to explain all of the DEIMOS archival data.

For a single science exposure of a given mask, we determine the flexure fitting parameters ($m_f$ and $b_f$) for all slits and plot these values as a function of slit position on the mask.  These values vary smoothly as a function of slitmask position (Figure~\ref{fig:flexure}).  To reduce noise, we fit a third-order 2D polynomial to each of these surfaces.   For any given slit, we use the 2D fits to look up the flexure fit parameters and apply this linear correction directly to the final 1D wavelength array.  

Across the masks presented in this paper, the median RMS for the flexure-corrected sky emission lines is 0.026\mbox{\AA} (1.0\kms at 7700\mbox{\AA}), comparable to the RMS in the original arc solutions presented in \S\,\ref{ssec:pypeit}.  We remove a handful of slits with sky RMS values larger than 0.4\,pixels that were not already removed by the RMS criteria for the wavelength solution.

\subsection{The Line Spread Function}\label{ssec:LSF}

In fitting the sky emission lines, we additionally measure the line broadening due to the spectrograph as a function of wavelength, which we refer to as the Line Spread Function (LSF).   We require an estimate of the LSF for our forward modeling procedure (\S\,\ref{sec:rvs}).   

In the right panels of  Figure~\ref{fig:skylines}, we plot the one-sigma Gaussian line width of the sky emission lines identified in \S\,\ref{ssec:flexure}.     The LSF varies slightly as a function of wavelength, usually increasing by a few percent at the edges of the DEIMOS field.  For ease of computation we adopt the weighted average of as a proxy for the LSF of each slit.    We plot the average LSF for each slit as a function of mask position in the rightmost panel of Figure~\ref{fig:flexure}.   The spatial variation in the LSF is due to variation in the instrument image quality across the field-of-view, with the minimum line widths occurring toward the center of the field.    We use the same procedure as for flexure above, fitting a third-order 2D polynomial to the LSF and use this to reduce noise in individual slits.   

The LSF determined from sky emission lines (which fill the slit) will be an overestimate in cases where the science target is unresolved relative to the slit width.  We use high-S/N science spectra across all DEIMOS masks to determine a correction function to the true absorption-line LSF, which depends on slit width and the seeing.  We determine the true LSF  by forward modeling the telluric absorption region using the synthetic telluric spectra (\S\,\ref{ssec:telluric}).  The seeing is determined from the median width of high S/N science spectra in each mask.  Th LSF correction is unity (no correction) when the seeing is larger than the slit width by 10\% or more.   As the seeing improves, the corrected LSFs are up to 20\% smaller for the best seeing conditions present in the DEIMOS archives.   We use this function to correct LSF values in our forward modeling procedures described below.

\subsection{Identifying Nonstellar Spectra}\label{ssec:marz}

Prior to running the stellar velocity code below, we identify extragalactic objects which are removed from the stellar analysis.   To identify extragalactic objects, we use the automated redshift software {\tt Marz}\footnote{https://github.com/Samreay/Marz} \citep{marz}.   {\tt Marz} determines redshifts via cross correlation using a wide range of galaxy and stellar templates.   Its front-end interactive  software allows for visually quality control.  We modified this software slightly, adding key stellar absorption features for display and changing the user classification scheme definitions.    For each DEIMOS mask, we generate a file formatted for the {\tt Marz} interface.    Guided by the automated {\tt Marz} classification, we visually flag and report redshifts for all extragalactic objects.  The results for extragalactic objects are reported in \S\,\ref{ssec:exgal}.    Spectra which cannot be automatically classified or quickly identified as extragalactic via visually inspection due to low S/N or other reasons are allowed to continue to the stellar velocity code.

%%%%%%%%%%%%%%%%%%%%%%%%%%%%%%%%%%
% Figure:  Telluric masks
\begin{figure}[th!]
\centering
\includegraphics[width=1.0\columnwidth]{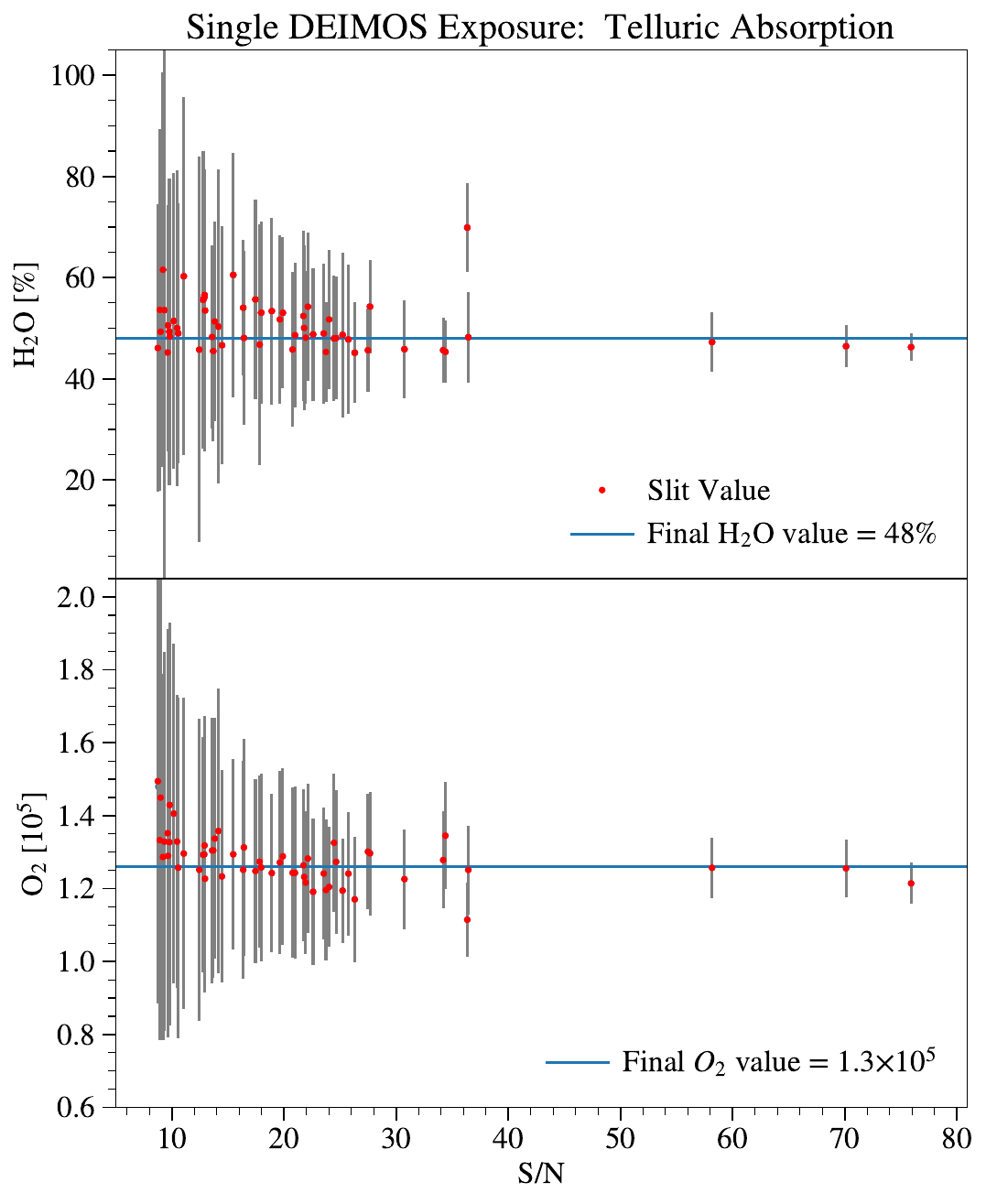}
\vskip -0.3cm
\caption{To determine the best atmospheric telluric absorption parameters for a single DEIMOS exposure, we determine values in all slits by searching over a grid of synthetic telluric spectra.   We plot the inferred humidity (H$_2$O) and the oxygen mixing ratio (O$_2$) as a function of the  median S/N in each slit.  The telluric values for each mask are found by the weighted average of the slit values (horizontal blue line in each panel). \label{fig:telluric_mask}}
\end{figure}
%%%%%%%%%%%%%%%%%%%%%%%%%%%%%%%%%%

\section{{\tt dmost}:  Line-of-Sight Velocity Measurements}\label{sec:rvs}

To determine a line-of-sight velocity and velocity error, we forward model each stellar spectrum combining synthetic telluric and stellar spectra.  This method, presented here, is called {\tt dmost} and can be adapted to any slit spectrograph\footnote{https://github.com/marlageha/dmost}. For each science exposure, we first determine the best synthetic telluric absorption spectrum for the full mask (\S\,\ref{ssec:telluric}).   We then determine the best-fitting synthetic stellar template for each spectrum (\S\,\ref{ssec:chi2}).   Using these two templates, we forward model each spectrum to determine the velocity in a single exposure (\S\,\ref{ssec:forward_model}).  If this method fails due to low S/N, we attempt to measure a velocity using the 1D coadded spectrum (\S\,\ref{ssec:coadd}).

%%%%%%%%%%%%%%%%%%%%%%%%%%%%%%%%%%
% Figure:  Telluric ALL
\begin{figure}[t!]
\centering
 \includegraphics[width=1.0\columnwidth]{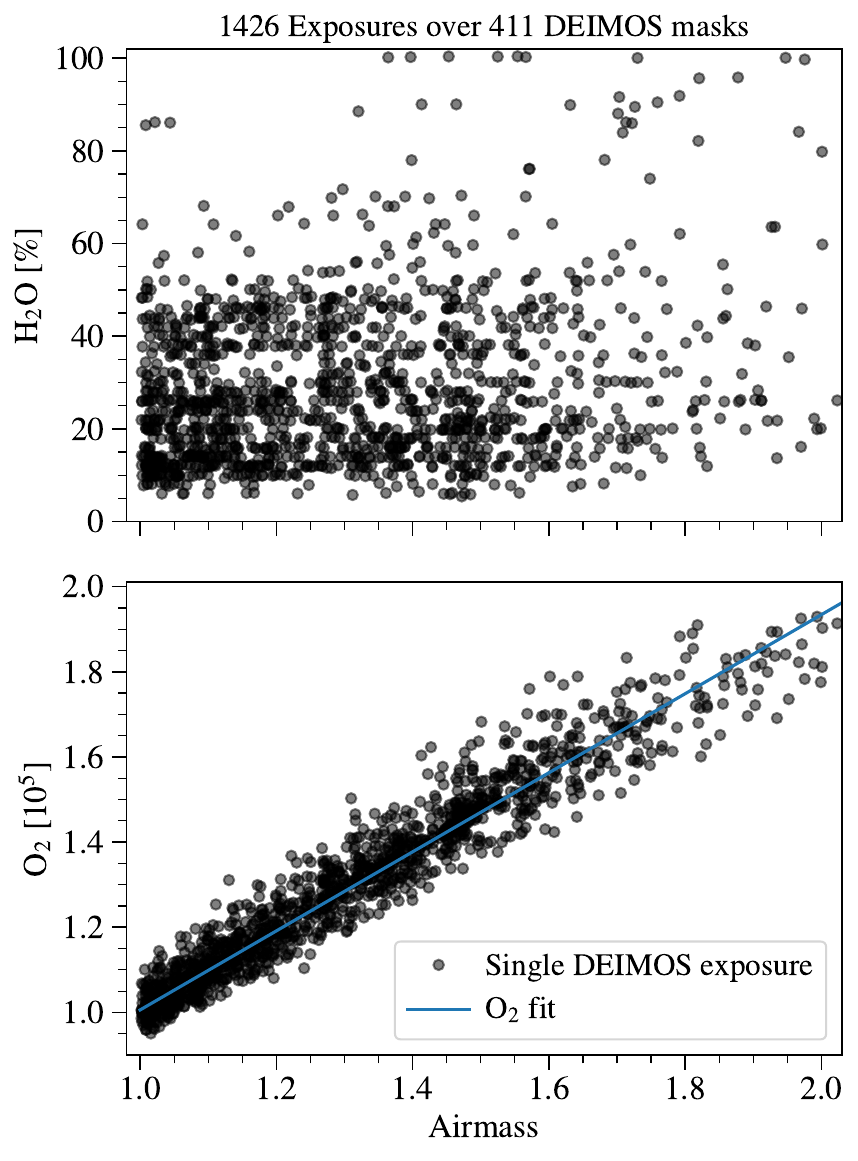}
\vskip -0.1cm
\caption{For each mask, we determine the best-fitting telluric absorption model.    The two atmospheric quantities which vary between science exposures and contribute to observable differences in the telluric absorption between exposures at the DEIMOS spectral resolution are the humidity (H$_2$O) and the oxygen mixing ratio (O$_2$).   We find a strong correlation between O$_2$ and airmass ({\it bottom}), while the humidity is uncorrelated ({\it top}). \label{fig:telluric_all}}
\end{figure}
%%%%%%%%%%%%%%%%%%%%%%%%%%%%%%%%%%

%%%%%%%%%%%%%%%%%%%%%%%%%%%%%%%%%%
% Figure:  Chi2 Template Example
\begin{figure*}[t!]
\centering
 \includegraphics[width=1.0\textwidth]{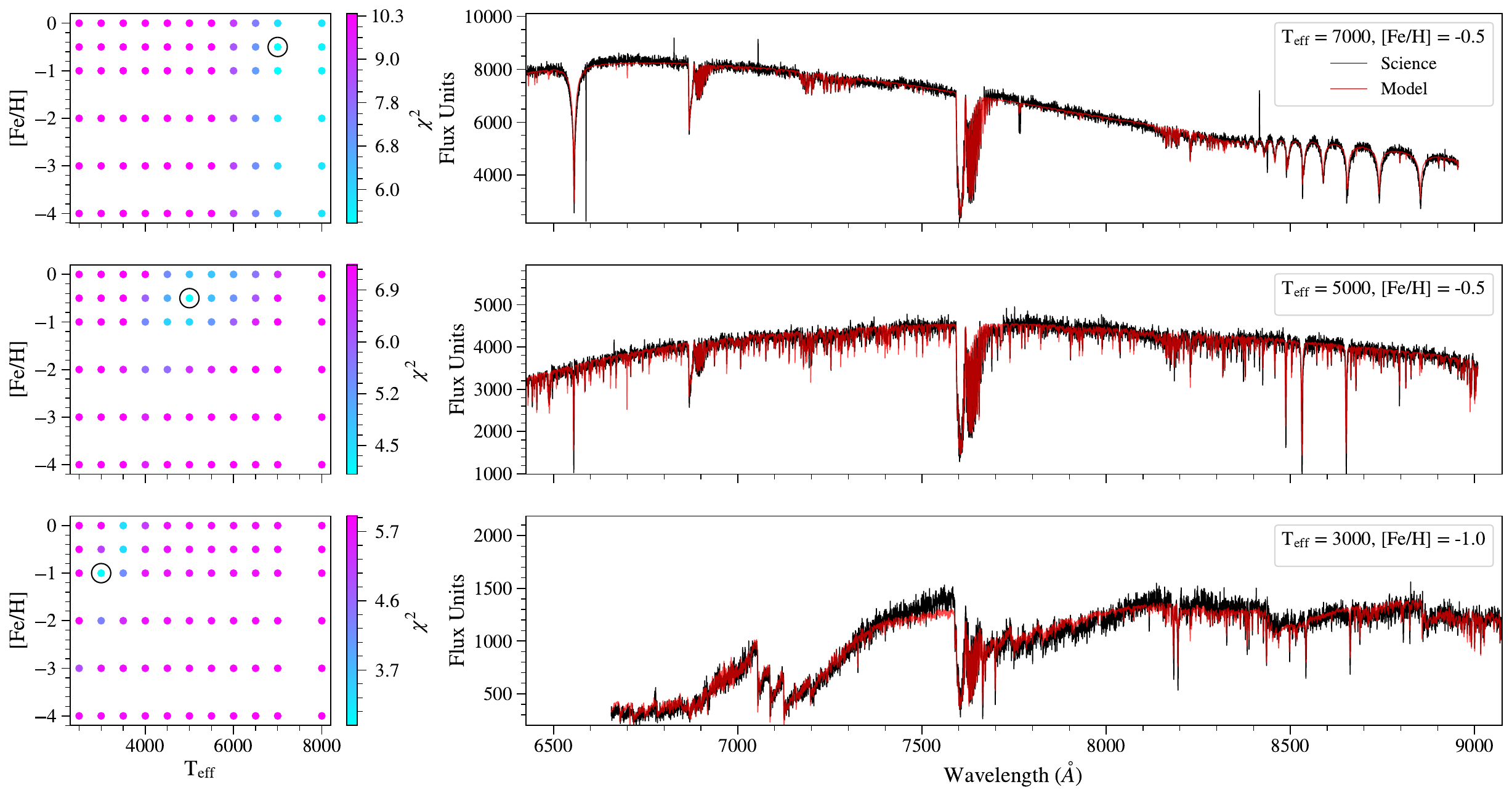}
\vskip -0.1cm
\caption{A synthetic stellar template is determined for each science
  target via $\chi^2$ minimization over a grid of synthetic templates.
  {\it Right:\/} We plot three stars which are representative of the range in stellar types found in the
  DEIMOS archive. {\it Left:\/}  The synthetic stellar template is determined via $\chi^2$ minimization over a grid of templates in effective temperature (T$_{\rm eff}$), metallicity ([Fe/H]) and surface gravity (not shown). We use this stellar template, the telluric template and the LSF to forward model the science spectra.  The reduced $\chi^2$ values are larger than unity and will be improved during the MCMC fitting method described in \S\,\ref{ssec:forward_model}. 
  \label{fig:chi2_template}}
\end{figure*}
%%%%%%%%%%%%%%%%%%%%%%%%%%%%%%%%%%

\subsection{Fitting a Telluric Absorption Spectrum}\label{ssec:telluric}

If an unresolved star is not perfectly centered along the width of a slit, the wavelength solution for the science target will be offset from that determined from the arc lamp wavelength calibration.   This will result in an apparent velocity offset of the object.    Sky {\it emission} lines fully fill the slit width and will not show this offset,  whereas the sky {\it absorption} lines will have the same offset as the science spectrum.  For the 1200G DEIMOS setup and $0.7''$ slitwidth, this offset can be up to $0.2\mbox{\AA}$ or roughly 10\kms.   This \textquotedblleft telluric" offset (sometimes called the A-band correction) can be determined and corrected by measuring the line shift from zero of the telluric absorption lines \citep{sohn2007}.   Past work determined the telluric offset using a single empirical telluric template, typically a hot rotating star so that telluric and stellar lines can be easily separated. A velocity offset is then applied to correct to the science spectra \citep[e.g.,][]{simon07a,martin2014,Kirby2015a}.   

In this work, we make two improvements to the telluric correction.   First, we use synthetic telluric templates, allowing us to match the telluric template spectrum to the atmospheric conditions during a given science exposure.  Second, rather than applying a velocity shift which is linear in log($\lambda$), we apply a linear wavelength shift ($w_{\rm tell}$) which better approximates the telluric shift \citep{cunningham2019}.

We first generate a grid of synthetic telluric spectra at high spectral resolution using {\tt TelFit}\footnote{ http://github.com/freddavies/Telluric-Fitter-py3} v.1.3.2 \citep{telfit2014}.  {\tt TelFit} is a Python wrapper around LBLRTM (Line-By-Line Radiative Transfer Model; \citet{LBLRTM2014}).  {\tt TelFit} assumes a standard atmospheric model and uses the temperature, pressure and atmospheric abundances at the altitude of the Keck Observatory to calculate a telluric spectrum.  The two atmospheric quantities which contribute to observable differences in the telluric absorption between science exposures at the DEIMOS spectral resolution (Figure~\ref{fig:spectrum}) are the humidity (H$_2$O) and the oxygen mixing ratio (O$_2$).  We use {\tt TelFit} to generate a grid of templates at high spectral resolution (0.02\mbox{\AA} per pixel) over the DEIMOS wavelength regime
covering the full expected range of these two parameters:  $1\%<\,$H$_2$O$\,<99\%$ and $0.7\times10^5 <~$O$_2\,< 2.3\times10^5$.  

For a given science exposure, we assume the atmospheric parameters are constant over the DEIMOS field-of-view and use the majority of slits in the mask to determine a best-fitting synthetic telluric spectrum.     To generate a single telluric spectrum model, we apply a linear wavelength shift ($w_{\rm tell}$) to the high-resolution telluric template and Gaussian convolve this synthetic model using the LSF for each slit determined during the flexure correction (\S\,\ref{ssec:flexure}).  We rebin this model onto the wavelength array of the science data and marginalize over the science continuum shape using a third-order polynomial.   For each slit, we generate a grid of model spectrum over H$_2$O, O$_2$ and $w_{\rm tell}$.   We evaluate the standard $\chi^2$ difference in the telluric wavelength windows shown in Figure~\ref{fig:spectrum}.   The best fitting telluric model for a given slit is determined by interpolating to find the minimum $\chi^2$ value in the H$_2$O and O$_2$ space.  This procedure is repeated for every slit in each exposure of a mask. 

In Figure~\ref{fig:telluric_mask}, we plot the inferred telluric parameters for each slit in an example exposure as a function of median per pixel S/N.  We determine the telluric parameters using slits with $5 < S/N < 150$.   We do not use slits with $S/N > 150$ as these tend to include saturated pixels, while slits with $S/N < 5$ do not have sufficient signal to constrain the telluric parameters.   We further exclude slits with large telluric $\chi^2$ values, which tend to be carbon or late M dwarf stars.  The final atmospheric parameters are determined as the weighted average of all good slits in a given science exposure, typically determined with 50 or more slits per mask.

%%%%%%%%%%%%%%%%%%%%%%%%%%%%%%%%%%
% Figure:  MCMC exam[le]
\begin{figure*}[t!]
\centering
 \includegraphics[width=1.0\textwidth]{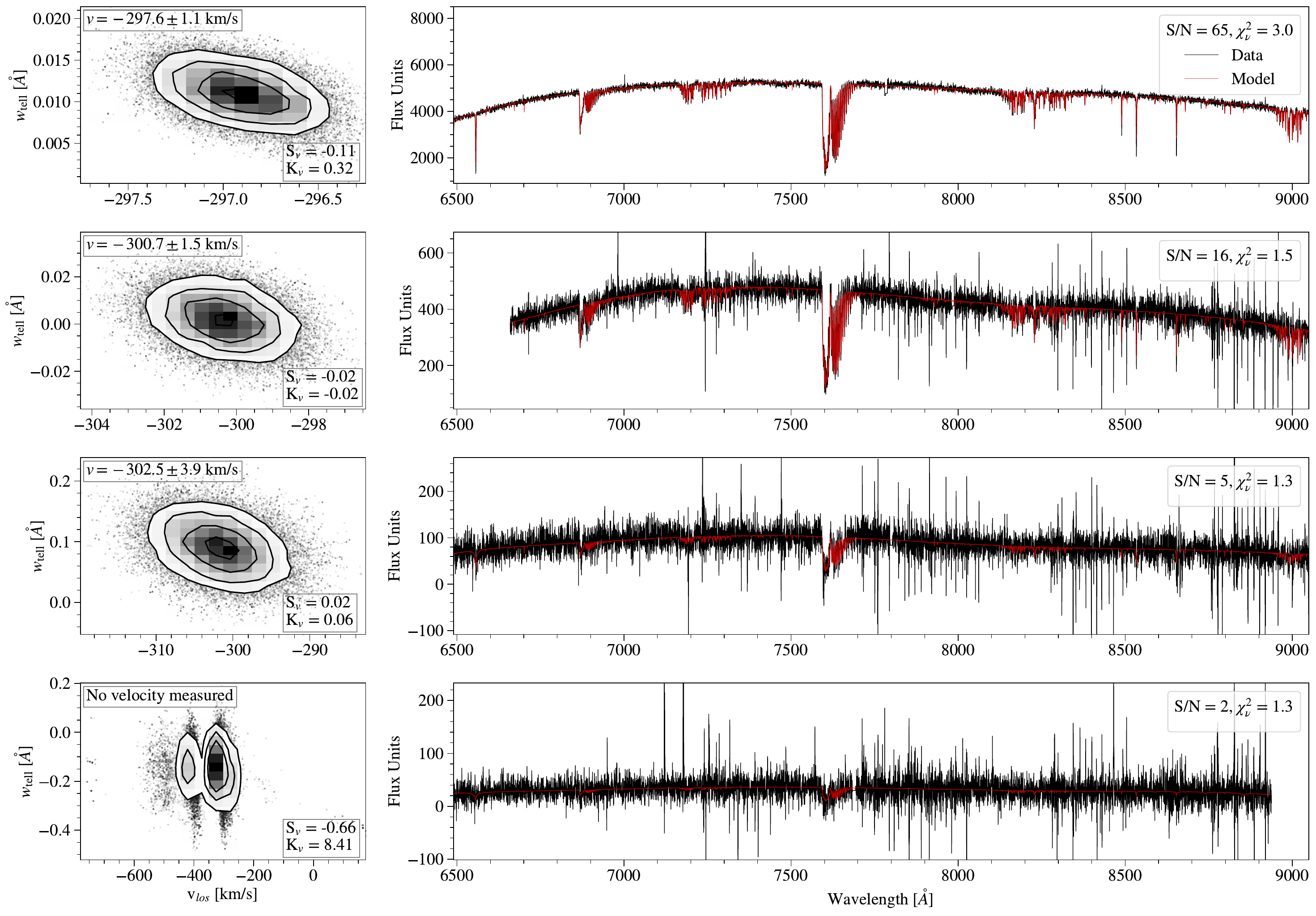}
\caption{Example DEIMOS spectra and best-fit {\tt dmost} models at varying S/N.  {\it Right:\/}  The best-fitting {\tt dmost} model spectrum (red) is plotted on top of the science data (black).   {\it Left:\/} The associated posterior distributions in velocity ($v_{\rm los}$) and telluric wavelength shift ($w_{\rm tell}$).  The line-of-sight velocity and velocity error (random component) are determined directly from these distributions.  The bottom row shows an example of a failed velocity measurement due to non-Gaussian posteriors in velocity. 
\label{fig:mcmc_example}}
\end{figure*}
%%%%%%%%%%%%%%%%%%%%%%%%%%%%%%%%%%

The above procedure produces the best-fitting atmospheric parameters (O$_2$ and H$_2$O) at the time of each science exposure.   The final values are interpolated from the original grid, thus for each exposure we generate a high -esolution (0.02\mbox{\AA} per pixel) synthetic telluric spectrum again using  {\tt TelFit}.   This telluric spectrum template (T$_{\rm tell}$) is used in the velocity forward modeling procedure described in \S\,\ref{ssec:forward_model} and  Equation~1.

In Figure~\ref{fig:telluric_all}, we plot atmospheric parameters for \nexp individual science exposures across \nmask DEIMOS masks as a function of airmass. Shown in the top panel, the humidity (H$_2$O) is largely below 50\%, as higher values indicate water vapor in the form of clouds or fog, and thus poor observing conditions.  The humidity is not correlated with airmass.   On the other hand, the oxygen mixing ratio (O$_2$, bottom panel) is a strong function of airmass.  We recognized this correlation in retrospect and therefore continue to fit for this quantity to account for the small scatter observed in this relationship.   However, given the strong correlation between airmass and O$_2$, we fit a linear function to this relation (0.928$\times$airmass + 0.078).  We replace O$_2$ with this fitted value in the rare cases where fewer than five slits have $S/N > 5$ in a given mask.

%%%%%%%%%%%%%%%%%%%%%%%%%%%%%%%%%%
%\begin{table}[h!] 
%\tablewidth{\textwidth}
%\begin{center}
%\caption{Synthetic Stellar Grids\label{tab:table1}}
%\begin{tabular}{l l r r r l}
%\hline 
%\hline
%Grid & S/N & T$_{\rm eff}\,[K]$ & $\Delta$T$_{\rm eff}$ & [Fe/H] & log\,$g$ \\
%      \hline
% 1 & $> 25$ & 2500-8000 & 500 & $-$4 to +0 & 1,3,5 \\
%2 & $10 - 25$ & 3000-8000 & 1000 & $-$4 to +0 & 1,3,5 \\
%3 & $< 10$ & 3000-8000 & 1000 & $-$3 to +0 & 2,5\\
%\hline
%\end{tabular}
%\tablecomments{Synthetic stellar spectra from \citet{Husser2013}.  The metallicity spacing is unity between [Fe/H] $= -4$ to $-1$ and 0.5 otherwise.}
%\end{center}
%\end{table}
%%%%%%%%%%%%%%%%%%%%%%%%%%%%%%%%%%

\subsection{Determining a Synthetic Stellar Template}\label{ssec:chi2}

We next determine the best-fitting synthetic stellar template for each science target, which we will use in the velocity analysis (\S\,\ref{ssec:forward_model}).   We assume the stellar template is the same for all science exposures of a given star and use the flexure-corrected coadded spectra determined in \S\,\ref{ssec:pypeit}.   This assumption will be wrong for some variable stars (e.g., RR Lyrae whose spectral properties change with pulsation phase); we will address such cases in future work.      

We use the synthetic stellar template atlas of \citet{Husser2013}.   These high-resolution (0.01\mbox{\AA}) spectra are based on the stellar atmosphere code PHOENIX and cover a wide range of stellar parameters.    Although we expect a range of alpha abundances in Milky Way stellar systems \citep{Kirby2011,Vargas2013}, we consider only models with solar alpha abundances ($[\alpha/M] = 0$), as detecting alpha abundance differences in this wavelength region is challenging \citep{kirby2020}.

To reduce computation, we search a restricted grid of possible stellar templates depending on the S/N of the each science spectrum.   We restrict in both log$\,g$ and T$_{\rm eff}$ to match the range of observed stars expected in Milky Way dwarf galaxies and globular clusters, and define three grids with parameter spacing matched to the information content of the spectrum at the DEIMOS spectral resolution and S/N (see Figure~\ref{fig:chi2_template}).   We note that log\,$g$ is particularly difficult to constrain given the DEIMOS wavelength coverage.    While the best-fitting templates are related to the physical properties of the science star, we use these templates primarily to determine velocities.  We convolve all template spectra to 0.02\mbox{\AA} to match the telluric template spectral resolution. 

Similar to the telluric template fitting in \S\,\ref{ssec:telluric}, we determine the best synthetic stellar template ($T_{\rm stellar}$) via $\chi^2$ fitting over the grid of models above.  For each slit, we use the corrected LSF to Gaussian convolve each synthetic model.  We rebin each model onto the wavelength array of the science data and marginalize over the science continuum using a third-order polynomial.   For each slit, we generate a grid of model spectrum over $T_{\rm eff}$, log\,$g$, [Fe/H] and a coarse grid in velocity.   We then evaluate the standard $\chi^2$ difference in the regions of the spectra which are not affected by telluric absorption (Figure~\ref{fig:spectrum}).   The stellar template which minimizes $\chi^2$ is recorded for the analysis in \S\,\ref{ssec:forward_model}, and the best-fitting velocity is used as a starting guess for the main velocity analysis.

In Figure~\ref{fig:chi2_template}, we show three stars representative of stellar types found in the DEIMOS archive.   The top panel shows a hot horizontal branch (HB) star with broad absorption lines;  as expected the [Fe/H] is largely unconstrained for this stellar type as seen by the $\chi^2$ values in the upper-left panel.  The middle panel shows a G-K-type star, which is the most typical stellar type found in the DEIMOS archives.  The bottom spectrum shows a cool M dwarf star.   In the left panels of Figure~\ref{fig:chi2_template}, the minimum reduced $\chi^2$ value is circled for each spectrum.  The reduced $\chi^2$ values are larger than unity and will be improved during the Markov Chain Monte Carlo (MCMC) fitting method described below.

\subsection{Forward Modeling Velocity Measurements}\label{ssec:forward_model}

For each star on a given DEIMOS mask, we have determined a synthetic telluric template (\S\,\ref{ssec:telluric}) and stellar template (\S\,\ref{ssec:chi2}).
We next determine the line-of-sight velocity and velocity error (random component) for each star in each science exposure using an MCMC method.   We determine systematic velocity errors and combine repeat velocity measurements in \S\,\ref{sec:errors_combine}.
  
Given the predetermined synthetic templates, the only free parameters in the MCMC are the line-of-sight velocity ($v_{\rm los}$), the telluric wavelength shift ($w_{\rm tell}$) and the continuum shape ($P_{\rm cont}$).  
All synthetic templates are on the same wavelength grid with a factor of 15 higher spectral resolution (0.02\mbox{\AA}) than the science spectra and are normalized to unity.  To generate a single model, $f_{\rm model}$ (Eqn.\,1), for a given choice of $v_{\rm los}$ and $w_{\rm tell}$, we begin with the stellar template and perform the following steps:
\begin{enumerate}
 \item Shift the synthetic stellar template ($T_{\rm stellar}$) by the choice of line-of-sight velocity $v_{\rm los}$ (a log-wavelength shift).
 \item  Multiply the shifted stellar template by the best-fit telluric template ($T_{\rm tell}$).
 \item Shift the combined spectrum linearly in wavelength by the choice of telluric wavelength shift $w_{\rm tell}$.  
 \item Convolve the spectrum by the corrected LSF determined for each slit in \S\,\ref{ssec:flexure} and \S\,\ref{ssec:LSF}.  
 \item Multiply by a fifth-order polynomial ($P_{\rm cont}$) marginalized over the wavelength range of the science spectrum to match the continuum shape. 
 \item Rebin this model onto the wavelength grid of the original science spectrum. 
\end{enumerate}

\noindent
 The above steps generate a model spectrum, $f_{\rm model}$, that can be directly compared pixel-by-pixel to the original science spectrum.   To summarize, a single model is determined as:
\begin{equation}
f_{\rm model} = ({\rm LSF}\circledast(T_{\rm tell}T_{\rm stellar}[v_{\rm los}])[w_{\rm tell}] )\,P_{\rm cont}. 
\end{equation}

\noindent
where the free parameters are $v_{\rm los}$, $w_{\rm tell}$ and the continuum shape $P_{\rm cont}$.    

For each spectrum, we run the MCMC sampler {\tt emcee} \citep{emcee} using 20 walkers initialized at the velocity and telluric wavelength shifts determined as part of the synthetic template fitting (\S\,\ref{ssec:chi2}).  We add a small random shift to these values when initializing each individual walker.  We run the sampler for 3000 steps.    Given a model spectrum, $f_{\rm model}$, we evaluate the goodness of fit by defining a log-likelihood function as:
\begin{equation}
\log \mathcal{L} = -\frac{1}{2}\sum_{i = \lambda_{\rm min}}^{\lambda_{\rm max}} \frac{(f_{{\rm sci},i} - f_{\rm model}[w_{\rm tell},v_{\rm los}]_i)^2}{\sigma_{{\rm sci},i}^2}.
\end{equation}
\noindent
where $f_{\rm sci}$ and $\sigma_{\rm sci}$ are the per pixel flux and error of the observed science spectrum, respectively.  The likelihood is evaluated over the wavelength range of the science spectrum ($\lambda_{\rm min}$-$\lambda_{\rm max}$), trimmed to account for vignetting at the edges of the DEIMOS mosaic.   We set a velocity prior slightly larger than the expected velocity range of Milky Way stellar satellites:  $-800 < v_{\rm los} < 800$\kms.  The telluric wavelength shift prior is set larger than the range of expected shifts for this DEIMOS setting: $-0.8 < w_{\rm tell} < 0.8$\AA\ (2.5\,pixels).

Shown in Figure~\ref{fig:mcmc_example}, we evaluate the shape of the resulting posterior distributions, measuring the skewness (S) and kurtosis (K).   Non-Gaussian posteriors are primarily driven by multiple peaks in velocity space.  We reject cases where the velocity posterior shape is far from Gaussian: $|S| > 1$ or $|K| > 1$, or the accepted number of samples is less than 70\%.    We evaluate the autocorrelation timescale to determine the number of burn-in samples (typically less than 100 samples). Removing the burn-in samples, we determine the stellar velocity and velocity error (random component) as the median and 16th and 84th percentiles of the posterior distribution.

%%%%%%%%%%%%%%%%%%%%%%%%%%%%%%%%%%
% Figure:  MCMC Chi2 values
\begin{figure*}[th!]
\centering
 \includegraphics[width=1.0\textwidth]{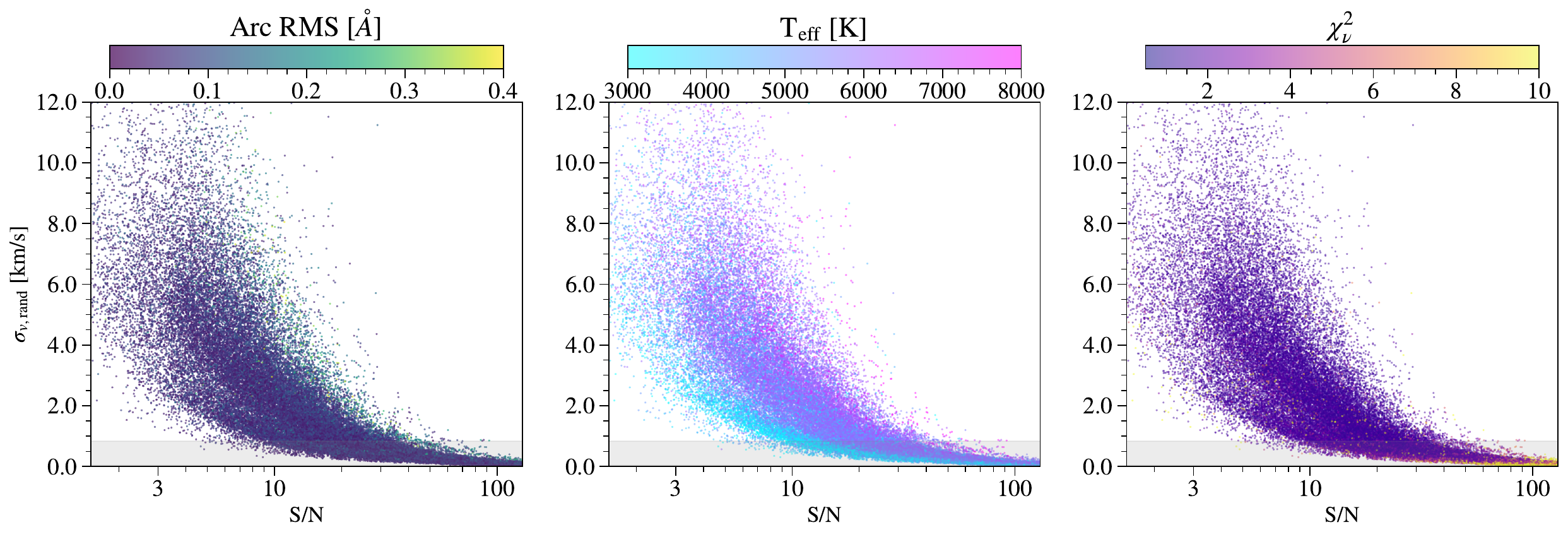}
\caption{The random component of the velocity error, $\sigma_{\rm v, rand}$, for individual stars on a given mask as a function of the median per pixel S/N.   As expected, the random velocity error decreases with increasing S/N.   Within this, additional trends are seen with the accuracy of the wavelength solution ({\it left}) and effective temperature of the star ({\it middle}).  The gray box at the bottom of each panel indicates the region where errors are dominated by the velocity error floor ($\sigma_{v,\rm floor}/k_v$).   In the rightmost panel, we color code by the goodness of the MCMC fit ($\chi_{\nu}^2$).   $\chi^2_{\nu}$ values in excess of unity are seen for S/N$>75$ as our models are not sufficient to fully describe all features seen in higher-S/N spectra (less than 5\% of the sample); our systematic velocity term dominates in this region. \label{fig:mcmc_chi2}}
\end{figure*}
%%%%%%%%%%%%%%%%%%%%%%%%%%%%%%%%%%

In Figure~\ref{fig:mcmc_example}, we show posterior distributions and best-fitting model spectra for four single exposures of varying S/N.   The posteriors are generally close to Gaussian.  In the bottom panel, the posterior values are non-Gaussian and the velocity fit is rejected by the criteria described above.  In such cases where we fail to measure a velocity from the single exposure spectrum due to low S/N (typically S/N$\sim2$), we coadd individual extracted 1D spectra and repeat the procedure on the coadded spectrum as described in \S\,\ref{ssec:coadd}.

We correct velocities measured in each science exposure to the heliocentric frame.   We then determine the final velocity for each star on a given DEIMOS mask by combining the velocities from individual science exposures via a weighted mean.   We note this combined velocity is more accurate than a velocity determined directly from a coadded spectrum, as the heliocentric and telluric corrections are tailored to each exposure.  

In Figure~\ref{fig:mcmc_chi2}, we plot the combined random velocity error for stars as a function of the median per pixel S/N.    Within the obvious trend of decreasing  error with increasing S/N, we see additional structure.   At a given S/N, poorer wavelength solutions (left panel) tend to toward higher velocity errors, while stars with hotter effective temperatures (middle panel) have smoother spectral features (see Figure~\ref{fig:chi2_template}), resulting in poorer velocity precision.  We do not see significant trends in other parameters such as the seeing, airmass, or magnitude of the telluric shift.    In the rightmost panel of Figure~\ref{fig:mcmc_chi2}, we color code by the goodness of the MCMC fit.   Our models are good fits ($\chi_{\nu}^2 \sim 1$) for S/N values up to 75 (see top panel of Figure~\ref{fig:mcmc_example}).  For higher-S/N spectra (less than 5\% of the sample), our models are not sufficient to fully describe the observed spectra, resulting in larger $\chi_{\nu}^2$ values.  Additional telluric parameters (particularly ozone which modifies the continuum shape) and increasing the stellar template grid spacing will improve the fits at higher S/N, but should not meaningfully affect the velocity values-- in this region the error is dominated by the systematic term (see \S\,\ref{ssec: esys_exp}). 

\subsection{Low Signal-to-Noise Ratio Velocity Forward Modeling}\label{ssec:coadd}

In cases where fewer than half of the individual science exposures for a given star on a mask produce a reliable velocity measurement,  we attempt to measure a velocity based on the coadded extracted 1D spectra.    We use the coadded spectrum from \S\,\ref{ssec:chi2}, in which the individual extracted spectra are corrected for flexure before coadding.  The majority of failed single exposure spectra have a median per pixel S/N below two (see bottom panel Figure~\ref{fig:mcmc_example}).  For the coadded spectra, we perform the same MCMC procedure and criteria for an acceptable velocity measurement described above in \S\,\ref{ssec:forward_model}.   In the forward model, we use the S/N-weighted average, across all individual science exposures, of the telluric parameters and LSF.  Of the velocities measured across all DEIMOS masks in this work,  13\% of measurements yielded only a coadded velocity.  These measurements are considered separately in the analysis of systematic velocity errors below, although in practice have the same velocity error correction.

\section{Determining Velocity Errors}\label{sec:errors_combine}

While our MCMC velocity analysis in \S\,\ref{ssec:forward_model}--\ref{ssec:coadd} determines the random component of the velocity error ($\sigma_{v,{\rm rand}}$, Figure~\ref{fig:mcmc_chi2}), we next determine the repeatability of our velocities and assess additional velocity error terms using thousands of repeat measurements across the DEIMOS archive (\S\,\ref{ssec: esys_exp}).  We then combine multiple velocity measurements and flag velocity variables (\S\,\ref{ssec:flag_var}).   

\subsection{Calculating Systematic Velocity Errors}\label{ssec: esys_exp}

To accurately determine the internal velocity dispersion, and thus dynamical mass, of a resolved stellar system, it is critical to determine both accurate velocities and velocity errors.  Underestimating (overestimating) velocity errors for individual stars translates directly into larger (smaller) values of the internal velocity dispersions and therefore the inferred dynamical mass of a system (\S\,\ref{ssec:results}).

We investigate the repeatability of our velocity measurements and assess the accuracy of our velocity errors from the {\tt dmost} pipeline using repeated observations of stars on different independent DEIMOS masks.   We will validate against larger spectroscopic surveys in \S\,\ref{sec:validation}.  The source of additional velocity error terms is both inherent to the DEIMOS instrument and the {\tt dmost} data analysis pipeline.  We note that calling this a \textquotedblleft systematic" error term is not entirely correct: we will include these terms in the reported velocity errors per mask and then treat them as statistical errors, adding these combined errors in quadrature when repeat measurements across masks are available.   

\begin{figure*}[t!]
\includegraphics[width=1.0\textwidth]{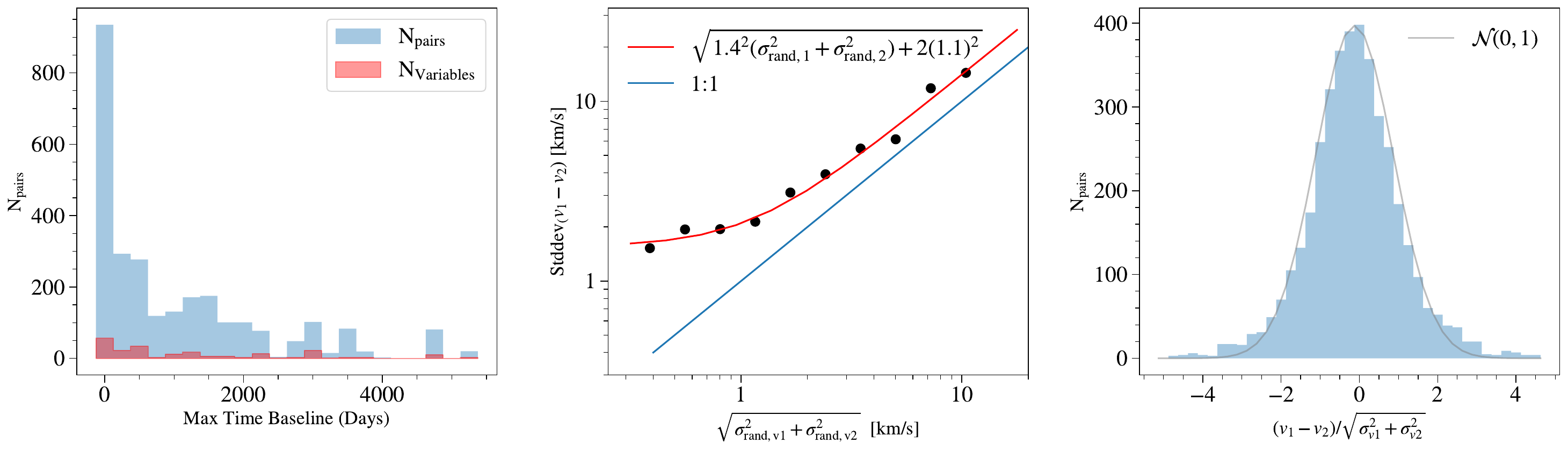}
\caption{{\it Left:}\/ Histogram of stars with measured velocities in two or more DEIMOS masks, plotted by the maximum temporal baseline between observations.    We use these repeated measurements to estimate systematic velocity error terms, restricting to pairs separated by less than 30 days and removing stars with significant velocity variation between measurements (red histogram).
{\it Middle:\/}  The quadrature combined pair-wise uncorrected (random-only) velocity errors vs.~the standard deviation of the pairwise velocity differences. 
 If the random velocity errors were correct, the binned values (black symbols) should follow the one-to-one line (blue).   Instead, the points confirm our model curve (red) in which the error correction in the legend has been applied.  {\it Right:\/}  After applying these velocity error terms, the normalize histogram of velocity differences follows a normal Gaussian distribution. \label{fig_sys_err}}
\end{figure*}

Independent repeat measurements from different DEIMOS masks will differ in their wavelength solution and differences due to a star's position on the mask and within its slit.  Past investigations with this DEIMOS setup have been limited to tens of repeated measurements and have asked only whether a velocity error floor exists (we compare to previous estimates in \S\,\ref{ssec:esys_exp_past}).  Our sample of repeat measurements reduced with {\tt dmost} is an order of magnitude larger than any prior estimates of the DEIMOS velocity errors.   There are \nrepeat unique stars observed in 2 or more DEIMOS mask in our Milky Way sample (left panel, Figure~\ref{fig_sys_err}), resulting in 4151 unique pairs of repeat measurements due to stars appearing in up to seven unique DEIMOS masks.  

Following \citet{li2019} and \citet{walker2023}, we model additional velocity error terms to include both a multiplicative factor and a velocity floor as:
\begin{equation}\label{eq:vel_err}
\sigma_v =  \sqrt{k_v^2\sigma_{v,\rm rand}^2 + \sigma_{v,\rm floor}^2}
\end{equation}
where $k_v$ is a scaling factor for the random velocity uncertainty, and $\sigma_{v,\rm floor}$ is the minimum velocity error floor.   We determine $k_v$ and $\sigma_{v,\rm floor}$ by modeling the pair-wise  velocity differences, $\delta v_{i,j} = v_i - v_j$, assuming a Gaussian model and including an outlier fraction ($f_{\rm out}$) as:
\begin{equation}
%\delta v_{i,j} = (1-f_{\rm out})\mathcal{N}(0,\sqrt{k_v^2(\sigma_{v,i}^2 +\sigma_{v,j}^2) + 2\sigma_{v,floor}^2} + f_{\rm out}\mathcal{N}(0,\sigma_{\rm outlier})
\delta v_{i,j} = (1-f_{\rm out})\mathcal{N}(0,\sqrt{\sigma_{v,i}^2 + \sigma_{v,j}^2}) + f_{\rm out}\mathcal{N}(0,\sigma_{\rm 
out})
\end{equation}
where $\sigma_{v,i/j}$ is given for the $i^{\rm th}$ or $j^{\rm th}$ star by Eqn.~\ref{eq:vel_err} and the free parameters are $k_v$, $\sigma_{v,{\rm floor}}$, $f_{\rm out}$, and $\sigma_{\rm out}$.   We fit this model to our sample of repeated observations, setting priors on the outlier distribution to be $\sigma_{\rm out} > 5$ and $f_{\rm out}$ as less than 25\% of the sample.  The outlier component should account for any true velocity variables in the sample; the best-fit outlier fraction is 10-15\% for the samples described below, consistent with the fraction of velocity variables identified in \S\,\ref{ssec:flag_var}.

By fitting this model to subsets of the repeat velocity sample, we explore the dependence of $k_v$ and $\sigma_{v,\rm floor}$ on various quantities.   We find no dependence on the slit width of observations, nor on whether the velocity was determined from the individual exposures or from the coadded spectra.   We do restrict the repeat pairs to temporal baselines less than 30 days.  Restricting the repeated sample in time reduces the inferred velocity error floor by 20\%.   While this may be due to true systemics in the system, its more likely this increase is due to undetected binary stars contributing to velocity variability beyond what is accounted for in our outlier model.  Based on all pairs with temporal baselines less than 30 days (\npair unique velocity pairs), we infer $k_v$ = 1.4 and $\sigma_{v,\rm floor} = 1.1$\kms.  The multiplicative constant is larger than unity due to covariance between pixels in the reduced spectra due to the resampling of the blue detector for the DEIMOS mosaic (\S\,\ref{ssec:pypeit}).   The velocity floor is comparable to the quadrature sum of the median RMS of our wavelength solution determined in \S\,\ref{ssec:pypeit}.   We compare these values to previous DEIMOS reduction methods in \S\,\ref{ssec:esys_exp_past}.

We internally confirm our inferred values of $k_v$ and $\sigma_{v,\rm floor}$ following \citet[][their Figure~11]{Cooper2023}, by plotting the quadrature combined pair-wise uncorrected (random-only) velocity errors versus the standard deviation of the pairwise velocity differences.  In the middle panel of Figure~\ref{fig_sys_err}, black points show the binned standard deviation estimate (16th to 84th percentiles) of the velocity differences  as a function of their combined random uncertainties.   If our random velocity errors were correct, the binned values should follow the blue one-to-one line.   Instead, the points confirm our model curve (red line) in which the error correction in Equation~\ref{eq:vel_err} has been applied.

As a side note, we have also explored velocity errors between independent single science exposures within a given mask.  Unlike the analysis above, these measurements share the same wavelength solution and synthetic stellar template, but differ in their flexure solution, telluric absorption template and heliocentric velocity correction.  Over \nmask masks, we have good velocity measurements for $\sim$43,000 unique pairs of measurements from exposures taken on the same mask.  We find a similar value of the multiplicative factor $k_{v, \rm single} = 1.3$, while the velocity error floor is much lower at $\sigma_{v,\rm floor, single} = 0.3$\kms.  This lower floor value, as compared to the velocity error floor between masks, is consistent with the latter being dominated by differences in the wavelength solution.

%%%%%%%%%%%%%%%%%%%%%%%%%%%%%%%%%%
% Figure:  velocity variables
%%%%%%%%%%%%%%%%%%%%%%%%%%%%%%%%%%

\subsection{Combining Velocities and Identifying Velocity Variables}\label{ssec:flag_var}

For a star observed in a single DEIMOS mask and reduced using the methods described in this paper, the total velocity error is given in Equation~\ref{eq:vel_err}.   If a star is observed in multiple DEIMOS masks, the velocity is determined as the weighted mean of velocities measured across masks, and the individual mask velocity errors are combined in quadrature.  In \S\,\ref{sec:validation}, we validate our velocity zero-point and error estimates by comparing to large all-sky kinematic surveys.

Combining velocity measurements taken at different times implicitly assumes that a given star has a constant velocity.  While true for the majority of this sample, we expect some fraction of stars  will be velocity variables, such as RR Lyrae stars or unresolved binary star systems.   There are usually too few epochs to determine a variable star's systematic velocity or the orbital solution of a binary system's center of mass (although see \citealt{koch2014} and \citealt{buttry2022}).  Here we simply identify and flag velocity variables.  We separately identify stars that are variable between exposures on a given DEIMOS mask (evaluating internal velocity differences with the lower velocity error floor), and stars that are variable between different masks.     

To identify velocity variables,  we evaluate whether or not velocities measured at different times are consistent with random fluctuations from a constant value.  For each star, we first calculate the weighted mean velocity, the equivalent of assuming a constant velocity model.   We calculate the $\chi^2$ value for this model, including random and systematic error components (Eqn.~\ref{eq:vel_err}).   Following \citet{maxted2001}, we then calculate the probability, $p$,  of obtaining the observed value of $\chi^2$ or higher from random fluctuations around a constant value.  $p$ is evaluated for the appropriate number of degrees of freedom (the number of observations minus one).   We flag a star as a velocity variable if log$_{10}p < -1$ ({\tt flag\_var == 1}, Table~\ref{table:schema}).

In left panel of Figure~\ref{fig_sys_err}, we show the maximum temporal baseline between observations for stars from multiple DEIMOS masks.   The baseline for measurements span between less than one hour to 14.3 years.   We compare the distribution of velocity variable and non-variable sources.     We find 708 (12\%) short timescale variables and 348 (4\%) longer timescale variables; 79 stars varied on both timescales.   These distributions include masks where variable targets were intentionally targeted (e.g., known RRL stars), thus deeper investigation is required to use these data as constraints on binary fractions or other science questions.

%%%%%%%%%%%%%%%%%%%%%%%%%%%%%%%%%%
% Figure:   velocity comparison
\begin{figure*}[th!]
\epsscale{1.1}
\plotone{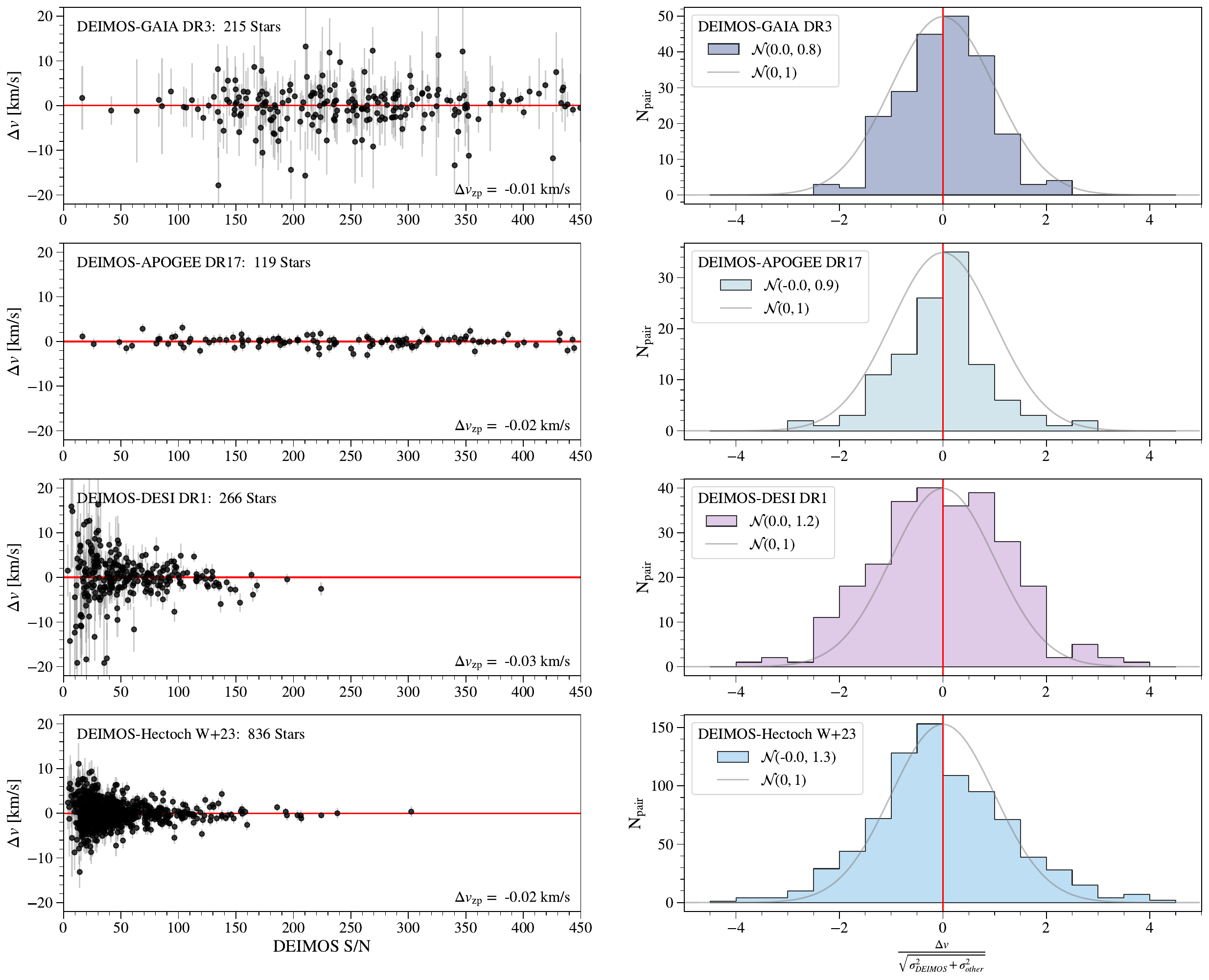}
\caption{Comparison of DEIMOS velocities and errors to matched stars in the Gaia DR3, APOGEE DR17, DESI DR1, and MMT/Hectochelle datasets (\S\,\ref{ssec:validate_samples}).  {\it Left:\/} We plot the velocity difference ($\Delta v = v_{\rm DEIMOS} - v_{\rm other}$) as a function DEIMOS S/N, with error bars as the quadrature sum from both surveys.  {\it Right:\/} Histogram of the velocity difference normalized by the quadrature sum of the velocity errors.  A normal distribution, $\mathcal{N}(0,1)$, is expected if the velocity zero-points are correct and the velocity errors from both surveys are Gaussian (gray line).  The DEIMOS zero-point agrees with these larger spectroscopic surveys, and the velocity error is close to unity agreement with all four literature surveys.\label{fig_v_validate}}
\end{figure*}
%%%%%%%%%%%%%%%%%%%%%%%%%%%%%%%%%%

\section{Velocity Validation}\label{sec:validation}

We next compare our DEIMOS velocities and velocity errors against large spectroscopic surveys (\S\,\ref{ssec:validate_samples}).   We use these surveys to validate our velocity zero-point and the accuracy of our velocity errors (\S\,\ref{ssec:validate_v}).   We then compare these quantities to previous DEIMOS pipelines (\S\,\ref{ssec:esys_exp_past}).   In \S\,\ref{sec: ew}, we repeat this exercise for our EW based [Fe/H] values.

\subsection{Literature Comparison: Selected Samples}\label{ssec:validate_samples}

In comparing our results against larger published datasets, we identify surveys with a significant  number of overlapping stars with comparable or better velocity resolution.  We primarily compare to four major datasets:  Gaia DR3, the Apache
Point Observatory Galactic Evolution Experiment(APOGEE) DR17, Dark Energy Spectroscopic Survey (DESI) DR1 and MMT/Hechochelle.  We chose a matching tolerance of $1.25''$ to allow for errors in the original DEIMOS targeting photometry. Before matching to each survey, we remove DEIMOS stars that are flagged as velocity variables and perform basic quality cuts on these literature datasets as follows:

\begin{itemize}

    \item {\it Gaia DR3}:  In addition to astrometry, the Gaia satellite \citep{GaiaDR3} DR3 provides spectra of 34 million stars taken with the RVS spectrograph (R=11,500) in the wavelength region around the Ca II triplet \citep{Katz2023}.  We filter sources to remove likely variable sources ({\tt astrometric\_excess\_noise}==0) and Gaia velocities with large error bars ({\tt gaia\_rv\_err}$<25$\kms).  We apply the velocity correction suggested by \citet[][their Eqn.~5]{Katz2023}.  We find 215 Gaia RVS matches within $1.25''$ of a DEIMOS stellar target.  

    \item {\it APOGEE DR 17}:   SDSS APOGEE DR17 \citep{sdss_dr17} is a high-resolution (R$\sim$22,500) infrared spectroscopic survey of half a million stars \citep{apogee}.   We filter APOGEE sources to remove likely variable sources ({\tt VSCATTER} $<1$\kms) and poor velocity measurements ({\tt STARFLAG}==0).  We correct the reported APOGEE velocity errors using Eqn.~1 of \citet{lewis2022}.   We find 119 APOGEE DR17 matches within $1.25''$ of a DEIMOS stellar target.  

    \item {\it DESI DR1}: DESI DR1 \citep[DESI,][]{Cooper2023} includes over 4,000,000 stars with  R$\sim 4000$ covering the full optical wavelengths \citep{koposov2025}.   We filter DESI sources, requiring {\tt RVS\_WARN} == 0 and {\tt RR\_SPECTYPE} == 'STAR'.   We remove stars from the DESI backup program and apply the recommended velocity error floors appropriate for other programs ('bright','dark') taken from Figure 13 of \citet{koposov2025}.  We further subtract a 0.4\kms\ zero-point offset from the DESI velocities to account for DESI's stated zeropoint offset from {\it Gaia} and APOGEE (see  \S\,4.3 in \citet{koposov2025}).    We find 266 DESI DR1 matches within $1.25''$ of a DEIMOS stellar target.  

    \item MMT/Hectochelle:  \citet{walker2023} presented data for stars in Milky Way  satellites using the MMT Hectochelle which has R$\sim$32,000 in a 150\mbox{\AA} wavelength range centered on 5200\mbox{\AA}.    We note little overlap with their M2FS dataset.   We filter sources to remove likely variable sources ({\tt f\_Vlosvar} == 0).  We find 836 MMT/Hectochelle  matches within $1.25''$ of a DEIMOS stellar target.  This sample is dominated by two bright dSphs (UMi and Leo\,I), but includes stars from 19 unique systems.

\end{itemize}
\noindent
  The resulting set of matched stars spans a wide magnitude range.  The majority of Gaia and APOGEE matches are bright ($r>16$), while overlapping DESI and MMT/Hectochelle stars are fainter ($r < 16$).  Thus, these samples test different DEIMOS S/N regions, reflected in the distribution of points along the $x$-axis in left column of Figure~\ref{fig_v_validate}.   Our comparison  additionally includes the compilation of kinematic surveys (the \textquotedblleft Survey of Surveys") from \citet{Tsantaki2022}, noting that the majority of overlapping stars (78 of 126) from this compilation are from APOGEE DR16.

\subsection{Literature Comparison:  Velocity Zero-point and Errors}\label{ssec:validate_v}

We compare our velocity zeropoint and velocity errors to the four spectroscopic surveys described in \S\,\ref{ssec:validate_samples}.   For each matching star, we determine the velocity difference between DEIMOS and the other survey ($\Delta\,v \equiv  v_{\rm DEIMOS} - v_{\rm other}$), plotting these differences as function of DEIMOS S/N in the left column of Figure~\ref{fig_v_validate}.  For all four surveys (Gaia, APOGEE, DESI, and MMT/Hectochelle), we initially found very small ($< 0.2$\kms) velocity zero-point offsets.  However, the zero-point difference was in the same direction in all cases.  We therefore chose to match our zero-point to Gaia by subtracting 0.1\kms\ from our measured velocities.   This zero-point is implemented in the {\tt dmost} pipeline in the same step as when the heliocentric velocity correction is applied.  The resulting zero-point differences, $\Delta v_{\rm zp}$, are listed in Figure~\ref{fig_v_validate}, and are less than 0.03\kms\ in all four cases.

To assess the accuracy of our velocity errors, we normalize $\Delta\,v$ by the quadrature combined velocity errors of DEIMOS and the matched survey.   If the velocity zero-points are correct and the velocity errors from both surveys are truly Gaussian, the normalized histograms shown in the right column of Figure~\ref{fig_v_validate} should be best fit by a normal distribution, $\mathcal{N}(0,1)$.  We report the values of the best-fit Gaussian in the right panels of this figure.  Our velocity errors are in agreement with the APOGEE DR17 errors, $\mathcal{N}_{\rm APOGEE}(0.0,1.0)$, while they are slightly overestimated compared to Gaia DR3, $\mathcal{N}_{\rm Gaia}(0.0,0.8)$, and underestimated compared to DESI DR1, $\mathcal{N}_{\rm DESI}(0.1,1.2)$, and MMT/Hectochelle, $\mathcal{N}_{\rm MMT}(0.0,1.3)$.   We have split the MMT/Hectochelle sample by S/N and see no difference in the resulting comparison.   It is unclear whether these small differences are due to our {\tt dmost} estimates, the other survey, or some combination.  We have also compared to \citet{Tsantaki2022} who homogeneously merged many large kinematic surveys, including older versions of Gaia and APOGEE.   Our velocity errors agree with this sample, $\mathcal{N}_{\rm SoS}(0.3,1.0)$, but we note a small +0.3\kms\ zero-point offset.

\subsection{Comparison to Previous DEIMOS Pipelines} \label{ssec:esys_exp_past}

Our estimates of the velocity error terms are specific to the {\tt PypeIt} and {\tt dmost} pipelines, and are dominated by errors in the wavelength solution.  The same data reduced using different 2D and 1D reduction steps will result in different velocity error floors.  Our velocity error floor of 1.1\,\kms\ is smaller than previously published values for the 1200G DEIMOS setup.   \citet{Martin2014ApJ} determined a systematic velocity floor
of 3.4\kms\ based on repeat measurements of 82 stars reduced with the data reduction pipeline described in \citet{ibata2011}.  \citet{simon07a} estimated a
velocity error floor of 2.2\kms\ based on 49 pairs of independent measurements reduced with the {\tt spec2d} pipeline \citep{newman2013a}.  \cite{Kirby2015a} used
52 stars observed with the same mask over consecutive nights.  These
authors also used the {\tt spec2d} pipeline, but determined velocities
using an improved set of stellar templates, finding a systematic error
term of 1.5\kms.  Our lower velocity error floor of 1.1\kms\ reflects both a far larger sample of repeat measurements, improved wavelength solutions, improved treatment of telluric corrections, and velocity measurements at the individual exposure level.

%%%%%%%%%%%%%%%%%%%%%%%%%%%%%%%%%%
% Figure:   [Fe/H] Calibration
\begin{figure*}[th!]
\epsscale{1.1}
\plotone{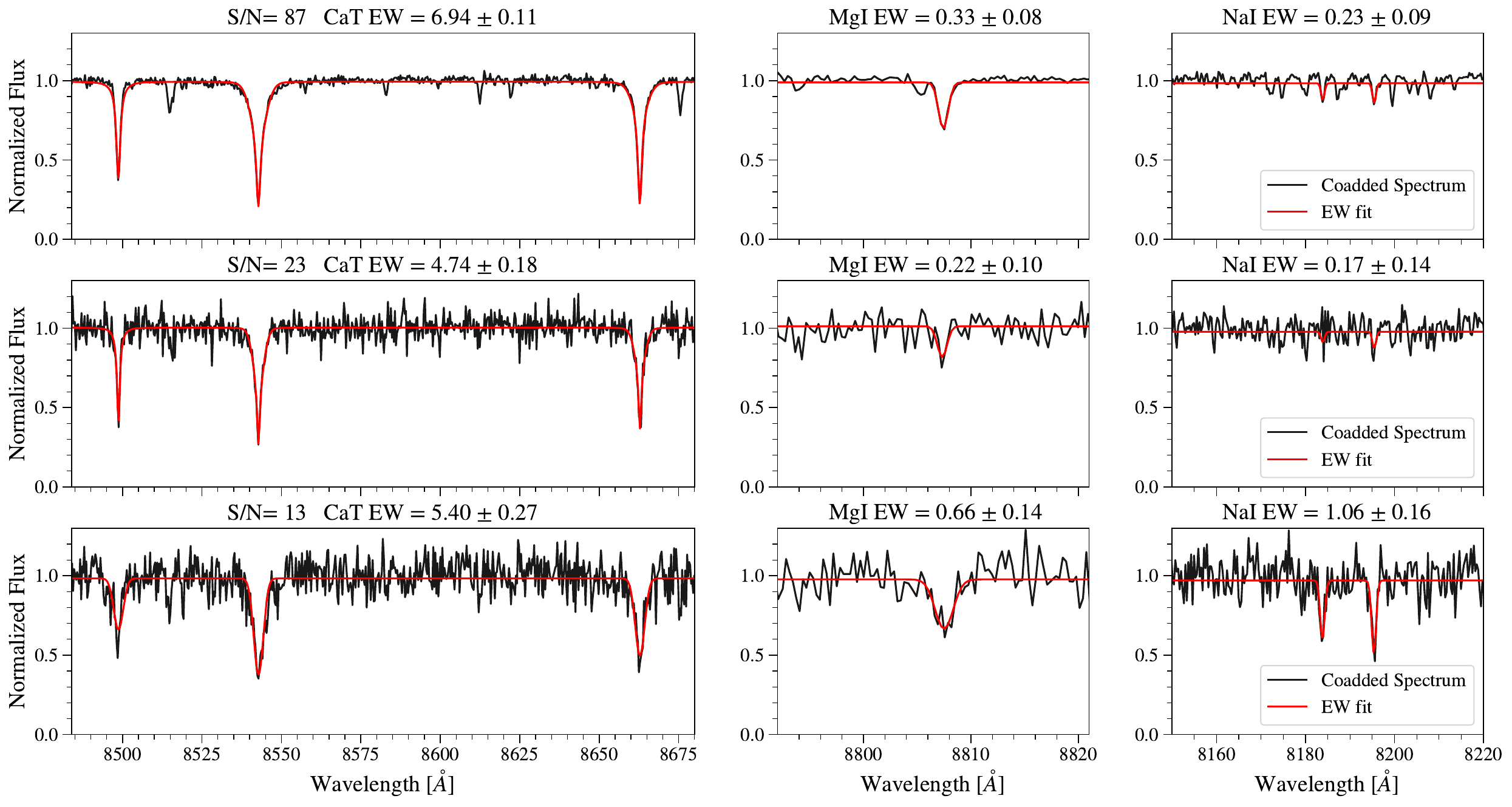}
\caption{Examples of DEIMOS stellar spectra in the wavelength regions used to determine, from left to right, EWs in the CaT, Mg\,I and Na\,I regions.  Coadded spectra are shown (black) for decreasing S/N values (top to bottom).   The EW fits (red) are integrated to determine the EW values for each set of lines.  We fit a single and double Gaussian to the Mg\,I and Na\,I EWs, respectively.   For CaT, we simultaneously fit a Gauss-Lorentz profile to all three Ca\,II lines at high S/N (top two rows) and a Gaussian profile below S/N = 15 (bottom row).  \label{fig_EW_fits}}
\end{figure*}
%%%%%%%%%%%%%%%%%%%%%%%%%%%%%%%%%%

\section{Equivalent Widths and Metallicities}\label{sec: ew}

For every star with a measured velocity, we calculate equivalent widths (EWs) for three sets of absorption lines: the Na\,I doublet and Mg\,I (\S\,\ref{ssec:ew_na_mg}), and the Ca\,II triplet (\S\,\ref{ssec: ew_cat}).   Where possible, we estimate a Ca\,II triplet-based [Fe/H] (\S\,\ref{ssec: cat_feh}) and validate our [Fe/H] metallicity estimates against literature values (\S\,\ref{ssec:compare_feh}).  These quantities are used to improve membership calculations (\S\,\ref{sec_membership}).

We measure EWs using the coadded 1D spectra described in \S\,\ref{ssec:coadd}.   In cases where a star appears on multiple DEIMOS masks, we combine EW measurements by taking the weighted mean of individual mask quantities.  We again use repeated measurements to assess the accuracy of our random errors using the method described in \S\,\ref{ssec:validate_v}.   Examples spectra and analytic fits used to determine EWs are shown in Figure~\ref{fig_EW_fits}.

\subsection{Na\,I and Mg\,I Equivalent Widths}\label{ssec:ew_na_mg}

To improve membership estimates calculated in \S\,\ref{sec_membership}, we measure the EWs of the Na\,I doublet (8183.3, 8194.8\mbox{\AA}) and Mg\,I line (8806.8\mbox{\AA}).   These transitions are sensitive to surface gravity and can be used to differentiate between foreground dwarf stars and more distant giant stars.   

For the Na\,I doublet, we adopt continuum definitions from \citet{Schiavon1997} and normalize the spectrum by fitting a linear function to the continuum passbands.   We simultaneously fit a double-Gaussian profile to the doublet region, fitting the heights, wavelength, and line width of the doublet using a nonlinear least-squares fit.  We fix the wavelength spacing between the lines and force the same line width for both lines.    We determine the Na\,I EW (EW$_{\rm NaI}$) by integrating the fitted profile.  We include the pixel variances in the fitting procedure and calculate the random component of the EW error analytically by propagating the errors in the fitted quantities. 

For the Mg\,I line, we adopt continuum definitions from \citet{Cenarro2001} and normalize the spectrum fitting a linear function to the continuum passbands.   We fit a single-Gaussian profile to the line region defined by \citet{Battaglia2012} and integrate the resulting profile to determine EW$_{\rm MgI}$.  The random component of the error is again determine analytically by propagating the errors on the fitted quantities. 

We assess the accuracy of the EW errors for both EW$_{\rm NaI}$ and EW$_{\rm MgI}$ using independent repeat measurements.   We adopt the same error model as for velocity (Eqn.~\ref{eq:vel_err}), determining both an error multiplier and error floor independently for each EW.   The repeat sample consists of $\sim 2000$ unique pairs.  We find no significant differences when restricting the sample by slit width, S/N, or time.   For both EWs, we find a multiplicative factor of $k_{NaI, MgI} = 0.7$ and an error floor of 0.05\mbox{\AA}.  

\subsection{The Ca\,II Triplet (CaT) Equivalent Width}\label{ssec: ew_cat}

We measure the EW of the Ca II triplet (CaT) lines (8498.0, 8542.1 and 8662.1\,\mbox{\AA}) by first normalizing the spectrum using a linear fit to the continuum regions defined by \citet{Cenarro2001}.   Following \citet{Carrera2013}, we simultaneously model all three Ca\,II lines with a Gaussian-plus-Lorentzian (GL) profile for stars with median S/N $>$ 15 per spectral pixel, and a Gaussian profile for stars at S/N $<$ 15 per pixel.   The GL profiles are usually better fits to the observed profile lines, as it accounts for the non-Gaussian wings of the Ca\,II lines.   However, at low S/N, we find that the GL fits perform poorly and thus switch to the more constrained Gaussian profiles in this regime.  For spectra with S/N $>$ 15,  we refit using a Gaussian line profile if the GL fit is poor as defined by an unphysical line ratio ($EW_{8542}/EW_{8662} <$ 0.2 or $EW_{8542}/EW_{8662} >$ 2.2) or large errors ($>$ 1\mbox{\AA}).  Examples of both profile types are shown in the left panels of Figure~\ref{fig_EW_fits}.

In both profile cases, the best-fit parameters are derived through a nonlinear least-squares fit.  The amplitudes of the three Ca\,II lines are free parameters, however,  we fix the relative line centers and fit for a single line width.   We determine the EWs of all three CaT lines by integrating the fitted profile.   We then sum the three EWs to determine the overall CaT EW.   We include the pixel variances in our fitting procedure.   The random errors on the CaT EWs are propagated from the fitted line parameters.   For the GL fits, we include the covariance terms for a given line between the individual Gauss and Lorentz line amplitudes but not between lines. In rare cases where one of the three CaT lines is not well measured, we replace the value and error of this single line with the fitting functions from \citet[][their Eqns.~2--7]{Heiger2024}, based on the average of the predicted value from other two measured CaT lines.

To assess our CaT EW error estimates, we again examine repeat measurements and adopt the same error model used in \S\,\ref{ssec: esys_exp}. For the Gaussian profile, we use $\sim$700 independent repeat measurements, finding a multiplicative factor of 1.2.  For the GL profile, we use $\sim$800 repeat pairs and determine a smaller multiplicative factor of 0.8 and an error floor of 0.05\mbox{\AA}.  Applying these error terms, we confirm that there is a smooth transition in both the CaT EW and CaT errors as a function of S/N.   

In cases where repeat measurements exist of the same star, we measure EWs in each unique mask and report the error-weighted mean of the individual values.  Unlike our velocity errors, we impose a minimum systematic error floor of 0.05\mbox{\AA} when coadding EW quantities across masks.   This represents the systematic uncertainty in estimating the true continuum of the lines which cannot be reduced via independent measurements.

\subsection{CaT-based Metallicity ([Fe/H]) Estimates}\label{ssec: cat_feh}

Our goal is to provide homogeneous metallicity estimates across a wide range of S/N, thus in this work we report EW-based stellar metallicities ([Fe/H]).  While our {\tt dmost} pipeline provides rough estimates of $T_{\rm eff}$ and [Fe/H], these values are determined on a coarse grid and are not sufficient for science.   Alternative template fitting methods can infer some additional metallicity information in these DEIMOS spectra \citep[e.g.,][]{Kirby2011, Vargas2013, Sandford2020}, however, this is only possible for high S/N $> 25$ spectra, a small fraction of the present sample.    DEIMOS gratings covering wavelengths blueward of 6500\mbox{\AA} (e.g., 1200B or 900ZD) are better suited for detailed elemental abundance analysis \citep[e.g.,][]{Henderson2025}, but at the expense of lower velocity resolution.

Using the CaT EWs measured in \S\,\ref{ssec: ew_cat}, we estimate [Fe/H] using the empirical calibration of \citet{Navabi2025} which is an update to the \citet{Carrera2013} calibration using the same functional form.  We adopt the calibration based on the absolute $V-$band magnitude of individual stars, transforming our $g-$ and $r-$band photometry (\S\,\ref{ssec:matching}) to $V-$band using the piecewise transformations of \citet[][their Appendix\,B.5]{des2021}.   Our [Fe/H] errors include: (1) errors on the calibration parameters from \citet{Navabi2025}, (2) the associated photometric and distance errors, and (3) the measured uncertainties on the CaT EWs.  Rather than directly propagating these three error terms analytically, we instead generate posterior samples for each error component, assuming each obeys either a normal distribution or, in the case of the CaT EW, a normal distribution which we truncate at zero (e.g., the CaT EW cannot be less than zero).  This sampling approach better preserves the possible asymmetry of the resulting [Fe/H] posterior which occurs when the [Fe/H] errors are large.   The metallicity and metallicity errors are determined by propagating the posterior samples from each of these inputs through the \citet{Navabi2025} relation.  Finally, we add a 0.1\,dex error floor to account for inherent scatter in the Navabi/Carrera et al.~calibration itself, noting this term dominates at S/N$>25$.  
 
\begin{figure}[t!]
%\epsscale{1.1}
\includegraphics[width=1.0\columnwidth,trim={0.25cm 0.1cm 0 0 0}]{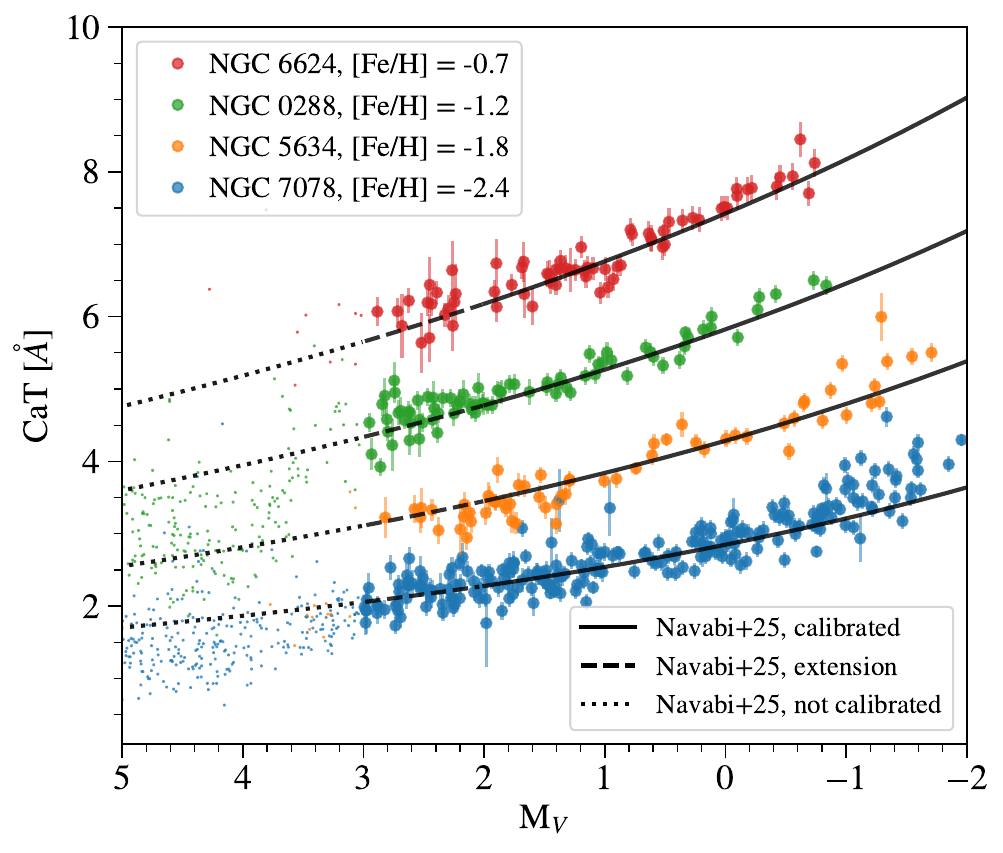}
\caption{Comparison of our CaT-based [Fe/H] measurements for four globular clusters spanning a range of metallicity.   We plot only stars identified as cluster members based on velocity and position in the CMD.  HB stars and stars fainter than $M_V >3$ are plotted as smaller symbols.   The black solid lines are the empirical relationship of \citet{Navabi2025}, plotted at the mean [Fe/H] determined for each cluster.   The relationship is calibrated brighter than $M_V < 2$ (solid lines), but we extend this to $M_V =3$ (dashed line) in this work.   The relationship appears to extend below this limit (dotted line) for metal-poor stars, but we do not report values in this region. \label{fig_FeH_GC}}
\end{figure}

The \citet{Navabi2025} empirical relationship between CaT and [Fe/H] is calibrated for red giant branch stars with absolute magnitudes brighter than $M_V = 2$.   Due to our larger and fainter dataset of globular cluster stars, we confirm the common practice of extending this relation to $M_V = 3$, one magnitude below the calibrated limit in Figure~\ref{fig_FeH_GC}.   We select four globular clusters with a range of [Fe/H] and plot CaT EW for all stars as a function of $M_V$.  The four lines in Figure~\ref{fig_FeH_GC} indicate the Navabi relationship plotted at the mean [Fe/H] for each globular cluster, determined from the measured data.  All four globular clusters agree with published values from \citet{pace2024} within 0.05\,dex.  The calibrated region is indicated by the solid black lines, but clearly extends into the dashed region.  Below the main sequence turn-off at high metallicities, the CaT EW values diverge below the linear Navabi calibration.   At low metallicities ([Fe/H]$ < -2$), the relation may extend below the main sequence turn-off ($M_V > 3$), but we leave this extension to future work.   We report [Fe/H] and the associated errors for stars with known distances (e.g., member stars identified in \S\,\ref{sec_membership}) that are brighter than $M_V < 3$.

%%%%%%%%%%%%%%%%%%%%%%%%%%%%%%%%%%
% Figure:   [Fe/H] Calibration
%%%%%%%%%%%%%%%%%%%%%%%%%%%%%%%%%%

%%%%%%%%%%%%%%%%%%%%%%%%%%%%%%%%%%
% Figure:   [Fe/H] Calibration
\begin{figure*}[th!]
\includegraphics[width=1.\textwidth]{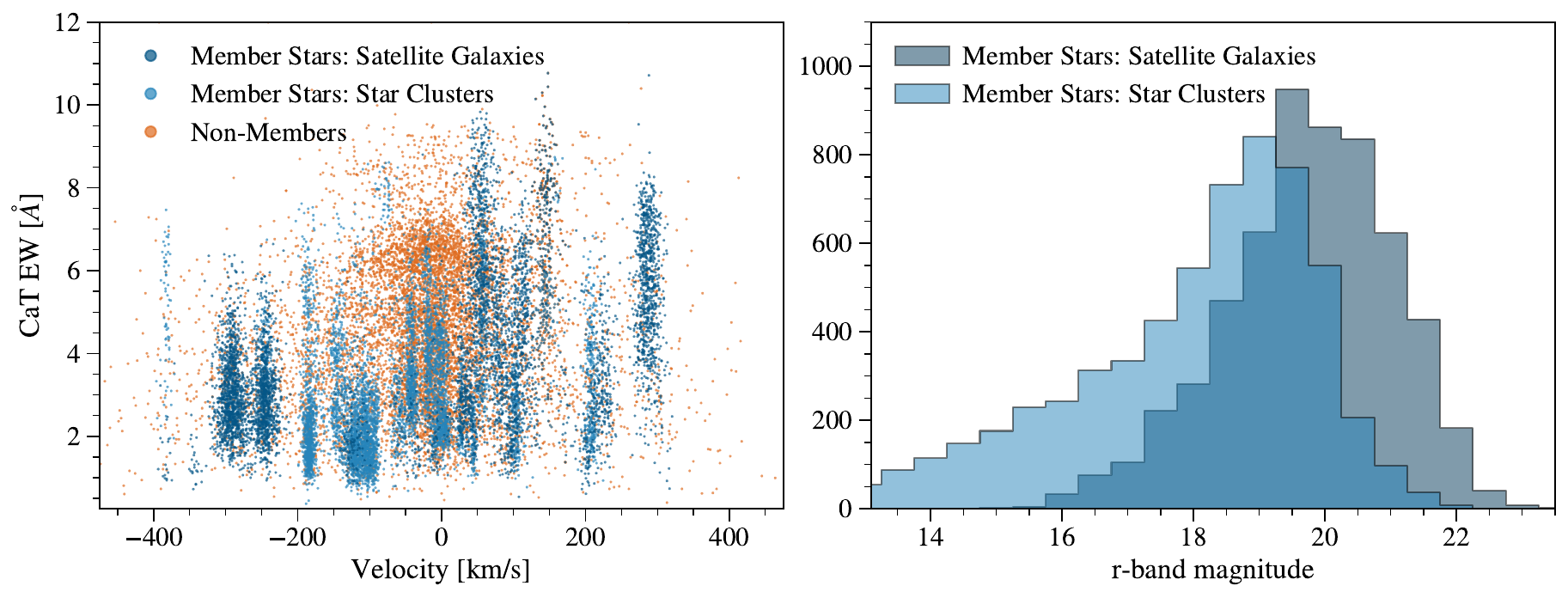}
\caption{{\it Left:\/} DEIMOS velocities versus the CaT EW  plotted for member (blue) and nonmember stars (orange).   Nonmember stars are centered neared zero velocity and show a large range of CaT EWs, as expected for a foreground Milky Way population.   Member stars in our \nobj satellite systems are seen at specific velocities, often with a smaller range of CaT EWs, roughly corresponding to a narrower [Fe/H] distribution. {\it Right:\/} The distribution of $r-$band magnitudes for member stars, split by members in stars clusters (light blue) and satellite galaxies (dark blue).   \label{fig_rmag}}
\end{figure*}
%%%%%%%%%%%%%%%%%%%%%%%%%%%%%%%%%%

\subsection{Validating [Fe/H] and [Fe/H] Errors} \label{ssec:compare_feh}

For stars with CaT-based [Fe/H] values (\S\,\ref{ssec: cat_feh}), we assess the accuracy of our error estimates and then evaluate the absolute zero-point of our [Fe/H] estimates against other methods in the literature.

As a sanity check on our [Fe/H] errors, we first estimate the internal metallicity spread of the two dozen Milky Way globular clusters available in the DEIMOS archive (Table~\ref{table_objects}). Given the accuracy of our CaT-based metallicities, we do not expect to detect metallicity spreads in these clusters.   Internal spreads are well known in globular clusters for light elements \citep{bastian2018}, and are sometimes observed in [Fe/H] \citep{bailin2019ApJS,bailin2022}.  While photometric estimates based on the width of the red giant branch suggest metallicity spreads up to 0.3\,dex \citep{Milone2017,Legnardi2022,Lardo2023}, spectroscopic studies tend to produce smaller values of order 0.05\,dex \citep{Carretta2009,Latour2025}.  In either case, these are largely below the sensitivity of our CaT-based measurements, and we thus expect to detect no metallicity spreads in these systems.  We evaluate the presence/absence of a spread by modeling the individual [Fe/H] measurements as a Gaussian distribution.  Our model assumes the distribution of measured metalicities is determined by both measurement errors and an intrinsic, internal metallicity spread.   We run the model twice, comparing the Bayesian evidence in the case where the internal spread is a free parameter and one in which the spread is set to zero.  We apply this to 24 globular clusters containing 20 or more DEIMOS stars with measured [Fe/H].  For 23 of 24 globular clusters, the distributions strongly prefer a model with zero intrinsic metallicity spread.   The exception is NGC~2419 which shows a metallicity spread of 0.19\,dex, consistent with published results from \citet{Cohen2010}, noting that \citet{larsen2019} interpret this as a spread in $\alpha$-elements rather than iron.  This exercise implies that we have not significantly underestimated our [Fe/H] errors, with the results from NGC~2419 providing some evidence that these errors are not overestimates.  We further explore internal metallicity spreads in Paper\,II.

We next evaluate both the [Fe/H] zero-point and accuracy of our [Fe/H] error estimates against literature values.  We use the same methods described in \S\,\ref{ssec:validate_v} when evaluating these quantities for velocity (i.e., Figure~\ref{fig_v_validate}).  We first compare to \citet{Tolstoy2023}, who measured CaT-based [Fe/H] metallicities via data from the VLT/FLAMES spectrograph.  Based on 233 overlapping stars in the Sculptor dSph, we see a small [Fe/H] zero-point offset of less than 0.1\,dex, and errors which are roughly in agreement.  Similarly, we compare to \citet{walker2023} whose MMT/Hectochelle estimates are performed at bluer wavelengths via spectral fitting.   Based on 569 overlapping stars, we find agreement in our [Fe/H] error estimates.  Walker et al.~note a 0.25\,dex offset between their [Fe/H] values and APOGEE.  Adding this offset, we also find good zero-point agreement.  Both comparison samples above include only metal-poor stars with [Fe/H] $<-1$.

We next compare to a more metal-rich sample from the SDSS APOGEE DR17 database \citep{sdss_dr17} with 86 stars with [Fe/H] estimates in common.   We find excellent agreement when comparing to both APOGEE [M/H] and [Fe/H] using the \citet{Navabi2025} CaT calibration.  We note that the widely used \citet{Carrera2013} calibration over-estimates [Fe/H] by up to 0.2\,dex above [Fe/H]$ > -1$.   Finally, we compare to the spectral fitting results from \citet{kirby08a} based on the same DEIMOS spectra.  We again find good agreement across the full range of [Fe/H], but note that the scatter between our CaT estimates and these spectral fitting results increases towards lower metallicities.  This is may be due to our EW fitting procedure, which imposes the same line widths for all three CaT lines.    We leave improvement of the CaT fitting method to future work.

%%%%%%%%%%%%%%%%%%%%%%%%%%%%%%%%%%
% Figure:   exGalaxies
\begin{figure*}[ht!]
 \includegraphics[width=1.01\textwidth]{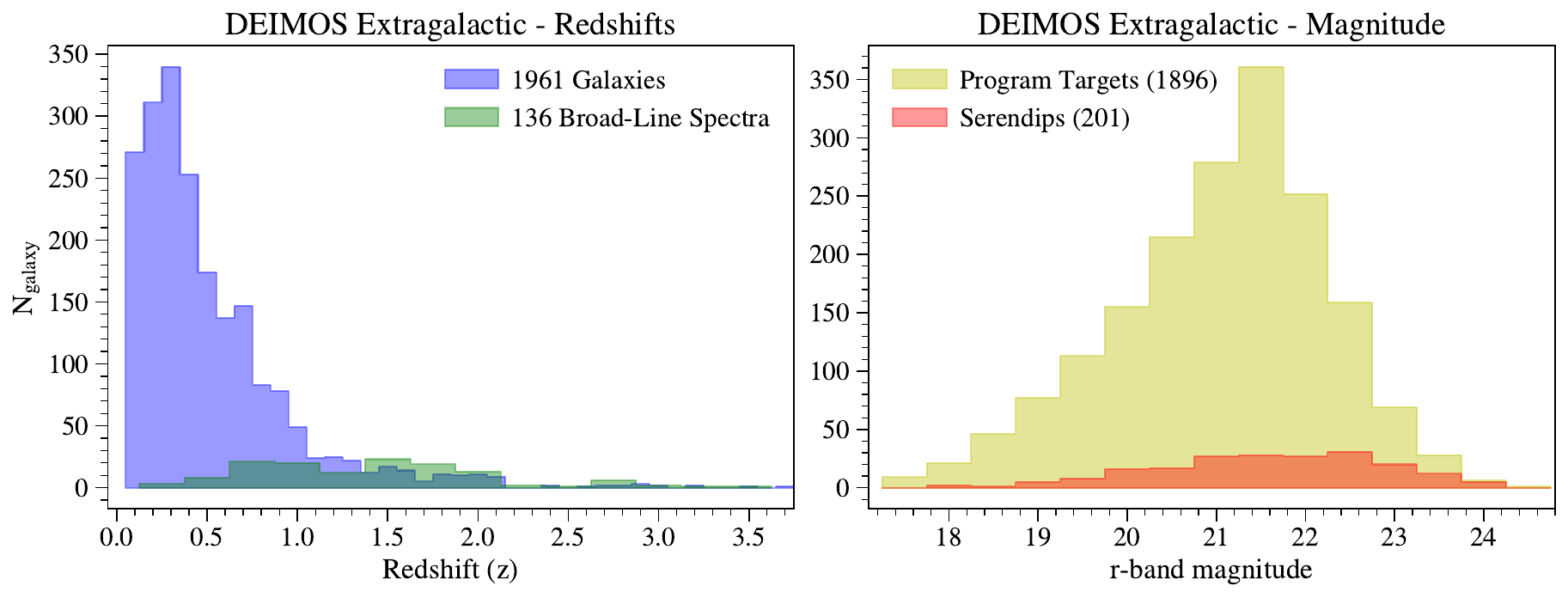}
 \caption{We identify \nexgal nonstellar, extragalactic objects and measure redshifts using the automated redshift software {\tt Marz} \citep{marz}.  {\it Left:\/}  Redshift distribution of extragalactic sources (both emission and absorption-line systems) shown in blue and broad-line QSO-like spectra in green.   {\it Right:\/}  Extragalactic objects which are program targets (yellow) were selected as point-like and are likely more compact as compared to both the average galaxy population and our serendipitous detections (orange).  }\label{fig_exgal}
\end{figure*}
%%%%%%%%%%%%%%%%%%%%%%%%%%%%%%%%%%

\section{Merged Catalogs and Memberships}\label{sec_membership}

For each of our \nobj Milky Way stellar systems, we create a merged catalog containing measured DEIMOS quantities and compiled literature information for each unique source (both targeted and serendipitous) within three effective radii of the system's center.   We present associated ground-based photometry and, where available, Gaia DR3 quantities for individual stars (\S\,\ref{ssec:matching}).     We briefly discuss measurements for extragalactic sources (\S\,\ref{ssec:exgal}).   For each star, we then determine a membership probability to a given Milky Way satellite galaxy or star cluster (\S\,\ref{ssec:member_cuts}).  Data access and description of the catalogs are covered in \S\,\ref{sec_catalogs}.

\subsection{Matching Photometry and Proper Motions}\label{ssec:matching}

For all spectroscopic targets, we match to both ground-based optical photometry and the Gaia DR3 catalog in order to assess membership and determine physical properties.   We use a matching radius of $1.25''$ to allow for errors in the original DEIMOS targeting catalogs.  In Tables~\ref{table_stars}--\ref{table:schema}, we report dereddened $g$- and $r$ magnitudes for all targeted sources in the DECam filter system.

Our primary photometry source is $g$- and $r$-band imaging from the DESI Legacy Imaging Survey DR10 \citep{Dey2019} which has a median limiting depth of $r\sim 23.3$.  We report their dereddened AB magnitudes ({\tt dered\_mag\_g/r}).  The Legacy survey applies Galactic extinction values derived from \citet{sfd98} with updated coefficients from \citet{Schlafly2011}.  

When Legacy imaging is not available, we use the photometric catalogs from the CFHT-Magellan MegaCam Survey \citep{munoz2018a}, which provide homogeneous $g$- and $r$-band photometry for Milky Way satellites beyond 25\,kpc with a limiting depth of $r\sim25.3$.   To maintain homogeneity with the Legacy DR10 photometry, we transform the Mu{\~n}oz et al.~SDSS-based magnitudes into the DECam system using the transformations from Appendix B of \citet{Dey2019}, and correct for extinction as described above.  Neither Mu{\~n}oz et al.~nor the Legacy Survey provide photometry in the crowded inner regions of nearby bright globular clusters, and in these cases we use the re-reduction of the SDSS DR7 photometry in crowded regions by \citet{An2008}, again transformed to the DECam system and corrected for Galactic extinction.   For objects which do not lie in any of these surveys, we use Pan-STARRS1 DR2 \citep{panstarrs} photometry, transformed into the DECam system. 

The source of photometry for each system is listed in Table~\ref{table_objects}.   Photometry ($g-$ and $r-$band) is provided  for all individual stars and extragalactic targets as AB magnitudes in the DECam filter system and corrected for Galactic extinction (Tables~\ref{table_stars}-\ref{table:schema}).  We find photometric matches for over 97\% of targeted sources with measured velocities.  Target stars without photometry are largely located in the inner regions of dense globular clusters.  For serendipitous sources (\S\,\ref{ssec:pypeit}, a few percent of the total sample), we increase the matching radius to $2''$, finding a lower matching rate of 70\%.   The median magnitude of the full sample is r = 19.7.  The magnitude distribution of sources identified as member stars is shown in Figure~\ref{fig_rmag}. For the subset of systems identified as a dwarf galaxy (dark blue histogram, right panel), the median magnitude of individual member stars is $r = 20.4$.

Each DEIMOS mask is associated with one unique Milky Way satellite (Figure~\ref{fig_intro}), with the exception of five masks centered on NGC\,6715 (M54), the central globular cluster of the Sgr dSph.  These five masks contain both NGC\,6715 and Sgr stars which are separated both spatially and in metallicity.  We impose a soft metallicity cut [Fe/H]$\sim-0.7$ inside $4'$ of NGC\,6715's center, and report values for member stars in each system separately.

The majority of DEIMOS stars in this work lie below the limiting magnitude of Gaia.  Nonetheless, we match to the Gaia DR3 catalog \citep{GaiaDR3} using a $1.25''$ matching length and report proper motions and parallaxes where available.   In addition to parallaxes and proper motions, we include in our extended catalogs (\S\,\ref{ssec:data_access}) the Gaia quantities {\tt astrometric\_excess\_noise} and {\tt phot\_variable\_flag}.  We find that these quantities are loosely correlated with our velocity variability flag (\S\,\ref{ssec:flag_var}) and are used in helping to identifying variable sources when available.

%%%%%%%%%%%%%%%%%%%%%%%%%%%%%%%%%%
% Figure:  Membership
\begin{figure*}[th!]
 \includegraphics[width=1.\textwidth]{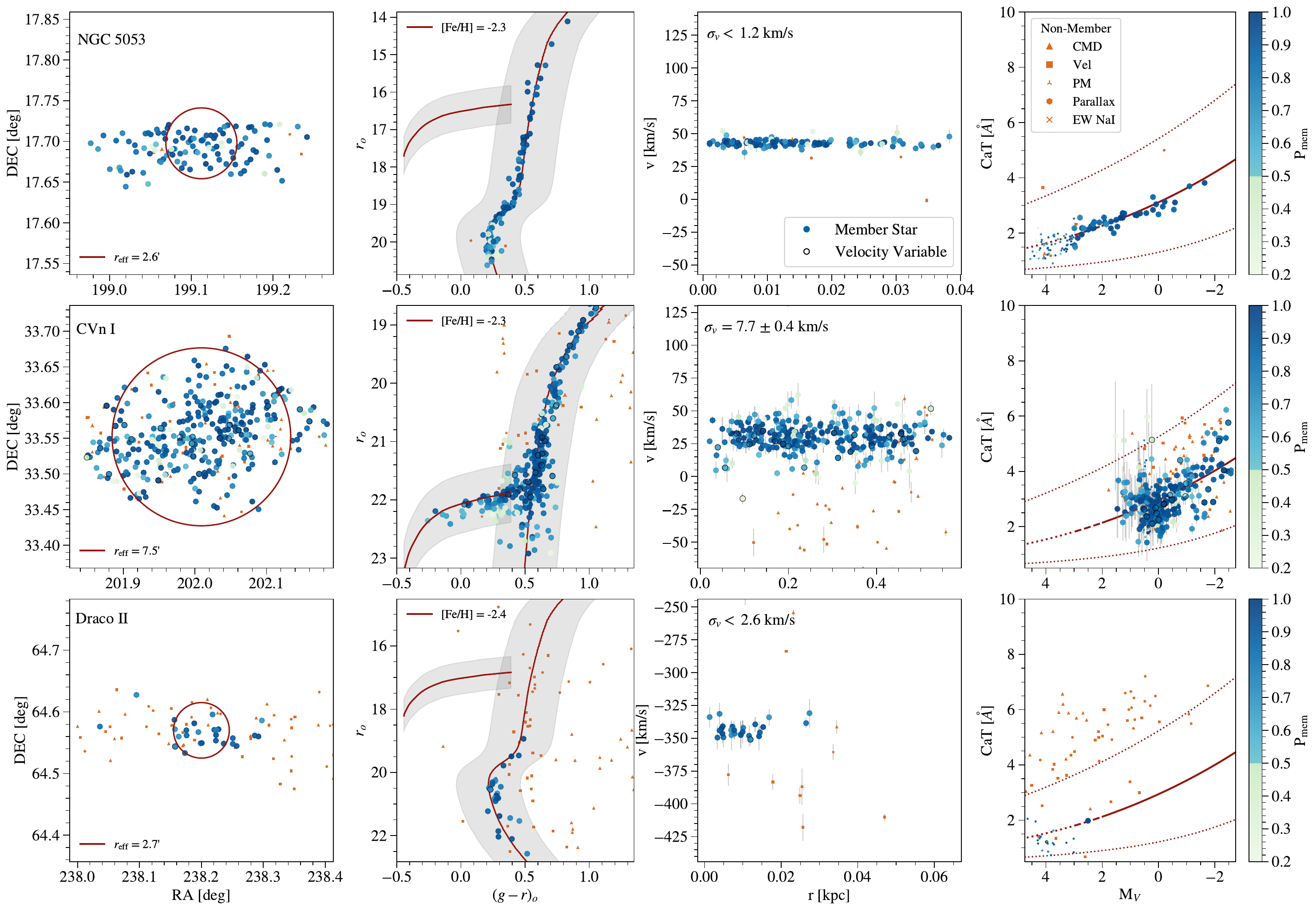}
\caption{Membership selection for three examples from the \nobj Milky Way stellar systems presented in this paper.    From left to right, we plot the spatial distribution of stars, the CMD, the velocity distribution as a function of projected radius, and the CaT EW as a function of absolute magnitude.  Nonmembers are shown as orange symbols; the symbol shape denotes the primary reason a star was classified as a nonmember. 
   \label{fig_members}}
\end{figure*}
%%%%%%%%%%%%%%%%%%%%%%%%%%%%%%%%%%

\subsection{Background Galaxies and Quasars}\label{ssec:exgal}

For the DEIMOS observations presented in this paper, the primary targets are resolved stars, however, a substantial number of spectra turned out to be extragalactic.     As an ancillary data product, we present redshifts for these extragalactic sources.  The redshift and magnitude distributions of these sources are shown in Figure~\ref{fig_exgal}.   Extragalactic objects which are program targets (right panel) were selected as point-like and are likely more compact as compared to both the average extragalactic population and our serendipitous detections.   We report the {\tt Marz}-determined redshifts as described in \S\,\ref{ssec:marz}.    The extragalactic redshift fits are  inspected and visually classified as narrow-line galaxy spectra or broad-line QSO-like spectra.      Of the \nexgal extragalactic sources, 136 sources are broad-line QSO-like spectra with line widths of 1000\kms\ or more.  Redshifts and classifications of extragalactic sources are presented on their own in Table~\ref{table_exgal}, and are included as part of the extended catalog described in \S\,\ref{ssec:data_access} and Table~\ref{table:schema}.

\subsection{Membership Estimates}\label{ssec:member_cuts}

In this work, we analyze DEIMOS data taken within three effective radii of all known Milky Way stellar satellites (Figure~\ref{fig_intro} and \S\,\ref{ssec:koa}). 
For the majority of the \nstar stars presented in this paper (Table~\ref{table_stars}), whether or not they belongs to a bound Milky Way stellar satellite is unambiguous.  That is, based on a star's measured properties, it can be securely identified as either a foreground star belonging to the Milky Way itself or as part of a known dwarf galaxy satellite,  globular or faint star cluster.  However for some fraction of stars, this identification is more ambiguous.  We next assess a  star's likelihood of being associated with a given satellite based on its position in the color-magnitude diagram (CMD), line-of-sight velocity, EW measurements, and, where possible, Gaia quantities.   

Given the wide range of physical properties of our systems ($M_{\rm stellar} = 10^{1.5}$ to $10^{7.5}\,M_{\odot}$, $M_V = 2$ to $-14$), we opt for a mixture of binary cuts and membership probability estimates.  Our generalized membership criteria are designed to adequately identify members across the range of systems, but is not optimized on a system-by-system basis.   Our membership estimates begin with rough guesses for a system's distance, radial velocity, and metallicity.   When available, we adopt these initial values from the \citet{pace2024} compilation.   If published quantities are not available, we iterate our membership method below to determine best first guesses.  

Detailed below, our membership estimates begin with basic quality cuts (1). Similar to \citet{collins2020}, we first determine a probability estimate based on a star's distance to a fiducial isochrone in color-magnitude space (2).  We then remove stars with a surface gravity-sensitive absorption line (EW$_{Na\,I}$) or Gaia parallax securely suggesting it is a foreground star in the Milky Way (3).    We then determine a probability estimate based on a star's proper motion from Gaia when available (4), CaT-based metallicity (5), and velocity (6).   The velocity and metallicity distributions are key observables.   Our probabilities are designed only to reject significant outliers in these distributions.   We multiply these six criteria together to determine $P_{\rm mem}$, the final membership probability for each star (7).  To determine the internal velocity dispersions of each system (\S\,\ref{ssec:results}), we impose a subjective membership threshold of $P_{\rm mem} > 0.5$ and remove stars flagged as velocity variables (8). We provide this final sample as a binary member/nonmember flag, $P_{\rm mem\_novar}$.   In detail, our membership selection criteria are as follows:
\vskip 0.2cm
\noindent
1. {\it Basic quality cuts ($P_{\rm basic}$)}:   We perform basic quality cuts, removing extragalactic sources and sources without a stellar velocity measurement ({\tt marz\_flag} $>1$ or {\tt v\_err $< 0$}).   We remove stars with large velocity errors ({\tt v\_err} $>$ 15\kms).  For serendipitous, non-targeted objects ({\tt serendip == 1}), we require a photometric match, but allow targeted stars to proceed in the rare cases when photometry is not available.

\vskip 0.2cm
\noindent
2. {\it CMD distance ($P_{\rm CMD}$)}:  We evaluate each star's distance to a matched single-age isochrone from the PAdova and tRieste Stellar Evolutionary Code \citep[PARSEC;][]{bressan2012}.  We use a 12\,Gyr isochrone, adopting the metallicity and distance for each Milky Way system from \citet{pace2024}.    We replace the HB region with a fiducial HB from M\,92 which provides a wider allowable region and better match to most systems.   We calculate the minimum perpendicular distance between this combined isochrone and the star's location on the CMD ($d_\perp$).   This CMD distance and the magnitude error of the star ($\delta_{\rm mag}$, the quadrature sum of the $r$- and $g$-band magnitude errors), are then used to evaluate a CMD probability:
\begin{eqnarray}
    P_{\rm CMD} = \exp{ \Biggl( - \frac{d_{\perp}^2}{2 \bigl[ \sigma_{\rm CMD}^2 + \delta_{\rm mag}^2 \bigr]} \Biggr) },
\end{eqnarray}

\noindent
where $\sigma_{\rm CMD}$ is set to 0.2\,mag to account for possible age/metallicity spreads and errors in distance.  However, for bright dSph ($M_V < -9$), we increase this threshold to $\sigma_{\rm CMD} = 0.3$ to better account for known age and metallicity spreads.  Examples of our CMD region are shown in the left-middle panels of Figure~\ref{fig_members}.

\vskip 0.2cm
\noindent
3. {\it Na\,I EW and Gaia parallax ($P_{\rm EW}$, $P_{\varpi}$}):  We  remove stars whose EW$_{Na\,I}$ (measured in \S\,\ref{sec: ew}) suggests it is a foreground dwarf star.  EW$_{Na\,I}$ is sensitive to a star's surface gravity and can be used to discriminate between dwarf and giant stars.  Following \citet{Schiavon1997}, we remove stars with (EW$_{Na\,I}$ - $\sigma_{{\rm EW}_{Na\,I}}) >$ 1\mbox{\AA} and $M_V < 4$.   While we also measure EW$_{Mg\,I}$, this line is redundant but less discriminating than our Na\,I criteria.  For those stars with a Gaia DR3 parallax, we remove stars based on a parallax criterion as:  $(\varpi - 3 \sigma_\varpi) > 2 \varpi_{\rm obj}$ where $\varpi$ is the Gaia parallax with associated uncertainty $\sigma_\varpi$, and $\varpi_{\rm obj}$ is the parallax of the system itself.  In most cases, the system parallax is effectively zero.

\vskip 0.2cm
\noindent
4. {\it Proper motion ($P_{\rm pm}$)}:   In systems with two or more stars passing the above three criteria with available Gaia DR3 proper motion measurements, we calculate a proper motion probability.   We first determine the proper motion of the system ($\mu_{\alpha*},\mu_\delta$), using the members identified via the above criteria and a crude velocity cut. We determine the $i^{\rm th}$ star's distance to the system proper motion as $d_{\rm pm} = ((\mu_{\alpha*} - \mu_{\alpha*,i})^2 + (\mu_\delta - \mu_{\delta,i})^2)^{1/2}$, along with its observational error $\delta_{\rm pm}$.  We determine a probability as:
\begin{eqnarray}\label{eq:mem_pm}
    P_{\rm pm} = \exp{ \Biggl( - \frac{d_{\rm pm}^2}{2 \bigl[ \sigma_{\rm pm}^2 + \delta_{\rm pm}^2 \bigr]} \Biggr) },
\end{eqnarray}
We set $\sigma_{\rm pm}$ to $1.0\,{\rm mas\,yr^{-1}}$ (corresponding to 100\kms\ at 20\,kpc), allowing for error in the system proper motion estimate, which is particularly important in cases where there are only a small number of stars above the Gaia magnitude limit.  We loosen this tolerance to $2.0\,{\rm mas\,yr^{-1}}$ inside two effective radii in the crowded regions of bright globular clusters and dSphs.
    
\vskip 0.2cm
\noindent
5.  {\it Metallicity ($P_{\rm [Fe/H]}$}):  The metallicity and velocity distributions of a given system are key observables; aggressive removal of outliers during membership selection can bias physical interpretation.   The metallicity distribution of dwarf galaxies is expected to be a skewed Gaussian, with longer tails toward low metallicity.   Using the combined memberships above and a crude velocity cut ($\pm 25$\kms), we first determine a mean metallicity (${\rm [Fe/H]}_{\rm  guess}$) and metallicity spread ($\sigma_{\rm [Fe/H], guess}$) for a given system, imposing a minimum spread of $\sigma_{\rm [Fe/H], guess} = 0.1$\,dex.   We then determine the metallicity probability, $P_{\rm [Fe/H]}$, for each star as:
\begin{eqnarray}\label{eq:mem_feh}
    P_{\rm [Fe/H]} = \exp \Biggl( -\frac{({\rm [Fe/H] - [Fe/H]_{\rm  guess}})^2}{2 \bigl[\sigma_{\rm [Fe/H]}^2 + \delta_{\rm [Fe/H]}^2 \bigr]} \Biggr),
\end{eqnarray}
where $\sigma_{\rm [Fe/H]} = 3 \sigma_{\rm [Fe/H],guess}$.    For stars with a lower than mean metallicity in dwarf galaxies, where we expect non-Gaussian tails, we increase this threshold to $\sigma_{\rm [Fe/H]} = 4 \sigma_{\rm [Fe/H],guess}$.  Stars with $M_V > 3$, where the CaT EW calibration is not valid, are set to unity for this criterion.

\vskip 0.2cm
\noindent
6.  {\it Velocity ($P_{\rm v}$)}:   We assume the velocities of member stars follow a Gaussian distribution and assign lower membership probabilities for $3\sigma$ outliers in this space.  We determine an initial guess of the velocity and velocity dispersion based on the above membership criteria and a crude velocity cut as above.  We then determine a star's distance from this initial guess.   For a star with velocity $v \pm \delta_v$, we calculate the following:
\begin{eqnarray}\label{eq:mem_vel}
    P_{\rm v} = \exp \Biggl( -\frac{(v - v_{\rm  guess})^2}{2 \bigl[\sigma_v^2 + \delta_v^2 \bigr]} \Biggr),
\end{eqnarray}
where $\sigma_v = 3\sigma_{\rm guess}$.   We iterate this step twice for systems with five or more member stars, replacing the crude velocity cut with $P_{\rm v}$ and re-determining $v_{\rm  guess}$ and $\sigma_{\rm guess}$.

\vskip 0.2cm
\noindent
7.  {\it Membership Probability ($P_{\rm mem}$)}:  We combine the criteria above to determine a final membership probability for each star.   Individual criteria range from zero (nonmember) to near unity (secure member).  If a star does not have a measured quantity for a given criterion, the associated probability is set to unity.  Our final membership probability is  determined as:
\begin{eqnarray}
    P_{\rm mem} = P_{\rm basic} ~ P_{\rm CMD}~  P_{\rm EW} ~P_{\varpi} ~ P_{\rm pm} ~ P_{\rm [Fe/H]}  ~P_{v}
\end{eqnarray}
We impose no criteria on the projected radial distance from the system's center.  For most science applications, we recommend a membership criteria threshold of $P_{\rm mem} > 0.5$.   For a more inclusive sample this can be loosened to $P_{\rm mem} > 0.2$.   Examples of membership selection for three systems are shown in Figure~\ref{fig_members}.  Stars with $P_{\rm mem} > 0.2$ are shown as larger color-coded circles, with a clear color difference for stars with $P_{\rm mem} > 0.5$.  Nonmembers are shown as orange symbols; the symbol shape denotes the primary reason a star was classified as a nonmember.

\vskip 0.2cm
\noindent
8.  {\it Velocity Member Sample ($P_{\rm mem\_novar}$)}:  Stars passing the membership probability criterion of $P_{\rm mem} > 0.5$ can be considered secure members of the system for most purposes.  However, for the purpose of determining the internal velocity dispersion of a system, we further refine the sample by removing stars identified as velocity variables (often RR Lyrae member stars or unresolved binary stars) using the {\tt flag\_var} flag set in \S\,\ref{ssec:flag_var}.     For clarity, we set $P_{\rm mem\_novar}$ to zero for nonmembers and unity for members, but note this estimate includes subjective choices which may not be appropriate on a case-by-case basis.

\vskip 0.2cm
\noindent
Based on these criteria, we identify \nmembers member stars across \nobj stellar Milky Way systems.   The \nonmembers stars identified as nonmember stars are largely Milky Way disk stars, although there are likely some thick disk and halo members. The full sample is shown in CaT-velocity space in the left panel of Figure~\ref{fig_rmag}.   Nonmember stars (orange symbols) are centered neared zero velocity with a wide roughly Gaussian distribution corresponding to the Milky Way itself.   Member stars in our \nobj satellite systems are seen at specific velocities, often with a smaller range of CaT EWs, roughly corresponding to a narrower [Fe/H] distribution.

\begin{figure}[t!]
 \includegraphics[width=1.05\columnwidth,trim={0.2cm 0cm 0 0 0}]{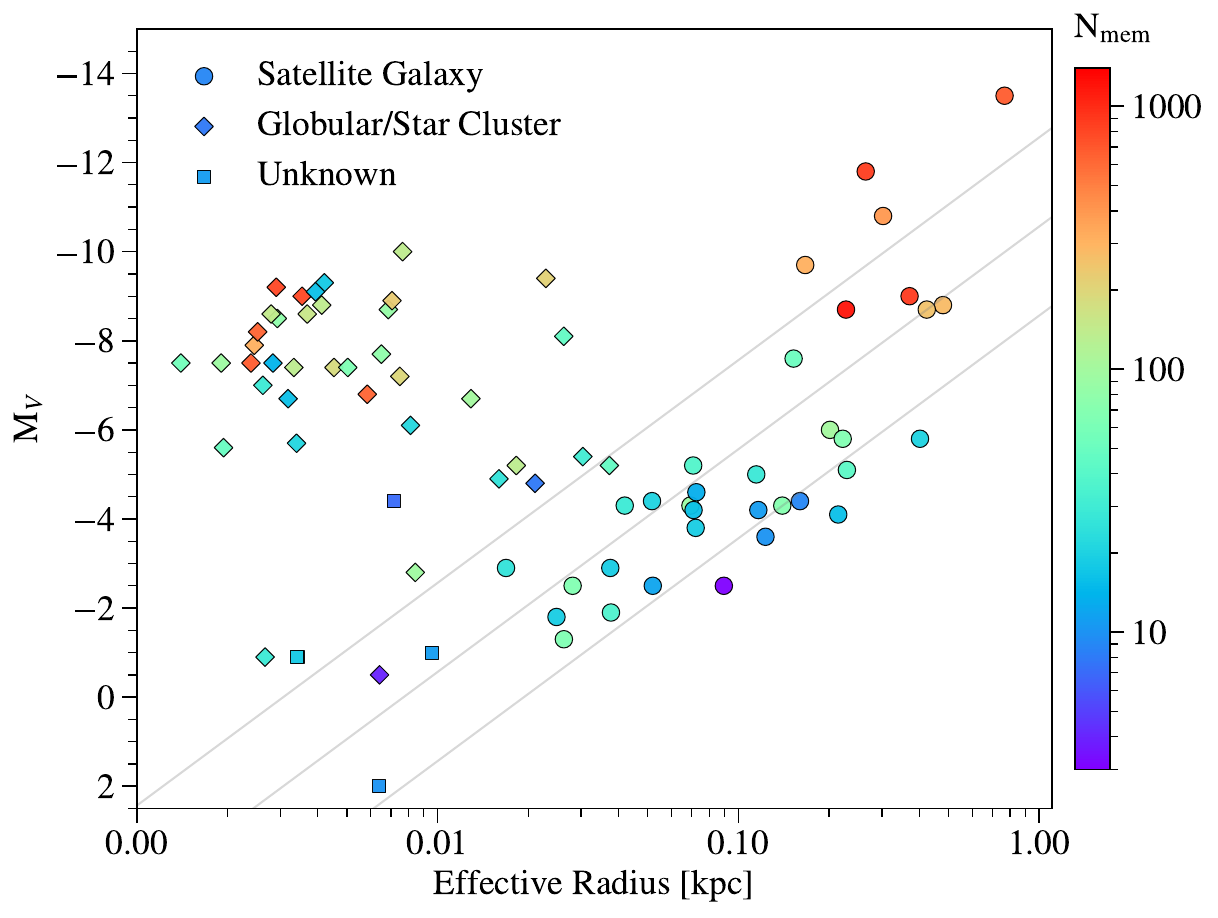}
%\vskip -0.5 cm
\caption{Absolute magnitude ($M_V$) vs.~effective radius, taken from the compilation of \citet{pace2024}, for the \nobj Milky Way systems presented in this paper. Systems are color-coded by the number of DEIMOS member stars with P$_{\rm mem} > 0.5$.  Thin diagonal lines of constant effective surface brightness are shown from top to bottom: 26, 28, 30\, mag/arcsec$^2$.  Satellite galaxies are shown as circles; systems identified as globular clusters or faint star systems are shown as diamonds.  Systems with ambiguous identifications are shown as squares.  
   \label{fig_nstars}}
\end{figure}

In Figure~\ref{fig_nstars}, we plot the absolute magnitude versus effective radius ($M_V$ vs.~$r_{\rm eff}$) of our \nobj stellar Milky Way systems, color-coded by the number of identified member stars (P$_{\rm mem} > 0.5$).  The most luminous systems, bright dSphs and globular clusters, contain up to 1200 DEIMOS member stars, while fainter systems contain as few as three.   In a companion paper \citep[Paper II;][]{geha_paper2}, we explore the physical properties of these systems, including dynamical mass estimates, mean metallicity and internal metallicity spreads.

%%%%%%%%%%%%%%%%%%%%%%%%%%%%%%%%%%
% Figure:  NSTARS
%%%%%%%%%%%%%%%%%%%%%%%%%%%%%%%%%%

\section{Data Access and Results}\label{sec_catalogs}

This paper describes a homogeneous dataset of stars observed with the Keck/DEIMOS 1200G grating across a large fraction of Milky Way stellar satellites, including dwarf galaxies and globular clusters.   In Paper\,II, we use this database to explore the physical properties of these systems.  To facilitate other science cases,  we describe catalogs and data access in \S\,\ref{ssec:data_access} for both reduced data and measured quantities.   A key goal of this work is improved velocity precision.  In \S\,\ref{ssec:results}, we demonstrate this improvement by comparing our estimates of the internal velocity dispersion to the literature.

\subsection{Data Access}\label{ssec:data_access}

For each Milky Way stellar system, we create a merged catalog containing measured DEIMOS quantities.  We provide data tables for the combined measured quantities in Tables~\ref{table_objects}--\ref{table:schema}.   We list these \nobj systems, along with literature and computed system properties in Table~\ref{table_objects}.   Observational details for each DEIMOS mask are listed in Table~\ref{table_obs}.  Measured quantities and compiled literature information for each unique star (both targeted and serendipitous) observed with DEIMOS within three effective radii of the system's center are listed in Table~\ref{table_stars}.  In cases where stars are measured across multiple DEIMOS masks, we present the weighted mean of properties measured in each mask.   The KOA Contributed Dataset includes access to the individual multiepoch measurements.   Objects which turned out to be extragalactic sources are provided with measured redshifts in Table~\ref{table_exgal}.   Finally, we provide an expanded set of measured quantities for both stars and extragalactic sources, listing the schema for this table in Table~\ref{table:schema}.   In the Appendix, we further describe the available data tables.   

Other data products described in this paper, including 2D reduced science images, 1D reduced spectra, and access to individual multiepoch measurements are available as a Contributed Dataset which can be downloaded from the Keck Observatory Archive (DOI: 10.26135/KOA7).  In a companion paper \citep[Paper II;][]{geha_paper2}, we provide an expanded set of integrated quantities for each Milky Way stellar system.

%%%%%%%%%%%%%%%%%%%%%%%%%%%%%%%%%%
% Figure:  Dispersion comparison
\begin{figure}[t!]
 \includegraphics[width=1.02\columnwidth,trim={0.2cm 0cm 0 0 0}]{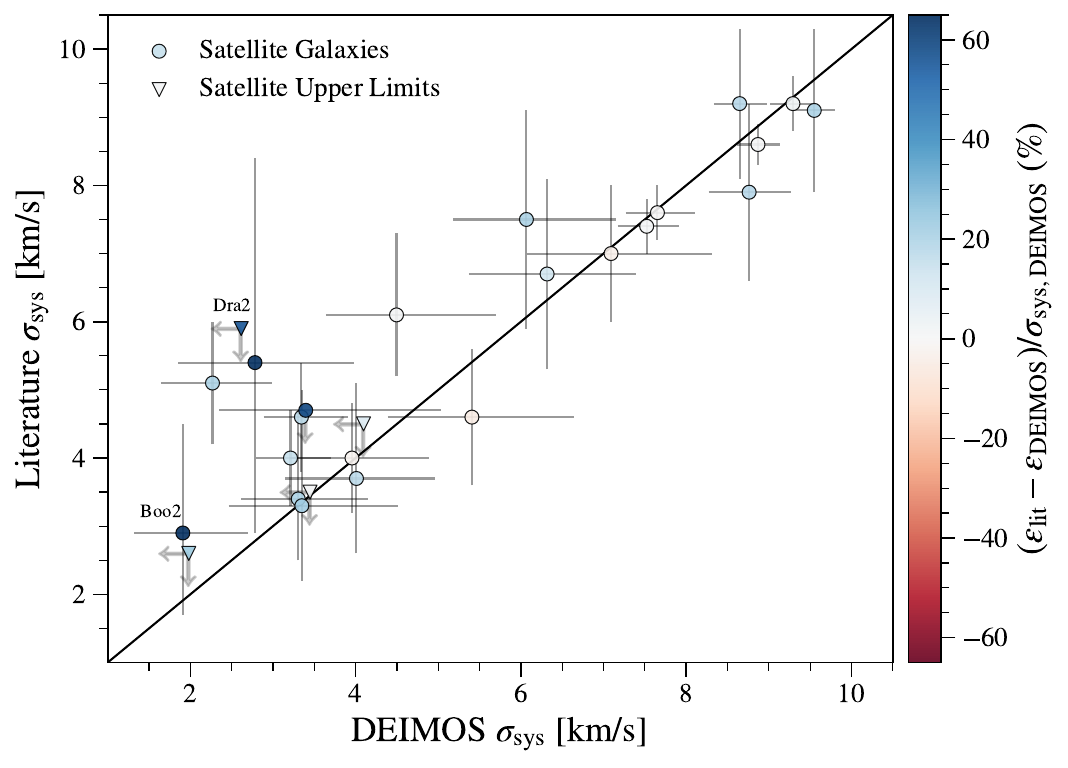}
\caption{The internal velocity dispersions for Milky Way stellar systems identified as dwarf galaxies determined from this work, compared to the literature compilation of \citet{pace2024}.  Included in this plot are systems with 10 or more DEIMOS member stars identified inside of 2 $r_{\rm eff}$.  Color coding is the difference in the velocity dispersion error ($\epsilon$), normalized by the DEIMOS velocity dispersion ($\sigma_{\rm sys, DEIMOS}$).  The labeled systems are discussed in the text.  The DEIMOS values are within $1\,\sigma$ of literature values, but tend to have smaller errors (bluer colors).  \label{fig_vdisp}}
\end{figure}
%%%%%%%%%%%%%%%%%%%%%%%%%%%%%%%%%%

\subsection{Velocity Dispersions of Milky Way Stellar Systems}\label{ssec:results}

A key goal of this work is improved measurement of the internal velocity dispersions, and thus dynamical masses, for a large homogeneous sample of Milky Way stellar satellites.   This requires both uniform velocity accuracy and velocity precision for individual stars, as well as careful treatment of unresolved binary star systems and membership criteria.   As noted in \S\,\ref{ssec: esys_exp}, underestimating (overestimating) velocity errors for individual stars translates directly into larger (smaller) values of the inferred dynamical mass of a system.  Including non-member stars or undetected binary stars will inflate, but sometimes deflate \citep{pianta2022}, these estimates.   Here we compare our DEIMOS-only estimates of internal velocity dispersions to those in the current literature.

To calculate the internal velocity dispersion, we use the $P_{\rm mem\_novar}$ sample described in \S\,\ref{ssec:member_cuts} in which we include stars with $P_{\rm mem} > 0.5$ and remove velocity variables.    We calculate the system radial velocity and internal velocity dispersion using the two-parameter Gaussian likelihood function described by \citet[][Eq 7-8]{Walker2006}.  This model assumes that the velocity measurements and associated errors are described by a mean system velocity and internal system velocity dispersion.   We use the {\tt dynesty} nested sampler algorithm \citep{DYNESTY2020,koposov_dynesty} to sample the posterior distributions.   We chose this sampler as it provides an estimate of the Bayesian evidence for each model.   In order to determine whether or not a system's internal velocity dispersion is resolved by the DEIMOS velocity sample, we compare the Bayesian evidence between our two-parameter Gaussian likelihood to a Gaussian model with zero internal dispersion (that is, the observed velocity dispersion is from error measurements only).  We consider the velocity dispersion as resolved (nonzero) if the log of the ratio of evidences is larger than unity \citep{Kass95}.   If resolved, we quote the 1-sigma errors of the posterior distributions.   In cases when the dispersion is not resolved, we report only an upper limit on the velocity dispersion as the 95th percentile of the posterior distribution.  

We determine the internal velocity dispersion for each system using member stars ($P_{\rm mem\_novar} =1$) within two half-light radii.   In Figure~\ref{fig_vdisp}, we show systems with 10 or more DEIMOS member stars.    We do not include globular clusters in this comparison, as their velocity dispersions are often a strong function of projected radius making literature comparison more difficult.  For the 30 systems passing these criteria, we compare our DEIMOS velocity dispersions to those from the literature compilation of \citet{pace2024} v.1.0.6.   We color code points in this figure by the difference in the dispersion error normalized by the DEIMOS dispersion:  $(\epsilon_{\rm literature} - \epsilon_{\rm DEIMOS})~/~\sigma_{\rm sys, DEIMOS}$.   Systems for which the velocity dispersion is more precisely determined by DEIMOS are in bluer colors.  The literature values come from a variety of sources and include both DEIMOS and non-DEIMOS kinematic measurements.   In general, our velocity dispersions agree with the literature values (1-$\sigma$ of the one-to-one line), but our DEIMOS values have smaller errors (bluer colors).

As specific examples, we highlight systems with small internal velocity dispersions.  The strictest velocity dispersion upper limit comes from the globular cluster NGC\,5053 in which we determine an upper 95\% limit of $\sigma_{\rm sys} < 1.2$\,kms from 83 member stars (see top panels, Figure~\ref{fig_members}).   While this object does not have a published velocity dispersion \citep{pace2024}, this is consistent with the predicted velocity dispersion for this system enclosed within 2\,$r_{\rm eff}$, assuming a stellar-only M/L = 1.7 \citep{Baumgardt2018}.   For dwarf galaxies, we highlight the two systems labeled in Figure~\ref{fig_vdisp}.   The system sitting furthest from the one-to-one line is Draco\,II (Dra\,2).  The Dra\,2 literature value from \citet{Longeard2018} is based on 14 member stars from two DEIMOS masks where they determined an upper limit of $<\,5.9$\kms\ (95\% confidence limit).  Based on the same dataset, our improved velocity errors provide a stronger upper limit of $<\,2.8$\kms\ (95\% confidence limit).  An additional unpublished DEIMOS mask improves on this value further, with 25 members stars providing an upper limit of $<\,2.6$\kms (see bottom panel, Figure~\ref{fig_members}).  The smallest velocity dispersion shown in Figure~\ref{fig_vdisp} is Bo\"otes\,II (Boo\,2).  The published velocity dispersion for Boo\,2 is based on 12 member stars with the Magellan IMACS spectrograph. \citet{Bruce2023} published a value of $2.9^{+1.6}_{-1.2}$\kms.  More recently, \citet{longeard2025} updated this value to $5.6^{+1.8}_{-1.1}$ based on 16 stars, combining the IMACS with additional VLT/FLAMES observations.  Our unpublished archival DEIMOS data based on 17 stars results in a more precise value of $1.9^{+0.8}_{-0.6}$\kms, in better agreement with the \citet{Bruce2023} value.   We further explore these values and determine the dynamical masses and mass-to-light ratios in Paper II.

\section{Summary and Future Work}

We present homogeneously reduced spectroscopic data from the Keck II telescope and DEIMOS spectrograph using the 1200G grating, focusing on resolved stars in the Milky Way's stellar satellites.  This paper include catalogs of \nall detected sources (\nstar stars and \nexgal extragalactic sources) within three effective radii of \nobj Milky Way dwarf galaxy satellites and globular clusters.   Based on thousands of repeat measurements, we evaluate the accuracy of our errors, finding a velocity error floor of 1.1\kms\ and a CaT-based metallicity floor of 0.1\,dex.  The median magnitude of the dwarf galaxy member sample is $r=20.4$, significantly fainter than most major radial velocity surveys. 

The Milky Way's dwarf galaxy satellites and globular clusters are critical tests for both galaxy formation and cosmology.  Old stellar ages and nearly pristine chemical compositions provide a unique probe of galaxy formation, while dark matter-dominated dwarf galaxies are compelling targets for indirect dark matter detection and tests of alternative dark matter models.    While a majority of the Keck/DEIMOS data in this work have been previously published \citep[e.g.,][]{simon07a, Martin2007, geha2009, kirby10a,willman2011, Kirby2011,collins2017, Longeard2020,2023A&A...672A.131A,Tan2025, cerny2024}, the data were reduced and analyzed using a variety of methods.  Our homogeneous reduction allows for a uniform, joint analysis across systems and leverages repeat measurements throughout the DEIMOS archives, providing a more accurate evaluation of error terms.   

In a companion paper \citep[Paper II;][]{geha_paper2}, we use these homogeneous measurements to determine dynamical mass estimates, mean metallicity and internal metallicity spreads for these systems, with the goal of further understanding the Milky Way satellites as a population.   To facilitate other investigations, we provide access to all data products, including stellar radial velocities, EW-based [Fe/H] metallicities, and membership estimates for over \nstar stars.  Data products are available both as tables in this contribution and as a KOA Contributed Dataset.   These data presented represent roughly one third of the existing 1200G DEIMOS archive.    In a future contribution, we will present a uniform data reduction for individual stars in M\,31 satellites (red symbols, Figure~\ref{fig_intro}) and low mass galaxies in the Local Volume.

\begin{acknowledgements}

This research has made extensive use of the Keck Observatory Archive (KOA), which is operated by the W. M. Keck Observatory and the NASA Exoplanet Science Institute (NExScI), under contract with the National Aeronautics and Space Administration. We thank Chris Gelino, Anastasia Laity and all the KOA and Keck staff for their support.  We also thank Yasmeen Asali, Megan Bedell, Emily Cunningham, Raja Guhathakurta, David W.~Hogg, Alex Ji, Marc Kassis, Evan Kirby, Sergey Koposov, Ting Li,  John O'Meara, August Muench, Andrew Pace, Luca Rizzi, and Josh Simon for helpful discussions.  We thank Ricardo Mu{\~n}oz for making his raw Megacam photometric catalogs available.  

M.G.~and W.C.~were supported in part by a grant~from the Howard Hughes Medical Institute (HHMI) through the HHMI Professors Program.   This material is based in part upon work supported by the National Science Foundation Graduate Research Fellowship Program under grant No.DGE-2139841. Any opinions, findings, and conclusions or recommendations expressed in this material are those of the author(s) and do not necessarily reflect the views of the National Science Foundation. 

Data presented herein were obtained at Keck Observatory, which is a private 501(c)3 non-profit organization operated as a scientific partnership among the California Institute of Technology, the University of California, and the National Aeronautics and Space Administration. The Observatory was made possible by the generous financial support of the W. M. Keck Foundation. 

The Legacy Surveys consist of three individual and complementary projects: the Dark Energy Camera Legacy Survey (DECaLS; Proposal ID \#2014B-0404; PIs: David Schlegel and Arjun Dey), the Beijing-Arizona Sky Survey (BASS; NOAO Prop. ID \#2015A-0801; PIs: Zhou Xu and Xiaohui Fan), and the Mayall z-band Legacy Survey (MzLS; Prop. ID \#2016A-0453; PI: Arjun Dey). DECaLS, BASS and MzLS together include data obtained, respectively, at the Blanco telescope, Cerro Tololo Inter-American Observatory, NSF’s NOIRLab; the Bok telescope, Steward Observatory, University of Arizona; and the Mayall telescope, Kitt Peak National Observatory, NOIRLab. Pipeline processing and analyses of the data were supported by NOIRLab and the Lawrence Berkeley National Laboratory (LBNL). The Legacy Surveys project is honored to be permitted to conduct astronomical research on Iolkam Du’ag (Kitt Peak), a mountain with particular significance to the Tohono O’odham Nation.

NOIRLab is operated by the Association of Universities for Research in Astronomy (AURA) under a cooperative agreement with the National Science Foundation. LBNL is managed by the Regents of the University of California under contract to the U.S. Department of Energy.

This project used data obtained with the Dark Energy Camera (DECam), which was constructed by the Dark Energy Survey (DES) collaboration. Funding for the DES Projects has been provided by the U.S. Department of Energy, the U.S. National Science Foundation, the Ministry of Science and Education of Spain, the Science and Technology Facilities Council of the United Kingdom, the Higher Education Funding Council for England, the National Center for Supercomputing Applications at the University of Illinois at Urbana-Champaign, the Kavli Institute of Cosmological Physics at the University of Chicago, Center for Cosmology and Astro-Particle Physics at the Ohio State University, the Mitchell Institute for Fundamental Physics and Astronomy at Texas A\&M University, Financiadora de Estudos e Projetos, Fundacao Carlos Chagas Filho de Amparo, Financiadora de Estudos e Projetos, Fundacao Carlos Chagas Filho de Amparo a Pesquisa do Estado do Rio de Janeiro, Conselho Nacional de Desenvolvimento Cientifico e Tecnologico and the Ministerio da Ciencia, Tecnologia e Inovacao, the Deutsche Forschungsgemeinschaft and the Collaborating Institutions in the Dark Energy Survey. The Collaborating Institutions are Argonne National Laboratory, the University of California at Santa Cruz, the University of Cambridge, Centro de Investigaciones Energeticas, Medioambientales y Tecnologicas-Madrid, the University of Chicago, University College London, the DES-Brazil Consortium, the University of Edinburgh, the Eidgenossische Technische Hochschule (ETH) Zurich, Fermi National Accelerator Laboratory, the University of Illinois at Urbana-Champaign, the Institut de Ciencies de l’Espai (IEEC/CSIC), the Institut de Fisica d’Altes Energies, Lawrence Berkeley National Laboratory, the Ludwig Maximilians Universitat Munchen and the associated Excellence Cluster Universe, the University of Michigan, NSF’s NOIRLab, the University of Nottingham, the Ohio State University, the University of Pennsylvania, the University of Portsmouth, SLAC National Accelerator Laboratory, Stanford University, the University of Sussex, and Texas A\&M University.

BASS is a key project of the Telescope Access Program (TAP), which has been funded by the National Astronomical Observatories of China, the Chinese Academy of Sciences (the Strategic Priority Research Program “The Emergence of Cosmological Structures” Grant \#XDB09000000), and the Special Fund for Astronomy from the Ministry of Finance. The BASS is also supported by the External Cooperation Program of Chinese Academy of Sciences (Grant \#114A11KYSB20160057), and Chinese National Natural Science Foundation (Grant \#12120101003, \#11433005).

\end{acknowledgements}

\software{This research made use of many community-developed or community-maintained software packages, including (in alphabetical order):
Astropy \citep{astropy:2022},
dynesty \citep{DYNESTY2020,koposov_dynesty}
emcee \citep{emcee},
IPython \citep{ipython},
matplotlib \citep{matplotlib},
NumPy \citep{numpy},
and SciPy \citep{scipy}.
This research has also made use of NASA's Astrophysics Data System.
}

\facility{Keck:II (DEIMOS)} 

\clearpage

\bibliographystyle{aasjournalv631} % old bib style without initials 
\bibliography{bib_deimos,software}

\appendix

\setcounter{equation}{0}
\setcounter{figure}{0} 
\setcounter{table}{0}
\renewcommand{\thefigure}{A\arabic{figure}}
\renewcommand{\thetable}{A\arabic{table}}

\label{sec:appendix}

As described in \S\,\ref{ssec:data_access}, we provide data tables for the combined measured quantities in Tables~\ref{table_objects}--\ref{table:schema}.   Additional data including access to reduced 1D spectra are provided in a KOA Contributed Dataset.  The data tables provided in this paper are as follows:

\begin{itemize}
    \item Table~\ref{table_objects} lists the Milky Way system names (e.g., Draco, NGC 4590), literature properties, and select measured system quantities (\nobj rows).
    \item Table~\ref{table_obs} lists  observational details for each DEIMOS mask included in this work (\nmask rows).   There are five additional rows in this table for overlapping masks in the Sgr and NGC\,6715 systems (see \S\,\ref{ssec:matching}).
    \item Table~\ref{table_stars} is the primary data table, one row per unique star observed with DEIMOS.  Measurements include velocity, CaT-based metallicity and membership estimates (\nstar rows).
    \item Table~\ref{table_exgal} lists identified extragalactic sources with measured redshifts (\nexgal rows).
    \item Table~\ref{table:schema} lists the schema for an expanded table of measurements for objects from both Tables~\ref{table_stars} and \ref{table_exgal} (\nall rows).  
\end{itemize}

All tables are available in a machine-readable format, both in the online journal article, as a KOA Contributed Dataset, and on request to the corresponding author.
\vskip 1cm

\startlongtable
\centerwidetable
%\movetableright=-1.5in
\begin{deluxetable}{rrrrrrrccrrrrrrrr}
\tablewidth{\textwidth}
\tabletypesize{\footnotesize}
\tablecaption{Observed Keck/DEIMOS Milky Way Satellites and Derived Properties\label{table_objects}}
\tablehead{\colhead{Full} & 
		\colhead{System} &
        \colhead{RA} & 
        \colhead{Dec} & 
        \colhead{Dist} & 
	    \colhead{$M_{V,o}$} & 
	    \colhead{$r_{\rm eff}$} & 
        \colhead{Type} &
        \colhead{Phot} &
        \colhead{$N$}&
	    \colhead{$N$}&
        \colhead{$v_{\rm sys}$}&
	    \colhead{$\epsilon_{v_{\rm sys}}$} & 
        \colhead{$\sigma_{\rm sys}^{2r_{\rm eff}}$}&
	    \colhead{$\epsilon_{\sigma_{\rm sys}}^{ll}$} & 
	    \colhead{$\epsilon_{\sigma_{\rm sys}}^{ul}$} &
        \colhead{$\epsilon_{\sigma_{\rm sys}}^{95}$}  \\
        \colhead{Name} & 
        \colhead{Name} & 
        \colhead{[deg]} & 
        \colhead{[deg]} &
		\colhead{[kpc]} & 
        \colhead{} & 
        \colhead{[arcm]} & 
        \colhead{} &
        \colhead{} & 
        \colhead{msk} &
        \colhead{mem} &
        \colhead{[km/s]} &
        \colhead{[km/s]} &
        \colhead{[km/s]} &
        \colhead{[km/s]} &
        \colhead{[km/s]} &
        \colhead{[km/s]}\\
        \colhead{(1)} & \colhead{(2)} & \colhead{(3)} & 
        \colhead{(4)} & \colhead{(5)} & \colhead{(6)} &
        \colhead{(7)} &\colhead{(8)} &\colhead{(9)}&
        \colhead{(10)} & \colhead{(11)} & \colhead{(12)} & 
        \colhead{(13)} & \colhead{(14)} & \colhead{(15)}& \colhead{(16)} & \colhead{(17)}  
        }
\startdata
Aquarius II  & Aqr2    & 338.481300& -9.327400 &  108   &  -4.4 &    5.1  & G      & LS     &        1 &       8 & -63.0 & 2.4     & 5.31  &       1.6  &      2.72 &   -999    \\
 Aquarius III & Aqr3    & 357.220000& -3.490000 &   85   &  -2.5 &    2.1  & G      & LS     &        1 &      11 & -13.2 & 1.0     & -999    &    -999    &   -999    &      3.65 \\
 Bootes I     & Boo1    & 210.020035& 14.513500 &   66   &  -6   &   10.5  & G      & LS     &        8 &      90 & 101.9 & 0.5     & 3.19  &       0.41 &      0.48 &   -999    \\
 Bootes II    & Boo2    & 209.514130& 12.855278 &   42   &  -2.9 &    3.07 & G      & LS     &        7 &      17 & -128.4& 0.7     & 1.92  &       0.6  &      0.77 &   -999    \\
 Bootes III   & Boo3    & 209.300000& 26.800000 &   46   &  -5.8 &   30    & G      & LS     &        9 &      16 & 189.9 & 1.8     & 5.27  &       1.67 &      2.1  &   -999    \\
 ComaBer      & CB      & 186.745422& 23.906918 &   42   &  -4.3 &    5.67 & G      & LS     &        8 &      82 & 93.5  & 0.6     & 3.33  &       0.46 &      0.58 &   -999    \\
 Cetus III    & Cet3    & 31.330833 & -4.270000 &  250   &  -2.5 &    1.23 & G      & LS     &        3 &       3 & 1.3   & 2.6     & -999    &    -999    &   -999    &   -999    \\
 Columba 1    & Col1    & 82.856960 & -28.042530&  182   &  -4.2 &    2.2  & G      & LS     &        1 &       9 & 148.8 & 2.3     & 5.96  &       1.81 &      2.78 &   -999    \\
 CVn I        & CVn1    & 202.009079& 33.552082 &  220   &  -8.8 &    7.48 & G      & LS     &        8 &     254 & 30.5  & 0.6     & 7.66  &       0.41 &      0.44 &   -999    \\
 CVn II       & CVn2    & 194.292664& 34.322640 &  160   &  -5.2 &    1.52 & G      & LS     &        5 &      25 & -131.8& 1.2     & 5.28  &       0.96 &      1.25 &   -999    \\
 Draco        & Dra     & 260.068451& 57.918472 &   81.5 &  -8.7 &    9.61 & G      & LS     &       26 &     995 & -292.2& 0.3     & 9.55  &       0.25 &      0.26 &   -999    \\
 Draco II     & Dra2    & 238.200000& 64.570000 &   21.5 &  -2.9 &    2.7  & G      & LS     &        3 &      25 & -342.3& 0.7     & -999    &    -999    &   -999    &      2.64 \\
 Eridanus     & Eri     & 66.185330 & -21.187555&   84.6 &  -4.9 &    0.65 & GC     & LS     &        2 &      24 & -18.4 & 0.3     & -999    &    -999    &   -999    &      0.89 \\
 Eridanus 4   & Eri4    & 76.438000 & -9.515000 &   76.5 &  -3.8 &    3.24 & G      & LS     &        1 &      19 & -34.4 & 1.3     & 4.55  &       0.89 &      1.15 &   -999    \\
 Fornax       & For     & 39.958336 & -34.499721&  142.5 & -13.5 &   18.5  & G      & LS     &        5 &     663 & 53.9  & 0.5     & 11.71 &       0.33 &      0.36 &   -999    \\
 Herc         & Herc    & 247.772201& 12.785194 &  131   &  -5.8 &    5.83 & G      & LS     &        9 &      43 & 45.3  & 0.6     & 2.25  &       0.61 &      0.67 &   -999    \\
 Hydra 2      & Hyd2    & 185.425125& -31.986029&  151   &  -4.6 &    1.65 & G      & M    &        1 &      11 & 304.4 & 1.1     & -999    &    -999    &   -999    &      4.11 \\
 Koposov 1    & K1      & 179.825333& 12.261528 &   48.3 &  -1   &    0.68 & U      & LS     &        1 &       7 & 5.9   & 1.3     & -999    &    -999    &   -999    &      5.95 \\
 Koposov 2    & K2      & 119.571495& 26.257398 &   24   &  -0.9 &    0.42 & U      & LS     &        3 &      13 & 108.2 & 1.3     & -999    &    -999    &   -999    &      5.23 \\
 Crater 1     & Lae1    & 174.066803& -10.877222&  145   &  -4.8 &    0.5  & GC     & LS     &        1 &       9 & 150.0 & 0.9     & -999    &    -999    &   -999    &      4.5  \\
 Laevens 3    & Lae3    & 316.729300& 14.984300 &   61.4 &  -4.4 &    0.4  & U      & LS     &        1 &       6 & -72.6 & 1.2     & -999    &    -999    &   -999    &      7.42 \\
 Leo I        & Leo1    & 152.114578& 12.305917 &  258   & -11.8 &    3.53 & G      & LS     &       19 &     686 & 285.9 & 0.4     & 9.29  &       0.26 &      0.26 &   -999    \\
 Leo II       & Leo2    & 168.362656& 22.152861 &  233   &  -9.7 &    2.46 & G      & LS     &        9 &     253 & 78.8  & 0.5     & 7.53  &       0.36 &      0.39 &   -999    \\
 Leo IV       & Leo4    & 173.240509& -0.543056 &  151   &  -5   &    2.61 & G      & LS     &        4 &      24 & 129.8 & 0.9     & 3.29  &       0.66 &      0.89 &   -999    \\
 Leo V        & Leo5    & 172.785675& 2.219389  &  169   &  -4.4 &    1.05 & G      & LS     &        2 &      10 & 170.8 & 1.5     & 3.37  &       1.04 &      1.72 &   -999    \\
 Leo VI       & Leo6    & 171.073000& 24.870000 &  111   &  -3.6 &    3.81 & G      & LS     &        1 &       9 & 170.3 & 1.8     & 3.78  &       1.37 &      2.15 &   -999    \\
 Leo T        & LeoT    & 143.729172& 17.048222 &  413   &  -7.6 &    1.27 & G      & LS     &        7 &      42 & 37.1  & 1.1     & 6.05  &       0.86 &      1.07 &   -999    \\
 Munoz 1      & Mun1    & 225.449036& 66.968193 &   45   &  -0.5 &    0.49 & GC     & LS     &        1 &       2 & -140.1& 6.4     & -999    &    -999    &   -999    &   -999    \\
 NGC 1904     & N1904   & 81.049580 & -24.524720&   13   &  -7.9 &    0.65 & GC     & LS     &        6 &      61 & 206.0 & 0.7     & 5.22  &       0.51 &      0.57 &   -999    \\
 NGC 2419     & N2419   & 114.535416& 38.881863 &   88.5 &  -9.4 &    0.89 & GC     & M    &        7 &      86 & -20.9 & 0.5     & 4.69  &       0.38 &      0.43 &   -999    \\
 NGC 0288     & N288    & 13.197706 & -26.589901&    9   &  -6.8 &    2.23 & GC     & LS     &        7 &     473 & -43.2 & 0.2     & 3.47  &       0.16 &      0.17 &   -999    \\
 NGC 4590     & N4590   & 189.866700& -26.743039&   10.3 &  -7.4 &    1.51 & GC     & LS     &        4 &     103 & -93.6 & 0.4     & 3.31  &       0.25 &      0.28 &   -999    \\
 NGC 5024     & N5024   & 198.230130& 18.169100 &   18   &  -8.7 &    1.31 & GC     & LS     &        1 &      33 & -64.8 & 0.8     & 3.77  &       0.57 &      0.66 &   -999    \\
 NGC 5053     & N5053   & 199.112400& 17.697700 &   17   &  -6.7 &    2.61 & GC     & LS     &        1 &      83 & 42.5  & 0.2     & -999    &    -999    &   -999    &      1.2  \\
 NGC 5272     & N5272   & 205.546770& 28.375450 &   10.5 &  -8.9 &    2.31 & GC     & LS     &        4 &      44 & -147.0& 0.7     & 3.85  &       0.46 &      0.53 &   -999    \\
 NGC 5634     & N5634   & 217.405335& -5.976430 &   26   &  -7.7 &    0.86 & GC     & LS     &        2 &      42 & -18.3 & 0.6     & 3.40  &       0.4  &      0.47 &   -999    \\
 NGC 5824     & N5824   & 225.994300& -33.068500&   32.1 &  -9.3 &    0.45 & GC     & LS     &        1 &       4 & -32.9 & 6.0     & 11.30 &       3.71 &      5.13 &   -999    \\
 NGC 5897     & N5897   & 229.351650& -21.010120&   12.5 &  -7.2 &    2.06 & GC     & LS     &        6 &     152 & 101.4 & 0.3     & 2.93  &       0.21 &      0.25 &   -999    \\
 NGC 5904     & N5904   & 229.640600& 2.082680  &    8   &  -8.8 &    1.77 & GC     & LS     &        1 &      50 & 55.2  & 0.7     & 4.49  &       0.47 &      0.56 &   -999    \\
 NGC 6205     & N6205   & 250.423450& 36.461300 &    7.5 &  -8.6 &    1.69 & GC     & GC     &        2 &      72 & -245.8& 0.7     & 5.74  &       0.49 &      0.6  &   -999    \\
 NGC 6218     & N6218   & 251.810484& -1.947821 &    5.1 &  -7   &    1.77 & GC     & LS     &        1 &      11 & -40.6 & 0.8     & -999    &    -999    &   -999    &      3.3  \\
 NGC 6229     & N6229   & 251.745249& 47.527790 &   30.1 &  -8.1 &    3    & GC     & LS     &        1 &      31 & -139.1& 0.6     & 2.38  &       0.45 &      0.59 &   -999    \\
 NGC 6254     & N6254   & 254.287461& -4.099326 &    5   &  -7.5 &    1.95 & GC     & LS     &        1 &      12 & 73.5  & 1.9     & 6.18  &       1.27 &      1.85 &   -999    \\
 NGC 6341     & N6341   & 259.280290& 43.136523 &    8.5 &  -8.2 &    1.02 & GC     & LS     &        6 &     110 & -120.0& 0.5     & 4.61  &       0.36 &      0.4  &   -999    \\
 NGC 6366     & N6366   & 261.934700& -5.076600 &    4   &  -5.7 &    2.92 & GC     & PS     &        1 &      21 & -120.6& 0.8     & 3.28  &       0.57 &      0.76 &   -999    \\
 NGC 6624     & N6624   & 275.919042& -30.361278&    8   &  -7.5 &    0.82 & GC     & PS     &        5 &      69 & 55.4  & 0.4     & 3.29  &       0.31 &      0.36 &   -999    \\
 NGC 6656     & N6656   & 279.100850& -23.903400&    3   &  -8.5 &    3.36 & GC     & PS     &        2 &      74 & -146.3& 0.9     & 7.05  &       0.62 &      0.69 &   -999    \\
 NGC 6715     & N6715   & 283.763875& -30.479861&   26.3 & -10   &    1    & GC     & PS     &        5 &      15 & 142.2 & 1.6     & 6.04  &       1.08 &      1.45 &   -999    \\
 NGC 6779     & N6779   & 289.147941& 30.184501 &   10.4 &  -7.4 &    1.1  & GC     & PS     &        3 &      74 & -133.6& 0.6     & 4.70  &       0.41 &      0.48 &   -999    \\
 NGC 6838     & N6838   & 298.442121& 18.778402 &    4   &  -5.6 &    1.67 & GC     & GC     &        2 &      17 & -21.8 & 0.6     & 2.07  &       0.47 &      0.58 &   -999    \\
 NGC 6864     & N6864   & 301.520167& -21.922222&   22   &  -8.6 &    0.46 & GC     & LS     &        3 &      36 & -186.3& 0.7     & 3.66  &       0.48 &      0.56 &   -999    \\
 NGC 7006     & N7006   & 315.372162& 16.187084 &   39.3 &  -7.4 &    0.44 & GC     & M    &        4 &      22 & -383.1& 0.8     & 3.25  &       0.55 &      0.69 &   -999    \\
 NGC 7078     & N7078   & 322.493200& 12.166800 &   10   &  -9.2 &    1    & GC     & LS     &       11 &     246 & -105.5& 0.4     & 6.18  &       0.3  &      0.33 &   -999    \\
 NGC 7089     & N7089   & 323.362552& -0.823318 &   11.5 &  -9   &    1.06 & GC     & LS     &       12 &     200 & -2.6  & 0.5     & 6.81  &       0.34 &      0.41 &   -999    \\
 NGC 7099     & N7099   & 325.091760& -23.179070&    8   &  -7.5 &    1.03 & GC     & LS     &       10 &     146 & -185.1& 0.4     & 4.01  &       0.28 &      0.3  &   -999    \\
 NGC 7492     & N7492   & 347.110199& -15.610805&   24.3 &  -6.1 &    1.15 & GC     & LS     &        1 &      21 & -175.4& 0.5     & 1.75  &       0.46 &      0.51 &   -999    \\
 Palomar 13   & Pal13   & 346.685812& 12.771250 &   23   &  -2.8 &    1.26 & GC     & LS     &        9 &      57 & 26.0  & 0.3     & -999    &    -999    &   -999    &      1.52 \\
 Palomar 14   & Pal14   & 242.754425& 14.958445 &   73.6 &  -5.4 &    1.42 & GC     & LS     &        2 &      25 & 73.1  & 0.4     & -999    &    -999    &   -999    &      1.56 \\
 Palomar 2    & Pal2    & 71.524795 & 31.381695 &   27   &  -9.1 &    0.5  & GC     & PS     &        1 &      15 & -111.7& 1.3     & 4.45  &       0.9  &      1.29 &   -999    \\
 Palomar 5    & Pal5    & 229.022000& -0.111300 &   23   &  -5.2 &    2.73 & GC     & GC     &        5 &      77 & -57.7 & 0.6     & 3.21  &       0.69 &      0.71 &   -999    \\
 Palomar 7    & Pal7    & 272.684441& -7.207595 &    4.6 &  -6.7 &    2.38 & GC     & PS     &        2 &      15 & 157.4 & 1.7     & 5.66  &       1.18 &      1.72 &   -999    \\
 Pegasus 3    & Peg3    & 336.100000& 5.410000  &  205   &  -4.1 &    3.6  & G      & LS     &        3 &      15 & -260.0& 1.1     & 2.76  &       0.91 &      1.24 &   -999    \\
 Pegasus 4    & Peg4    & 328.539000& 26.620000 &   90   &  -4.3 &    1.6  & G      & LS     &        5 &      21 & -271.4& 1.1     & 3.33  &       0.86 &      1.19 &   -999    \\
 Pisces 2     & Pisc2   & 344.636458& 5.955544  &  182   &  -4.2 &    1.34 & G      & LS     &        6 &       9 & -227.7& 1.7     & 3.72  &       1.33 &      1.98 &   -999    \\
 Sculptor     & Scl     & 15.018333 & -33.718613&   83.9 & -10.8 &   12.4  & G      & LS     &        5 &     379 & 110.7 & 0.5     & 8.66  &       0.33 &      0.33 &   -999    \\
 Segue 1      & Seg1    & 151.750366& 16.075582 &   23   &  -1.3 &    3.93 & G      & LS     &       13 &      53 & 203.1 & 0.9     & 3.97  &       0.86 &      0.97 &   -999    \\
 Segue 2      & Seg2    & 34.822582 & 20.162500 &   36   &  -1.9 &    3.6  & G      & LS     &       11 &      29 & -41.2 & 0.4     & -999    &    -999    &   -999    &      2.06 \\
 Segue 3      & Seg3    & 320.379486& 19.117805 &   17   &  -0.9 &    0.54 & GC     & LS     &        3 &      17 & -165.5& 0.8     & -999    &    -999    &   -999    &      3.27 \\
 Sextans      & Sext    & 153.262833& -1.613306 &   86   &  -8.7 &   16.9  & G      & LS     &       13 &     239 & 221.9 & 0.6     & 8.77  &       0.47 &      0.49 &   -999    \\
 Sgr          & Sgr     & 284.095166& -30.549887&   26   & -13.5 &  234    & G      & PS     &        5 &      60 & 142.7 & 1.1     & 8.31  &       0.74 &      0.86 &   -999    \\
 Sgr 2        & Sgr2    & 298.170000& -22.070000&   69   &  -5.4 &    2    & GC     & LS     &        4 &      28 & -176.4& 0.6     & 2.32  &       0.43 &      0.53 &   -999    \\
 Terzan 5     & Ter5    & 267.002080& -24.779166&    6.6 &  -7.5 &    0.73 & GC     & PS     &        6 &      15 & -85.4 & 2.7     & 10.27 &       1.74 &      2.45 &   -999    \\
 Tri 2        & Tri2    & 33.315500 & 36.169100 &   36.5 &  -1.8 &    2.34 & G      & PS     &        9 &       7 & -382.2& 1.1     & -999    &    -999    &   -999    &      5.54 \\
 UMa 1        & UMa1    & 158.770584& 51.947979 &   97   &  -5.1 &    8.13 & G      & LS     &        6 &      36 & -58.5 & 1.4     & 7.16  &       1.04 &      1.21 &   -999    \\
 UMa 2        & UMa2    & 132.872635& 63.133530 &   34.6 &  -4.3 &   13.9  & G      & LS     &       13 &      64 & -118.0& 1.1     & 6.30  &       0.95 &      1.06 &   -999    \\
 Ursa Minor   & UMi     & 227.241959& 67.222138 &   70   &  -9   &   18.2  & G      & LS     &       25 &     827 & -246.1& 0.3     & 8.86  &       0.24 &      0.27 &   -999    \\
 Unions 1     & UNI1    & 174.708000& 31.071100 &   10   &   2   &    2.2  & U      & LS     &        1 &      10 & 89.6  & 0.9     & -999    &    -999    &   -999    &      4.35 \\
 Willman 1    & W1      & 162.343628& 51.050083 &   38.5 &  -2.5 &    2.51 & G      & M    &       12 &      47 & -12.5 & 0.9     & 3.97  &       0.72 &      0.88 &   -999    \\
\enddata
\tablecomments{Literature and computed properties of the Milky Way stellar satellites included in this work.  (1)-(2) Name and abbreviated name of system, (3)-(7) position, distance, absolute magnitude and half-light radius, $r_{\rm eff}$, taken from \citet{pace2024}. (9) Source of photometry, LS = Legacy Survey DR10, \citet{Dey2019}, MZ = \citet{munoz2018a}, GC = \citet{An2008}, PS = Pan-STARRS1 DR2, \citet{panstarrs}, (10) Number of DEIMOS masks observed within 3 $r_{\rm eff}$, (11) number of member stars identified.    (12)-(13) The systematic velocity and associated error determined for the system.  (14)  The velocity dispersion determine using stars inside of 2 $r_{\rm eff}$, and (15)-(16) lower and upper error bars determined for the system where the velocity dispersion is resolved.  (17)  For systems with an unresolved velocity dispersion, the upper 95\% confidence limit on the internal velocity dispersion inside 2 $r_{\rm eff}$.   The full table, consisting of \nobj rows, is available in a machine-readable format. }
\end{deluxetable}

\newpage

\begin{deluxetable*}{l r C r C C C C C C}
\tablewidth{\textwidth}
\tabletypesize{\footnotesize}
\tablecaption{Summary of DEIMOS Mask Observations\label{table_obs}}
\tablehead{\colhead{System} & 
		\colhead{DEIMOS} &
	    \colhead{Date Obs.}& 
		\colhead{PI}&
		\colhead{Slit width}&
        \colhead{N$_{\rm exp}$} & 
        \colhead{$t_{\rm exp}$} &
        \colhead{Seeing} & 
        \colhead{$N_{\rm slits}$} &
        \colhead{ $N_{\rm good}/N_{\rm slits}$}\\
       \colhead{Name} & 
       \colhead{Mask} & 
       \colhead{[YYYYMMDD]} & 
     \colhead{} & 
	    \colhead{[arcsec]}&
	    \colhead{} &
        \colhead{[sec]} & 
        \colhead{[arcsec]} &
        \colhead{}&
        \colhead{} \\
           \colhead{(1)} & \colhead{(2)} & \colhead{(3)} & 
        \colhead{(4)} & \colhead{(5)} & \colhead{(6)} &
        \colhead{(7)} &\colhead{(8)} &\colhead{(9)}&\colhead{(10)} }
\startdata
Aqr2   & VB1      &   20150912 & Collins      &      0.75 &       3 &        3600 &     0.59 &       20 &     0.85 \\
 Aqr3   & aqd3r1   &   20231006 & Geha         &      0.7  &       9 &       15300 &     1.19 &       87 &     0.26 \\
 Boo1   & BooI-5   &   20130413 & J.Simon      &      0.7  &       3 &        3360 &     0.93 &       51 &     0.63 \\
 Boo1   & BooI-1   &   20130311 & Geha         &      0.7  &       4 &        4800 &     1.07 &       60 &     0.57 \\
 Boo1   & BooI-2   &   20130311 & Geha         &      0.7  &       4 &        4800 &     0.87 &       67 &     0.6  \\
 Boo1   & 601BoS   &   20100616 & Brooks       &      0.7  &       3 &        3100 &     1.16 &       64 &     0.77 \\
 Boo1   & 208BoSB  &   20060528 & Chapman      &      0.7  &       4 &        4800 &     1.11 &       62 &     0.71 \\
 Boo1   & 207BoSB  &   20060527 & Chapman      &      0.7  &       3 &        3600 &     0.76 &       68 &     0.68 \\
 Boo2   & Boo2-6   &   20150718 & Geha         &      0.7  &       4 &        4800 &     0.87 &       54 &     0.06 \\
 Boo2   & booiicB  &   20130708 & Bullock      &      0.7  &       8 &       10433 &     1.16 &       63 &     0.32 \\
.. & .. & .. &.. & .. & .. &.. & .. & .. & ..
\enddata
\tablecomments{List of DEIMOS masks present in this work.   (1) Name of associated Milky Way stellar system (e.g., Column 1 of Table~\ref{table_objects}), (2) DEIMOS mask name, (3) date observed (first date if multiple) as YYYYMMDD, (4) Principal Investigator (PI) name, (5) slit width, (6) number of exposures reduced in this work, (7) total integrated exposure time, (8) mean value of the seeing across all exposures, (9) number of slits in the mask, (10) fraction of targeted objects in the mask with a measured velocity.  While there are \nmask unique DEIMOS masks, this table includes five additional rows for overlapping masks in the Sgr and NGC\,6715 systems (see § 8.1).  The full table, consisting of \nmask + 5 rows, is available in a machine-readable format.}
\end{deluxetable*}

\startlongtable
\centerwidetable
\begin{deluxetable*}{l C C  C C C c r c c c c c c c r}
\tablewidth{\textwidth}
\tabletypesize{\footnotesize}
\tablecaption{DEIMOS Measured Quantities for Individual Stars\label{table_stars}}
\tablehead{\colhead{System} & 
        \colhead{RA} & 
        \colhead{Dec} & 
        \colhead{$r_0$}&
        \colhead{$(g-r)_0$}&
        \colhead{$n_{\rm mask}$}&
        \colhead{$t_{\rm exp}$}&
	    \colhead{S/N}&
        \colhead{$v_{\rm helio}$}&
        \colhead{$\epsilon_{v_{\rm helio}}$}&
	    \colhead{EW$_{\rm CaT}$}& 
	    \colhead{$\epsilon_{EW_{\rm CaT}}$}&
        	    \colhead{[Fe/H]}& 
	    \colhead{$\epsilon_{\rm [Fe/H]}$}&
        \colhead{Var}&
        \colhead{P$_{\rm mem}$} \\
        \colhead{Name} & 
       \colhead{[deg]} & 
       \colhead{[deg]} & 
        \colhead{}&
        \colhead{}&
        \colhead{}&
        \colhead{[sec]} & 
        \colhead{} &
        \colhead{[\kms]} &
        \colhead{[\kms]}&
        \colhead{[\mbox{\AA}]}&
        \colhead{[\mbox{\AA}]}& 
        \colhead{[dex]}&
                \colhead{[dex]}& 
                \colhead{Flag}&
        \colhead{} \\
        \colhead{(1)} & \colhead{(2)} & \colhead{(3)} & 
        \colhead{(4)} & \colhead{(5)} & \colhead{(6)} &
        \colhead{(7)} &\colhead{(8)} &\colhead{(9)}&
        \colhead{(10)} & \colhead{(11)} & \colhead{(12)} & 
        \colhead{(13)} &\colhead{(14)}& \colhead{(15)} &\colhead{(16)}
}
\startdata
Aqr2     & 338.478833 & -9.287944 & 18.1 &  0.81 &       1 &    3600 & 56.7 &   49.48 &   1.13 &  6.63 &     0.17 & -999    &   -999    &  -999 &   0    \\
 Aqr2     & 338.469542 & -9.285833 & 18.5 &  0.72 &       1 &    3600 & 44.8 &  -56.56 &   1.22 &  2.85 &     0.15 &   -2.56 &      0.14 &  -999 &   0.85 \\
 Aqr2     & 338.460833 & -9.315222 & 19.5 &  0.6  &       1 &    3600 & 23.9 &  -68.37 &   1.55 &  2.28 &     0.24 &   -2.69 &      0.17 &  -999 &   0.75 \\
 Aqr2     & 338.499458 & -9.360583 & 20.2 &  0.58 &       1 &    3600 & 16.5 & -127.28 &   2.06 &  3.75 &     0.33 & -999    &   -999    &  -999 &   0    \\
 Aqr2     & 338.450667 & -9.361722 & 20.3 &  0.53 &       1 &    3600 & 16.4 &  -63.06 &   2.25 &  1.29 &     0.87 &   -3.12 &      0.76 &  -999 &   0.91 \\
 Aqr2     & 338.454958 & -9.372556 & 20.5 &  0.64 &       1 &    3600 & 15.8 & -105.49 &   1.76 &  5.89 &     0.48 &   -1.1  &      0.22 &  -999 &   0.21 \\
 Aqr2     & 338.449083 & -9.269389 & 20.5 &  0.53 &       1 &    3600 & 13.2 &  -58.05 &   3.04 &  2.66 &     0.37 &   -2.38 &      0.21 &  -999 &   0.92 \\
 Aqr2     & 338.492875 & -9.407889 & 20.6 & -0.14 &       1 &    3600 & 11.7 &  -48.46 &   4.07 &  2.78 &     0.58 & -999    &   -999    &  -999 &   0.3  \\
 Aqr2     & 338.491667 & -9.321861 & 20.6 & -0.11 &       1 &    3600 & 10.6 &  -63.42 &   5.63 &  2.8  &     0.61 & -999    &   -999    &  -999 &   0.99 \\
 Aqr2     & 338.521042 & -9.381056 & 20.9 &  0.4  &       1 &    3600 &  9.3 & -142.57 &   3.12 &  3.33 &     0.5  & -999    &   -999    &  -999 &   0.01 \\
 .. & .. & .. &.. & .. & .. &.. & .. & .. & .. & .. & .. & .. &..& .. &..
\enddata
\tablecomments{One row per unique star with measured DEIMOS properties.  (1) Associated Milky Way stellar system (from column 1 of Table~\ref{table_objects}), (2)-(3) RA and Declination, (4)-(5) extinction-corrected $g-$ and $r-$band photometry, (6) number of DEIMOS masks contributing to the measurements, (7)-(8) total exposure time and combined S/N, (9)-(10) heliocentric velocity and associated error, (11)-(12) EW of the calcium triplet (CaT) and error, (13)-(14) CaT-based [Fe/H] and error are populated only for member stars with $M_V < 3$, (15) flag set to unity if source is a velocity variable as defined in \S\,\ref{ssec:flag_var}, this flag is set to -99 if there are no repeat measurements, (16) membership probability, based on the criteria in \S\,\ref{sec_membership}.   The full table, consisting of \nstar rows, is available in a machine-readable format.  }
\end{deluxetable*}

\begin{deluxetable*}{l C C  C C C c r r c c}
\tablewidth{\textwidth}
\tabletypesize{\footnotesize}
\tablecaption{DEIMOS Measured Quantities for Extragalactic Background Galaxies and Quasars\label{table_exgal}}
\tablehead{\colhead{System} & 
        \colhead{RA} & 
        \colhead{Dec} & 
        \colhead{$r_0$}&
        \colhead{$(g-r)_0$}&
        \colhead{$n_{\rm mask}$}&
        \colhead{$t_{\rm exp}$}&
	    \colhead{S/N}&
        \colhead{{\tt marz}}&
        \colhead{{\tt marz}}&
	    \colhead{Serendip}\\
        \colhead{Name} & 
        \colhead{[deg]} & 
        \colhead{[deg]} & 
        \colhead{}&
        \colhead{}&
        \colhead{}&
        \colhead{[sec]}&
        \colhead{}&\colhead{redshift}&
        \colhead{flag}&\colhead{flag}\\
        \colhead{(1)} & \colhead{(2)} & \colhead{(3)} & 
        \colhead{(4)} & \colhead{(5)} & \colhead{(6)} &
        \colhead{(7)} &\colhead{(8)} &\colhead{(9)}&
        \colhead{(10)} & \colhead{(11)}
}
\startdata
Aqr2     & 338.499625  & -9.361752   &   21.9 &   0.88 &       1 &    3600 &  1.4 & 0.3045 &           4 &          1 \\
 Aqr2     & 338.492700  & -9.406653   &   22.9 &   1.32 &       1 &    3600 &  1.2 & 0.5794 &           4 &          1 \\
 Aqr3     & 357.096900  & -3.463888   &   18.2 &   0.79 &       1 &   15300 & 45.5 & 0.1707 &           4 &          0 \\
 Aqr3     & 357.071880  & -3.458887   &   19.3 &   0.46 &       1 &   15300 & 19.4 & 0.0991 &           4 &          0 \\
 Aqr3     & 357.281175  & -3.470348   &   20.4 &   0.3  &       1 &   15300 & 18.2 & 1.4299 &           4 &          0 \\
 Aqr3     & 357.279345  & -3.456365   &   20.2 &   0.26 &       1 &   15300 & 17.6 & 0.0782 &           4 &          0 \\
 Aqr3     & 357.126480  & -3.442731   &   19.7 &   0.74 &       1 &   15300 & 16.9 & 0.3111 &           4 &          0 \\
 .. & .. & .. &.. & .. & .. &.. & .. & .. & .. & .. 
\enddata
\tablecomments{Objects identified as extragalactic.   Program targets were primarily selected as stellar candidates and are likely to be more compact as compared to the overall extragalactic population. 
Columns (1)-(8) are the same as described in Table~\ref{table_stars}.  (9) Redshift determined by the {\tt marz} package described in \S\,\ref{ssec:exgal}.   (10) Flag set via visual identification:  3 = Possible Galaxy, 4 = Secure Galaxy, 6 = Broad-line extragalactic source.  (11) This flag is set to zero if object is a program target, and unity for a serendipitous source. The full table, consisting of \nexgal rows, is available in a machine-readable format.}
\end{deluxetable*}

\startlongtable
\centerwidetable
\begin{deluxetable*}{llcl}
\tabletypesize{\scriptsize}
\tablecaption{Schema of Expanded DEIMOS Quantities for All Targets\label{table:schema}}
\tablehead{\colhead{No.} &  \colhead{Label} & \colhead{Units} &\colhead{Explanations}}
\startdata
1   & system\_name & $\cdots$ & Name of Milky Way Satellite \\
2   & objname & $\cdots$ & Target name taken from DEIMOS design file\\
3   & RA   & deg & Right Ascension in decimal degrees (J2000) \\
4   & DEC  & deg & Declination in decimal degrees (J2000) \\
5   & nmask & $\cdots$ & Number of unique DEIMOS masks \\
6   & nexp  & $\cdots$ & Number of unique exposures across all masks \\
7   & t\_exp & seconds & Total exposure time across all masks\\
8   & masknames & $\cdots$ & Names of DEIMOS masks.  If multiple masks, names are joined with '+'\\
9  &  slitwidth & arcseconds & Width of slits.  If multiple masks, this is an average across masks\\
10  &  mean\_mjd & days &   The mean MJD using all individual exposures \\
11 &  SN        & $\cdots$ & Mean per pixel signal-to-noise\\
12 & serendip	& $\cdots$ & Flag is zero if object is a program target, unity otherwise \\
13 & marz\_flag  & 	$\cdots$ & 1 = Star, 3 = Possible Galaxy, 4 = Secure Galaxy, 6 = Broad-line extragalactic source\\ 
14 & v & \kms & Heliocentric velocity \\
15 & v\_err & \kms & Heliocentric velocity error\\
16 &v\_chi2 & $\cdots$ &  Reduced $\chi^2$ of stellar velocity template fit\\
17 & phot\_source & $\cdots$ &  Source of photometry:  LS = Legacy Survey DR10, \citet{Dey2019},\\
   &              &          &  MZ = \citet{munoz2018a}, GC = \citet{An2008}, PS = Pan-STARRS1 DR2, \citet{panstarrs}\\
18 & gmag\_o & mag &  $g$-band magnitude, corrected for Galactic extinction\\
19 & rmag\_o & mag &  $r$-band magnitude, corrected for Galactic extinction\\
20 & gmag\_err & mag &  $g$-band magnitude error\\
21 & rmag\_err & mag &  $r$-band magnitude error\\
22 & MV\_o      & mag & Absolute V-band magnitude, assuming distance of system\\
23 & rproj\_arcm& arcmin & Projected radial distance from center of system \\
24 & rproj\_kpc & kpc & Projected radial distance from center of system \\
25 & ew\_cat	& \mbox{\AA} & EW of the combined Calcium II triplet (CaT) lines\\
26 & ew\_cat\_err & \mbox{\AA} & CaT EW error \\
27 & ew\_naI	  & \mbox{\AA} & EW of the Na\,I doublet lines\\
28 & ew\_naI\_err & \mbox{\AA} & Na\,I EW error \\
29 & feh          & dex & CaT-based [Fe/H],  only for member stars brighter than $M_V < 3$ \\
30 & feh\_err     & dex & CaT-based [Fe/H] error \\
31 & ew\_w1        & \mbox{\AA} & EW of first CaT line ($\lambda = 8498.0\,\AA$)\\
32 & ew\_w2        & \mbox{\AA} & EW of second CaT line  ($\lambda = 8542.1\,\AA$) \\
33 & ew\_w3 & \mbox{\AA} & 	EW of third CaT line ($\lambda =  8662.1\,\AA$)\\
34 & ew\_gl & $\cdots$  & 1-2=Gaussian fit for CaT EW, 3-4=Gauss-Lorentz fit; 1/3 = fit with all 3 CaT lines, \\
   &        &           & 2/4 = fit with 2 CaT lines and model described in \S\,\ref{ssec: ew_cat}\\
35 & gaia\_source\_id	& $\cdots$  & Gaia DR3 source identifier\\
36 & gaia\_pmra	      & mas yr$^{-1}$ &  Gaia DR3 proper motion in right ascension direction\\
37 & gaia\_pmra\_err  & mas yr$^{-1}$  &  Gaia DR3 standard error of proper motion in right ascension direction\\
38 & gaia\_pmdec	  & mas yr$^{-1}$  & Gaia DR3 proper motion in declination direction\\
39 & gaia\_pmdec\_err & mas yr$^{-1}$  & Gaia DR3 standard error of proper motion in declination direction\\
40 & gaia\_pmra\_pmdec\_corr & $\cdots$  &  Gaia DR3 correlation between proper motion in right ascension and proper motion in declination\\
41 & gaia\_parallax	& mas & {\it Gaia} DR3 parallax\\
42 & gaia\_parallax\_err	& mas  & {\it Gaia} DR3 standard error of parallax\\
43 & gaia\_aen& $\cdots$  & {\it Gaia} DR3 astrometric excess noise \\
44 & gaia\_aen\_sig& $\cdots$  &  {\it Gaia} DR3 significance of astrometric excess noise\\
45 & flag\_coadd &  $\cdots$ & Flag set if velocity is from coadded DEIMOS spectra\\
46 & flag\_var & $\cdots$ & Flag set if multiepoch velocities measurements are variable, 0=not variable, -99 = no multiepoch data\\
47 & flag\_gaia & $\cdots$ & Flag set if star has a Gaia DR3 match\\
48 & flag\_HB & $\cdots$ & Flag set if star is on the HB\\
49 &  Pmem & $\cdots$ & Membership probability, suggested secure members have Pmem $> 0.5$\\
50 &  Pmem\_novar & $\cdots$ &  Binary membership (1=member, 0=nonmember), velocity variables removed
\enddata
\tablecomments{Schema for all targeted objects presented in this paper.  This table includes all data presented in Table~\ref{table_stars} and Table~\ref{table_exgal} with additional information.  The full contents of this table (\nall rows) are available in a machine-readable format. }
\end{deluxetable*}
%%%%%%%%%%%%%%%%%%%%%%%%%%%%%%%%%%%%%%%%

\end{document}